\documentclass[fleqn,usenatbib]{mnras}

\usepackage{newtxtext,newtxmath}

\usepackage[T1]{fontenc}

\DeclareRobustCommand{\VAN}[3]{#2}
\let\VANthebibliography\thebibliography
\def\thebibliography{\DeclareRobustCommand{\VAN}[3]{##3}\VANthebibliography}

\usepackage{graphicx}	
\usepackage{amsmath}	
\usepackage{amsmath}
\usepackage{braket}
\usepackage{lipsum}
\usepackage{float}
\usepackage{xcolor}
\usepackage{epsfig}
\usepackage{float}
\usepackage{placeins}
\usepackage{mwe}
\usepackage{ulem}
\usepackage{appendix}
\usepackage{lscape}

\title[PKS 0402-362]{Gamma-ray flares and broadband spectral study of PKS 0402-362}

\author[Das et al.]{
Avik Kumar Das,$^{1}$\thanks{E-mail:avikdas@iisermohali.ac.in}
Sandeep Kumar Mondal,$^{2}$ 
and Raj Prince$^{3}$
\\
$^{1}$Department of Physical Sciences, Indian Institute of Science Education and Research Mohali, Knowledge City, Sector 81, SAS Nagar, Punjab 140306, India\\
$^{2}$Raman Research Institute, C. V. Raman Avenue, Sadashivnagar, Bangalore: 560080, India\\
$^{3}$Center for Theoretical Physics, Polish Academy of Sciences, Al.Lotnikow 32/46, 02-668, Warsaw, Poland \\
}

\date{Accepted XXX. Received YYY; in original form ZZZ}

\pubyear{2023}

\begin{document}
\label{firstpage}
\pagerange{\pageref{firstpage}--\pageref{lastpage}}
\maketitle

\begin{abstract}
We study the long-term behavior of the bright gamma-ray blazar PKS 0402-362. We collected approximately 13 years of Fermi-LAT data between Aug 2008 to Jan 2021 and identified three bright $\gamma$-ray activity epochs. The second was found to be the brightest epoch ever seen in this source. We observed most of the $\gamma$-ray flare peaks to be asymmetric in profile suggesting a slow cooling time of particles or the varying Doppler factor as the main cause of these flares. The $\gamma$-ray spectrum is fitted with PL and LP spectral models, and in both cases, the spectral index is very steep. The $\gamma$-ray spectrum does not extend beyond 10 GeV energy suggesting the emission is produced within the BLR. The absence of time lags between optical-IR and $\gamma$-ray suggest one zone emission model. Using the above information, we have modeled the broadband SED with a simple one-zone emission model using the publicly available code `GAMERA'. The particle distribution index is found to be the same as expected in diffusive shock acceleration suggesting it as the main mechanism of particle acceleration to very high energy up to 4 - 6 GeV. Throughout the various flux phases, we noticed that the optical emission is dominated by the thermal disk, suggesting it to be a good source to examine the disk-jet coupling. We postulate that the observed broadband flares could be linked with perturbation produced in the disk, which propagates to the jet and interacts with the standing shock. However, a more detailed examination is required. 
  
\end{abstract}




\section{Introduction}
With the recent EHT observations of M87 and SgA* in the Milky-way, It is proven that each galaxy hosts a supermassive black hole (SMBH) at its center. In some cases the SMBH is active and it accretes the matter from the surrounding classified as an active galaxy. The central part of these active galaxies is known as active galactic nuclei or AGN. AGN consists of three main parts, the accretion disk, a broad-line region (BLR), and a molecular or dusty torus.
In some cases, AGN also possesses bipolar relativistic jets propagating to a kpc scale. These sources are randomly oriented in the universe and the sources with highly collimated relativistic jets pointing towards the earth are classified as blazars \citep{Urry1995Sep} through an AGN unification scheme. 
Blazars have been observed throughout the entire electromagnetic spectrum (EM) with various ground-based and space-based telescopes covering the energy range of low-frequency radio to TeV energy $\gamma$-ray. 
The observational property of blazar reveals a high flux variability with a variability time scale of the order of minutes to years \citep{Goyal2018Mar}. In addition to this, many blazars also show high optical polarization variability and significant change in polarization angle (\citealt{Blinov2015Oct}, \citealt{Avachat2016Nov}, \citealt{Kiehlmann2016Jun}). 
\begin{figure*}

\centering
\includegraphics[height=2.6in,width=5.3in]{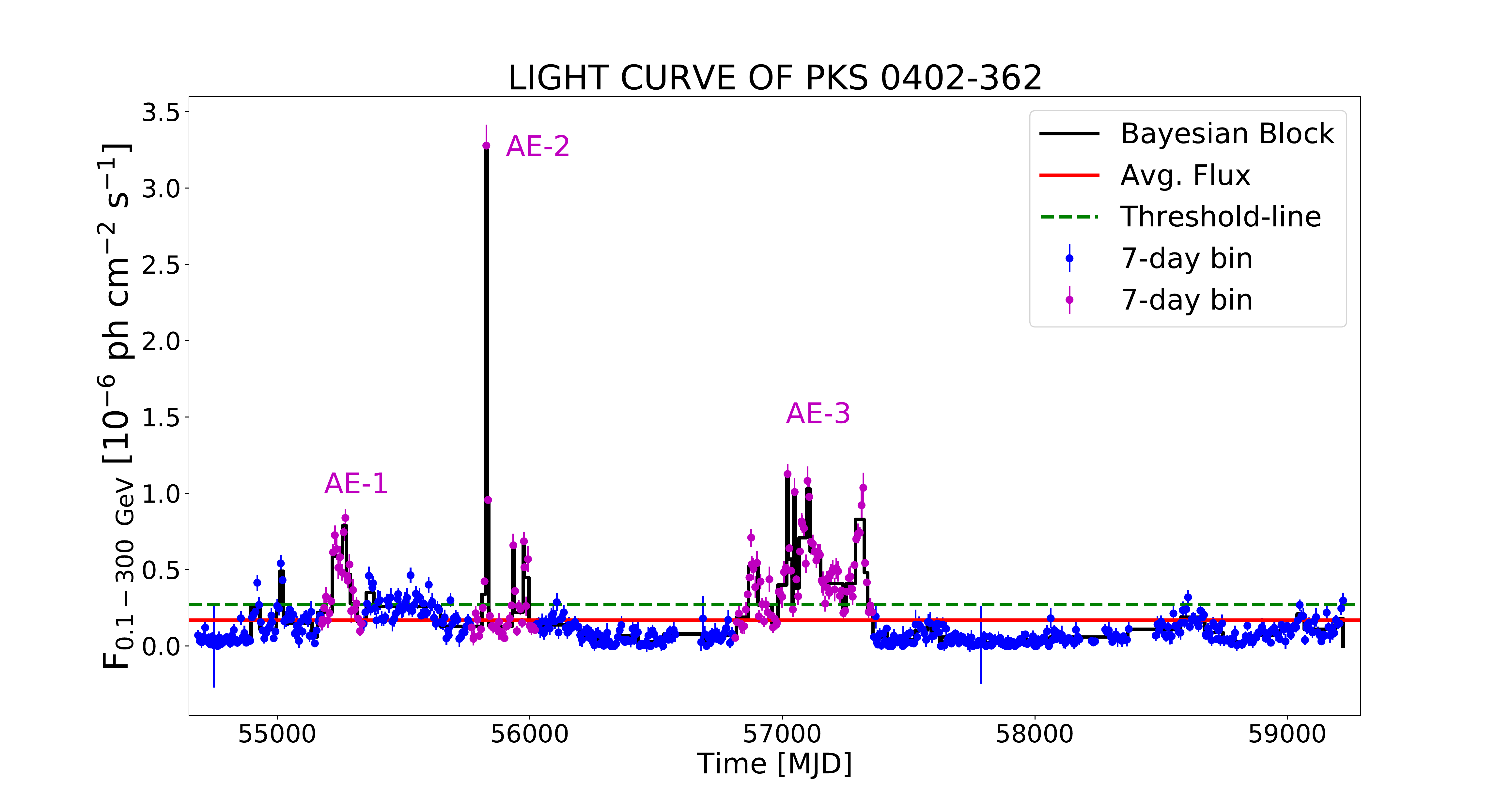}
\caption{Fermi-LAT light curve of PKS 0402-362 (MJD 54682 - 59220) in seven-day binning. Three major Flares have been identified, which are shown by magenta color points.}
\label{fig:1}
\end{figure*}
 The observed broadband spectral energy distribution (SED) has two characteristics hump peaking in low and high energy bands. The low-energy hump covers the infrared to soft X-ray part of the EM spectrum which is produced by the synchrotron emission. The high-energy hump spread over the soft X-ray to TeV energy in gamma-ray and can be explained by the leptonic or the hadronic processes. In the leptonic scenario, the high-energy leptons upscattered the low-energy photons which can be the synchrotron photons themselves, the process is named Synchrotron Self-Compton \citep{1992ApJ...397L...5M} or can be supplied from outside the jet such as from the broad-line region (BLR) or dusty torus (DT) or accretion disk and commonly recognized as external Compton (\citealt{1993ApJ...416..458D}, \citealt{1994ApJ...421..153S}). However, the hadronic model involves 
 proton synchrotron (\citealt{2000NewA....5..377A}, \citealt{2001APh....15..121M}) or proton-photon  (\citealt{1992A&A...253L..21M}, \citealt{1999MNRAS.302..373B}, \citealt{2002PhRvD..66l3003N}) or proton-proton interactions (\citealt{PhysRevD.47.5270}, \citealt{1997MNRAS.287L...9B}, \citealt{PhysRevD.80.083008}). The important thing to note here is that in the pp or p$\gamma$ interaction along with high energy $\gamma$-ray, the neutrinos are also expected to produce. Therefore, the hadronic process is most suitable for sources where neutrinos are expected. A single zone leptonic SED modeling has been done on many sources to understand the broadband emission (\citealt{Prince_2018, Prince2019, Prince_2020, 2020ApJS..248....8D, 2021A&A...654A..38P}). In some cases, the one-zone model fails to produce the broadband SED, and hence two-zone emission model is preferred (\citealt{Prince_2019, Das_2021}). \\
Blazars are also classified into two broad classes namely flat spectrum radio quasars (FSRQ) and BL Lacertae (BL Lacs) objects based on the presence or absence of strong emission lines in their optical spectra, respectively. 

In the last 1.5 decades, the $\gamma$-ray astronomy has grown extensively thanks to the Fermi-LAT instrument. The recently published Fermi catalog (4FGL DR3) reports more than 5000 sources as $\gamma$-ray emitters and 90$\%$ of them are identified as blazars. This provides a unique opportunity to study these sources and answer some basic questions that astronomers are puzzled about such as what is the main mechanism of producing the $\gamma$-ray, how the particles are accelerated to very high energy, where the emission site is located in the jet, how the emissions are correlated with each other 
and many more. We make an effort to answer some of the questions and present a detailed temporal and spectral study on blazar PKS 0402-362. Blazar, PKS 0402-362 is an FSRQ type source located at redshift, $z$ = 1.4228 \citep{Jones2009Oct} with RA = 60.975, Dec = -36.087. It has been continuously monitored by Fermi-LAT since the beginning of the operation and has been a consistent gamma-ray emitter. In the 4FGL DR3 catalog, it has been recognized as 4FGL J0403.9-3605.

  
In this paper, we have analyzed the Fermi-LAT data collected over a period of 12 years from this source (ref section \S\ref{sec:2}) and identified the flares in its long-term light curve (ref section \S\ref{sec:3}). The gamma-ray SED and correlation for all the episodes are presented in section \S\ref{sec:4} and section \S\ref{sec:5}, respectively. Subsequently, we discussed the broadband SED modeling in section \S\ref{sec:6} and detailed discussions and conclusions are presented in section \S\ref{sec:7} and section \S\ref{sec:8}.

\section{DATA ANALYSIS} \label{sec:2}

\subsection{FERMI-LAT ANALYSIS}
Fermi-LAT (Fermi-Large Area Telescope) is an imaging pair conversion $\gamma$-ray telescope, which covers the energy range from 50 MeV to 1 TeV \citep{https://doi.org/10.48550/arxiv.2005.11208} with a wide field of view of $>$ 2.4 sr \citep{2009ApJ...697.1071A}. It has an orbital period of 96 minutes and scans the entire sky every two orbits ($\sim$ 3.2 hours). The complete characteristics of LAT are provided on the Fermi Webpage \footnote{\url{https://fermi.gsfc.nasa.gov/ssc/data/analysis/software/}}. 

We have used Fourth Source Catalog (4FGL-DR2) for the Fermi-LAT analysis. We have downloaded\footnote{\url{https://fermi.gsfc.nasa.gov/cgi-bin/ssc/LAT/LATDataQuery.cgi}} the Pass8R2 data of an FSRQ, namely, PKS 0402-362 (4FGL J0403.9-3605) from August 2008 to Jan 2021 ($\sim$ 12.5 years) and analyzed it with the help of Fermi science tool (version- 1.0.10). We have used `evclass=128' and `evtype=3' to analyze the data with energies ranging from 100 MeV to 300 GeV. A maximum zenith angle cut of 90$^\circ$ (standard value provided by the LAT team) has been used to avoid the $\gamma$-ray contamination from the earth's limb. Filter expression ``\texttt{DATA\_QUAL}$>$0 \&\& \texttt{LAT\_CONFIG}==1 \&\& \texttt{ANGSEP}(\texttt{{}RA\_SUN},\texttt{DEC\_SUN},60.974,-36.0839)$>$15" is implemented to select the good time interval data and also to ignore the time interval when the Sun is within the 15$^\circ$ from the source. The galactic diffuse emission model (\texttt{gll\_iem\_v07.fits}) and isotropic diffuse emission model (\texttt{iso\_P8R3\_SOURCE\_V3\_v1.tx}) are used as background models along with other point sources 
We have used the preferred method, ``Unbinned likelihood analysis" and kept fixed all the source parameters that are away from the Region of Interest or ROI (10$^{\circ}$) for the analysis. We have further followed the same procedure for the analysis as given in \cite{2020ApJS..248....8D}.

\begin{table*}
\caption{Results of Swift-XRT fit. The first and second column represent the Obs-ID and statistic used in fitting to calculate the model flux, respectively. $\chi^{2}$ or C-value of the fit (for improved/best-fitted model) with d.o.f is also given in the second column. Third and Fourth column represent the fit results for log parabola ($\alpha$ and $\beta$) and power-law ($\Gamma$) models, respectively. The last two columns (fifth and sixth) represent the value of X-ray flux (calculated using `cflux' model with energy range 0.3 - 10.0 keV) and F-test probability, respectively. Only the `power-law' model has been used for `C-stat' cases to compute the flux (see text for more details).} 
\label{tab:1}
\centering
\begin{tabular}{ccccc rrrr}   
\\
\hline\hline                        
Observation-id & Statistics & \multicolumn{2}{c}{Log-parabola fit results}  & Power-law fit results & X-ray Flux & F-test prob.\\ [0.8ex] 
& & $\alpha$ & $\beta$ & $\Gamma$ & [$10^{-11}$ erg cm$^{-2}$ s$^{-1}$]\\
\hline
\hline
00033354001 & $\chi^{2}$ (26.94/29) & 1.28$^{+0.16}_{-0.18}$ & 0.77$^{+0.38}_{-0.34}$ & 1.55$^{+0.09}_{-0.09}$ & 0.66$^{+0.06}_{-0.06}$ & 0.001\\ 
\hline
00033354002 & $\chi^{2}$ (27.49/30) & 1.42$^{+0.19}_{-0.21}$ & 0.35$^{+0.36}_{-0.33}$ & 1.59$^{+0.10}_{-0.09}$ & 0.60$^{+0.06}_{-0.05}$ & 0.07\\
\hline
00036504001 & $\chi^{2}$ (27.18/21) & 1.56$^{+0.17}_{-0.19}$ & 0.06$^{+0.36}_{-0.33}$ & 1.58$^{+0.12}_{-0.12}$ & 0.47$^{+0.05}_{-0.05}$ & 0.80\\ 
\hline
00036504002 & $\chi^{2}$ (27.79/53) & 1.64$^{+0.56}_{-0.88}$ & -0.67$^{+1.96}_{-1.58}$ & 1.44$^{+0.55}_{-0.53}$ & 0.54$^{+0.18}_{-0.13}$ & 0.38\\
\hline
00036504003 & $\chi^{2}$ (11.00/20) & 2.09$^{+1.73}_{-2.09}$ & -0.15$^{+0.15}_{-2.50}$ & 2.05$^{+1.27}_{-0.94}$ & 0.15$^{+0.06}_{-0.05}$ & 1.00\\
\hline
00036504005 & $\chi^{2}$ (19.51/23) & 1.56$^{+0.17}_{-0.19}$ & -0.16$^{+0.35}_{-0.32}$ & 1.49$^{+0.12}_{-0.12}$ & 2.88$^{+0.30}_{-0.28}$ & 0.42\\
\hline
00036504006 & $\chi^{2}$ (94.23/72) & 1.40$^{+0.09}_{-0.10}$ & 0.07$^{+0.18}_{-0.17}$ & 1.43$^{+0.06}_{-0.06}$ & 2.60$^{+0.16}_{-0.16}$ & 0.55\\
\hline
00036504007 & $\chi^{2}$ (63.70/52) & 1.42$^{+0.10}_{-0.11}$ & 0.15$^{+0.22}_{-0.21}$ & 1.47$^{+0.07}_{-0.07}$ & 2.21$^{+0.17}_{-0.17}$ & 0.60\\
\hline
00036504008 & $\chi^{2}$ (36.55/38) & 1.46$^{+0.13}_{-0.14}$ & -0.21$^{+0.28}_{-0.26}$ & 1.54$^{+0.08}_{-0.08}$ & 1.99$^{+0.17}_{-0.17}$ & 0.17\\
\hline
00036504009 & $\chi^{2}$ (99.84/92) & 1.17$^{+0.09}_{-0.11}$ & 0.34$^{+0.19}_{-0.18}$ & 1.32$^{+0.06}_{-0.06}$ & 1.56$^{+0.09}_{-0.09}$ & 0.003\\
\hline
00036504010 & $\chi^{2}$ (37.50/36) & 1.35$^{+0.15}_{-0.17}$ & 0.21$^{+0.33}_{-0.30}$ & 1.43$^{+0.10}_{-0.10}$ & 1.11$^{+0.12}_{-0.12}$ & 0.26\\
\hline
00049661001 & C (97.21/127) & 1.75$^{+0.74}_{-0.86}$ & -0.35$^{+1.31}_{-1.13}$ & 1.54$^{+0.28}_{-0.26}$ & 2.80$^{+0.59}_{-0.49}$ & - \\
\hline
00049661002 & $\chi^{2}$ (15.32/10) &  1.31$^{+0.44}_{-0.62}$ & 0.09$^{+1.14}_{-0.92}$ & 1.35$^{+0.24}_{-0.23}$ & 1.76$^{+0.53}_{-0.45}$ & 0.92\\
\hline
00049661003 & C (16.74/21) & 2.32$^{+0.52}_{-0.59}$ & -1.27$^{+1.49}_{-1.35}$ & 2.07$^{+0.68}_{-0.56}$ & 0.33$^{+0.11}_{-0.07}$ & -\\
\hline
00049661004 & C (25.69/35) & 1.64$^{+0.58}_{-0.70}$ & -0.76$^{+1.24}_{-1.14}$ & 1.31$^{+0.43}_{-0.44}$ & 0.60$^{+0.24}_{-0.15}$ & -\\
\hline
00049661005 & C (17.75/29) & 1.76$^{+0.57}_{-0.72}$ & -0.65$^{+1.64}_{-1.38}$ & 1.58$^{+0.53}_{-0.53}$ & 0.40$^{+0.20}_{-0.12}$ & -\\
\hline
00049661007 & C (7.72/9) & 0.72$^{+1.26}_{-0.72}$ & -0.70$^{+4.03}_{-0.70}$ & 0.50$^{+1.15}_{-0.49}$ & 0.95$^{+19.5}_{-0.69}$ & -\\
\hline
00049661008 & C (10.60/16) & 1.50$^{+0.83}_{-1.13}$ & -0.53$^{+3.66}_{-0.53}$ & 1.37$^{+0.81}_{-0.79}$ & 0.48$^{+0.41}_{-0.19}$ & -\\
\hline
00049661009 & C (31.95/30) & 1.66$^{+0.60}_{-0.78}$ & 0.48$^{+1.65}_{-1.34}$ & 1.81$^{+0.46}_{-0.46}$ & 0.22$^{+0.12}_{-0.08}$ & -\\
\hline
00049661011 & C (78.42/64) & 1.42$^{+0.31}_{-0.35}$ & -0.03$^{+0.62}_{-0.56}$ & 1.40$^{+0.22}_{-0.21}$ & 0.46$^{+0.10}_{-0.09}$ & -\\
\hline
00049661012 & C (38.62/45) & 1.28$^{+0.54}_{-0.63}$ & 0.48$^{+1.18}_{-1.07}$ & 1.47$^{+0.35}_{-0.35}$ & 0.44$^{+0.20}_{-0.15}$ & -\\
\hline
00049661013 & C (51.53/64) & 0.43$^{+0.67}_{-0.43}$ & 1.91$^{+1.34}_{-1.27}$ & 1.27$^{+0.29}_{-0.30}$ & 0.25$^{+0.13}_{-0.10}$ & -\\
\hline
00049661014 & C (40.90/60) & 1.67$^{+0.54}_{-0.66}$ & -0.30$^{+1.07}_{-0.95}$ & 1.53$^{+0.37}_{-0.37}$ & 0.36$^{+0.12}_{-0.09}$ & -\\
\hline
00049661015 & C (9.25/15) & 2.96$^{+0.94}_{-1.09}$ & -2.39$^{+2.05}_{-2.39}$ & 1.94$^{+0.90}_{-0.70}$ & 0.72$^{+0.17}_{-0.10}$ & -\\
\hline
00049661016 & $\chi^{2}$ (9.03/7) & 1.11$^{+0.53}_{-0.69}$ & 0.39$^{+1.38}_{-1.14}$ & 1.28$^{+0.25}_{-0.25}$ & 0.41$^{+0.12}_{-0.10}$ & 0.67\\
\hline
00049661017 & $\chi^{2}$ (2.01/6) & 1.47$^{+0.44}_{-0.53}$ & 0.27$^{+0.96}_{-0.81}$ & 1.60$^{+0.22}_{-0.21}$ & 0.33$^{+0.06}_{-0.06}$ & 0.41\\
\hline
00080950001 & $\chi^{2}$ (16.15/12) & 1.74$^{+0.26}_{-0.29}$ & -0.28$^{+0.51}_{-0.47}$ & 1.61$^{+0.18}_{-0.17}$ & 0.42$^{+0.06}_{-0.05}$ & 0.45 \\
\hline
00080950002 & $\chi^{2}$ (8.52/7) & 1.28$^{+0.35}_{-0.42}$ & 0.33$^{+0.93}_{-0.79}$ & 1.40$^{+0.22}_{-0.22}$ & 0.32$^{+0.08}_{-0.07}$ & 0.58\\
\hline
00035523001 & $\chi^{2}$ (7.91/8) & 1.63$^{+0.24}_{-0.29}$ & -0.66$^{+0.65}_{-0.59}$ & 1.44$^{+0.24}_{-0.24}$ & 0.54$^{+0.09}_{-0.07}$ & 0.09\\
\hline
00035523002 & C (79.75/86) & 1.67$^{+0.37}_{-0.42}$ & -0.16$^{+0.86}_{-0.79}$ & 1.61$^{+0.28}_{-0.27}$ & 0.36$^{+0.09}_{-0.07}$ & -\\

\hline   
\hline
\end{tabular}
\end{table*}

\begin{figure*}
\centering

\includegraphics[height=2.6in,width=5.3in]{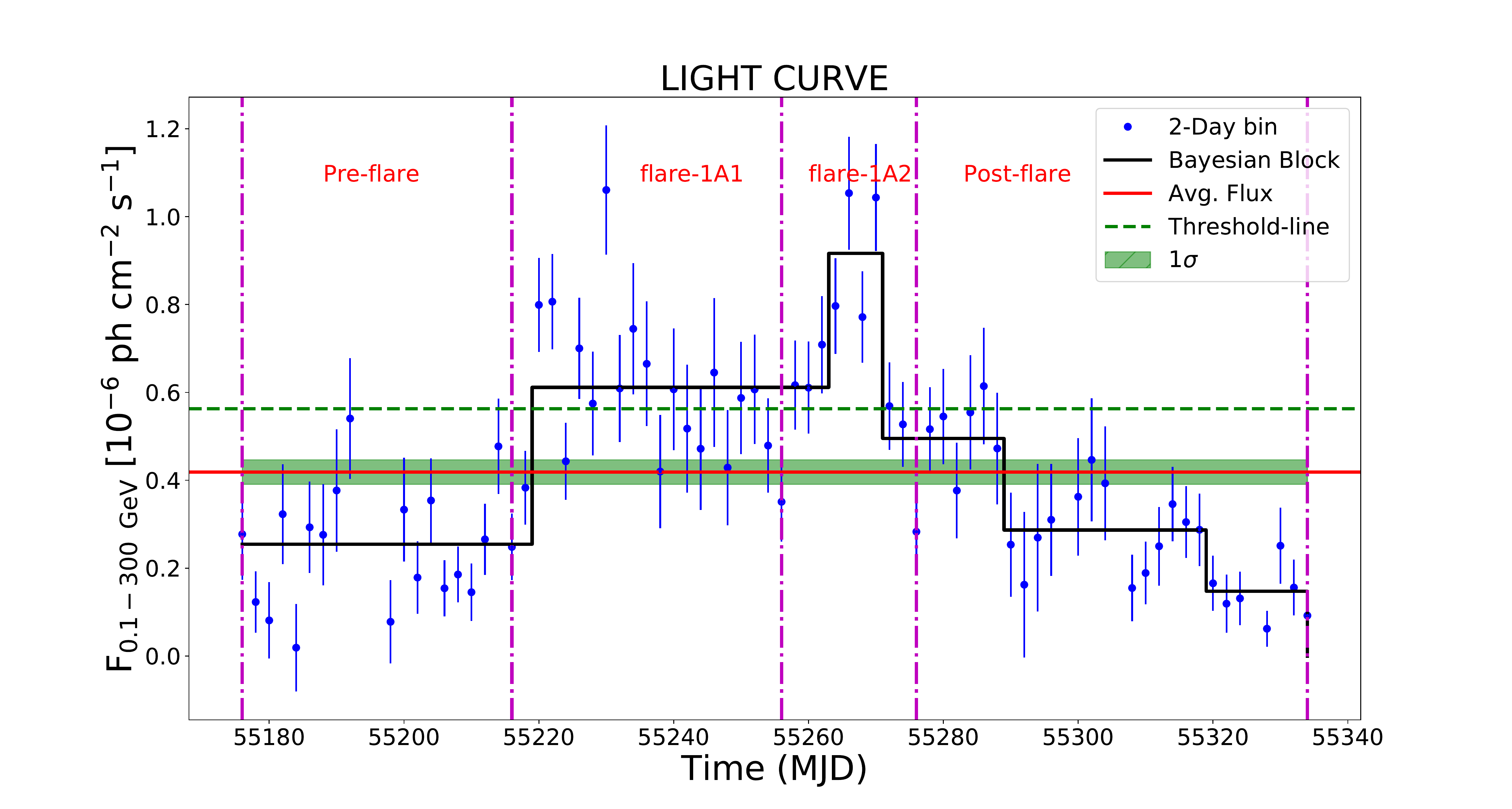}

\includegraphics[height=2.6in,width=5.3in]{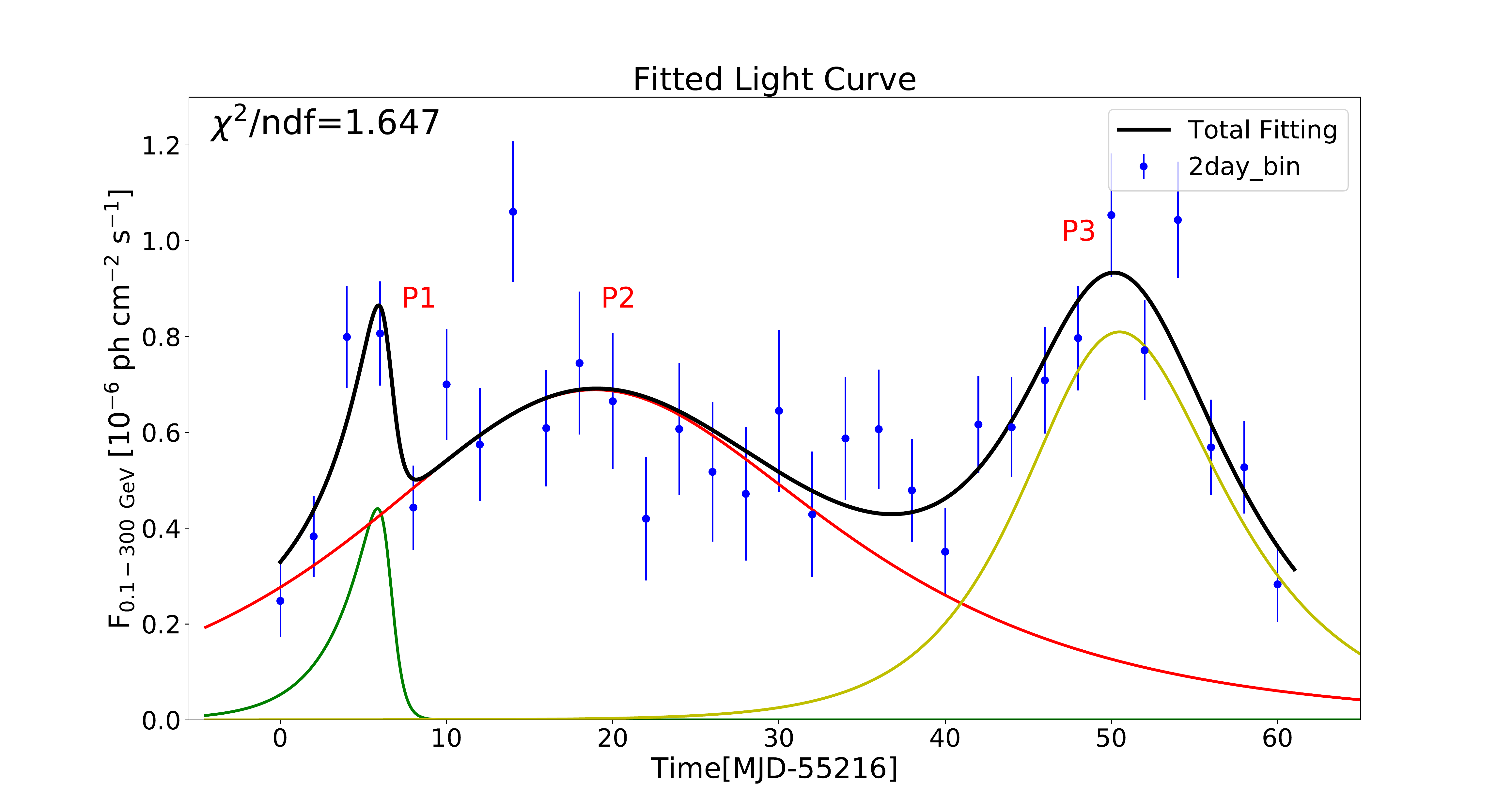}
\caption{\textbf{Upper panel}: Two-day binning light curve of AE-1A. The time duration of the different phases is MJD 55176-55216 (Pre-flare), MJD 55216-55256 (flare-1A1), MJD 55256 - 55276 (flare-1A2), and MJD 55276-55334 (Post-flare). These are shown by a dash-dot magenta line. \textbf{Lower panel}: The fitted light curve of flare-1A1 and flare-1A2 with a time span of 40 days (MJD 55216 - 55256) and 20 days (MJD 55256 - 55276), respectively. Here, the light curve of both phases is shown in one figure.}
\label{fig:2}
\end{figure*}
 
\subsection{SWIFT-XRT/UVOT}
We have analyzed the X-ray and Ultraviolet-Optical data of the Neil Gehrels Swift Observatory (Swift-XRT/UVOT), which are taken during the same period as $\gamma$-ray observation from HEASARC webpage\footnote{\url{https://heasarc.gsfc.nasa.gov/cgi-bin/W3Browse/swift.pl}}. The task `$xrtpipeline$' (version - 0.13.2.) is used to process the XRT-data \citep{2005SSRv..120..165B} with an energy range from 0.3 to 10.0 keV for each observation-id. The calibration files with version 20160609 and standard screening criteria have been implemented in this process. We have chosen a circular radius of 25 arc seconds and the background region as an annular ring around the source to analyze these data. `$xrtmkarf$' and `$grppha$' has been used to create the ancillary response file and group the spectra, respectively. Next, these grouped spectra have been fitted in XSPEC (version - 12.11.0) with two models - `$powerlaw$' and `$logparabola$'.  The log parabola model was chosen as the best-fitted model if the F-test probability $\leq$ 0.05. Due to low photon counts, we grouped the spectra with at least one count per bin for a few observation ids. We employed C-statistics and used the `$powerlaw$' model to fit those cases' spectra. To account for galactic absorption, we used the model `$tbabs$' with the neutral hydrogen column density ($n_{H}$) of $  6.92\times10^{19} cm^{-2}$ as given in HEASARC webpage \footnote{\url{https://heasarc.gsfc.nasa.gov/cgi-bin/Tools/w3nh/w3nh.pl}} (i.e., HI 4 Pi Survey). The detailed results of the fit are shown in Table-\ref{tab:1}.

The observation was also made for this source by the UVOT \citep{2005SSRv..120...95R}. The source and background have been chosen as a circular region of 5 arc seconds and an annular region around the source, respectively, for all six filters: V, B, U, W1, M2, and W2. The task called `$uvotsource$' is used to extract the source magnitudes. After that, `$uvotimsum$' is used, to sum up, more than one observation. These results are then corrected for galactic reddening 
\citep{2011ApJ...737..103S}, and the atmospheric extinction. Finally, the magnitudes are converted into flux by using the zero points \citep{2011AIPC.1358..373B} and proper conversion factors \citep{2016MNRAS.461.3047L}.

\subsection{SMARTS and Steward Data}
We have also used publicly available archival data of SMARTS\footnote{\url{http://www.astro.yale.edu/smarts/glast/home.php}} Optical-IR observation and Steward Optical Observatory, Arizona \footnote{\url{http://james.as.arizona.edu/ psmith/Fermi/}}, which are a part of {\it Fermi} Blazars monitoring program\citep{Bonning2012Aug}. SMARTS provides photometric data in the BVRJ band. Steward observatory provides both photometric (in V and R band) and polarimetric data \citep{2009arXiv0912.3621S}. However, there is no photometric data available on the steward's website. We have collected BVRJ data of this source during MJD 55838 - 57297.

\begin{table*}
\caption{Different fitting parameters value for the flaring phases. $F_{0}$ is the flux at time $t_{0}$. $T_{r}$, and $T_{d}$ are the rising and decay timescale of the peaks, respectively. The sixth column represents the time ($t_{m}$) at maximum flux. The last column represents the asymmetry parameter of the peaks.} 
\label{tab:2}
\centering
\begin{tabular}{ccccc rrrr}   
\\
\hline\hline                        
Peak & $t_0$ & $F_0$ & $T_r$ & $T_d$ & $t_m$ & K \\ [0.8ex] 
 & MJD & [10$^{-6}$ ph cm$^{-2}$ s$^{-1}$] & [day] & [day] & [MJD]\\
\hline
& & & AE-1A\\
\hline
P1 & 6.50 &  0.33$\pm$0.21 & 2.57$\pm$1.67 &  0.42$\pm$0.78 & 5.85$\pm$0.48 & -0.72$\pm$0.48\\
P2 & 18.00 &  0.69$\pm$0.07 & 11.62$\pm$3.91 &  13.46$\pm$4.29 & 18.91$\pm$2.87 & 0.07$\pm$0.23\\
P3 & 49.80 &  0.80$\pm$0.11 & 4.78$\pm$1.35 &  6.18$\pm$1.15 & 50.49$\pm$0.86 & 0.13$\pm$0.17 \\
\hline
& & & AE-2A\\
\hline
P1 & 5.09 &  6.07$\pm$0.60 & 1.35$\pm$0.28 &  0.82$\pm$0.20 & 4.84$\pm$0.32 & -0.24$\pm$0.15 \\
\hline
& & & AE-2B\\
\hline
P1 & 4.98 &  0.77$\pm$0.09 & 1.51$\pm$0.26 &  3.71$\pm$0.53 & 5.95$\pm$0.20 & 0.42$\pm$0.09 \\
P2 & 20.21 & 1.47$\pm$0.26 & 0.48$\pm$0.11 &  1.20$\pm$0.35 & 20.52$\pm$0.08 &  0.43$\pm$0.15\\
\hline
& & & AE-3A\\
\hline
P1 & 8.75 &  0.88$\pm$0.12 & 2.47$\pm$0.38 &  0.77$\pm$0.19 & 8.07$\pm$0.12 & -0.52$\pm$0.11 \\
P2 & 19.50 &  0.57$\pm$0.09 & 6.68$\pm$1.81 &  0.96$\pm$0.34 & 17.87$\pm$0.35 & -0.75$\pm$0.10\\
\hline
& & & AE-3B\\
\hline
P1 & 3.94 &  1.62$\pm$0.23 & 2.03$\pm$0.34 &  1.26$\pm$0.30 & 3.57$\pm$0.20 & -0.23$\pm$0.14 \\
P2 & 14.24 &  0.86$\pm$0.00 & 3.34$\pm$0.85 &  3.48$\pm$0.73 & 14.31$\pm$0.56 & 0.02$\pm$0.16 \\
\hline
& & & AE-3C\\
\hline
P1 & 1.76 &  1.04$\pm$0.20 & 0.43$\pm$0.12 &  3.26$\pm$0.42 & 2.53$\pm$0.10 & 0.77$\pm$0.06 \\
P2 & 6.84 &  0.74$\pm$0.17 & 1.91$\pm$0.89 &  0.29$\pm$0.11 & 6.37$\pm$0.16 & -0.74$\pm$0.14\\
\hline
& & & AE-3D\\
\hline
P1 & 0.63 &  0.70$\pm$0.09 & 0.94$\pm$0.36 &  7.60$\pm$2.71 & 2.38$\pm$0.46 & 0.78$\pm$0.10 \\
P2 & 30.98 &  1.54$\pm$0.24 & 0.60$\pm$0.15 &  3.09$\pm$0.64 & 31.10$\pm$1.01 & 0.67$\pm$0.08 \\
\hline   
\hline
\end{tabular}
\end{table*}

\section{Gamma-ray Light curves} \label{sec:3}
We have analyzed the 12.5 years (MJD 54682 - 59220) Fermi-LAT data of PKS 0402-362 in 7 days of binning. Depending on the simultaneous observation in Swift-XRT/UVOT, three major Activity Epochs (AE) have been identified for study, namely, AE-1, AE-2 and AE-3 (shown in Figure-\ref{fig:1}). These epochs are analyzed in one day binning for further study. We have used Bayesian Block (BB) \citep{2013ApJ...764..167S} representation \footnote{\url{https://docs.astropy.org/en/stable/api/astropy.stats.bayesian\_blocks.html}} in this finer binned (one-day or two-day) light curve to identify different sub-structures: AE-2A, AE-2B, AE-3A, etc. We have restricted our analysis to two-day time bins for four sub-structures because of significant flux errors observed in the one-day time bin light curve. Finally, these sub-structures are divided into different phases (e.g. Pre-flare, flare, Post-flare, etc.) depending on the BB flux level. In order to identify the time duration of the flare phase, we have made use of Ivan Kramarenko's algorithm. This method uses an iterative approach to divide the light curve data points into two sets, low state set and high state or anti set. The threshold value is defined as $<F> + 2 \times F_{Disp}$, where $<F>$ is the mean flux value of the light curve and $F_{Disp}$ is the true dispersion value of the low flux state. More details of the algorithm can be found in \citet{Geng2020Nov}. The high flux or flare phase is considered when the flux level goes above the threshold value (e.g., dashed green line in upper panel of Figure-\ref{fig:2}).

Each flare phase's light curve consists of several peaks. We fitted these peaks with the following (sum of exponential) function \citep{2010ApJ...722..520A} to compute the rising ($T_r$) and decay time ($T_d$) of each peak:

\begin{equation} \label{eq:1}
F(t)=2\sum_{i}^{n}F_{0,i}\Big[\exp \Big(\frac{t_{0,i}-t}{T_{r,i}}\Big)+\exp \Big(\frac{t-t_{0,i}}{T_{d,i}}\Big)\Big]^{-1} 
\end{equation}

Where, $i$ runs over the number of major peaks (n).$F_0$ is the photon flux at time $t_{0}$ for a particular peak. The time ($t_{m}$) at the maximum flux value is given by:

\begin{equation} \label{eq:2}
t_{m} = t_{0} + \frac{t_{r} t_{d}}{t_{r}+t_{d}} \ln({\frac{t_{d}}{t_{r}}}) 
\end{equation}

We have used Bayesian Information Criteria or BIC value to choose the reasonable number of exponential functions (Equation-\ref{eq:1}) to fit the light curve. It is defined as \citep{Meyer2019May}:

\begin{equation}\label{eq:3}
    BIC = f\ln{N} + \chi^{2}
\end{equation}

Where f and N are the numbers of free parameters of the fit and the total number of data points, respectively. There are four parameters per exponential function in our case. We have chosen one, two, or three exponential functions together to fit the light curve of the flare phase for which the combination of the $BIC$ value is minimum.  

The results of our fitting and value of $t_{m}$ have been described in Table-\ref{tab:2}. In this paper, all the reported $\gamma$-ray photon fluxes are in the unit of $10^{-6}$ ph cm$^{2}$ s$^{-1}$. 

\begin{figure*}
\centering

\includegraphics[height=1.90in,width=2.6in]{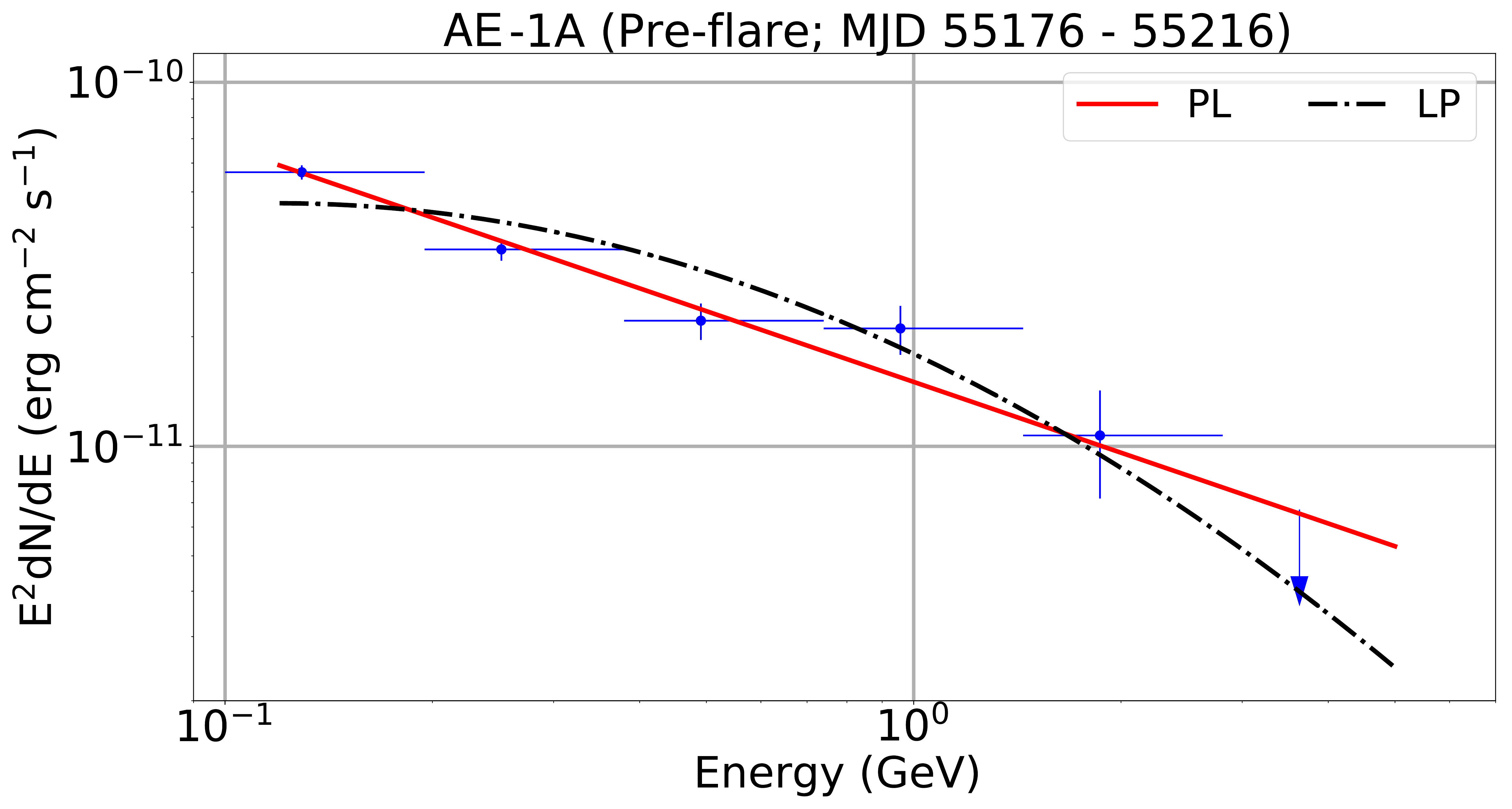}
\includegraphics[height=1.90in,width=2.6in]{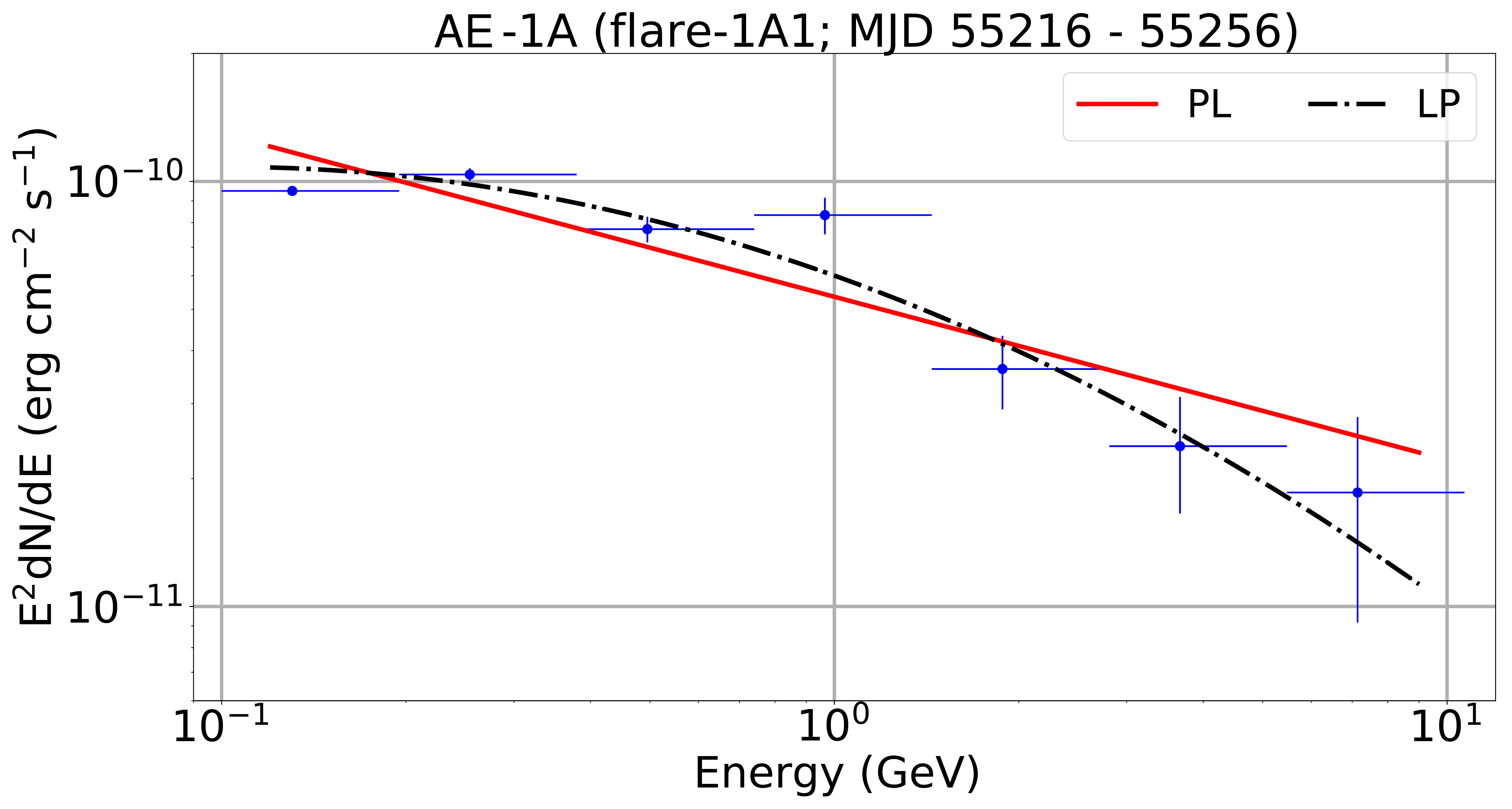}
\includegraphics[height=1.90in,width=2.6in]{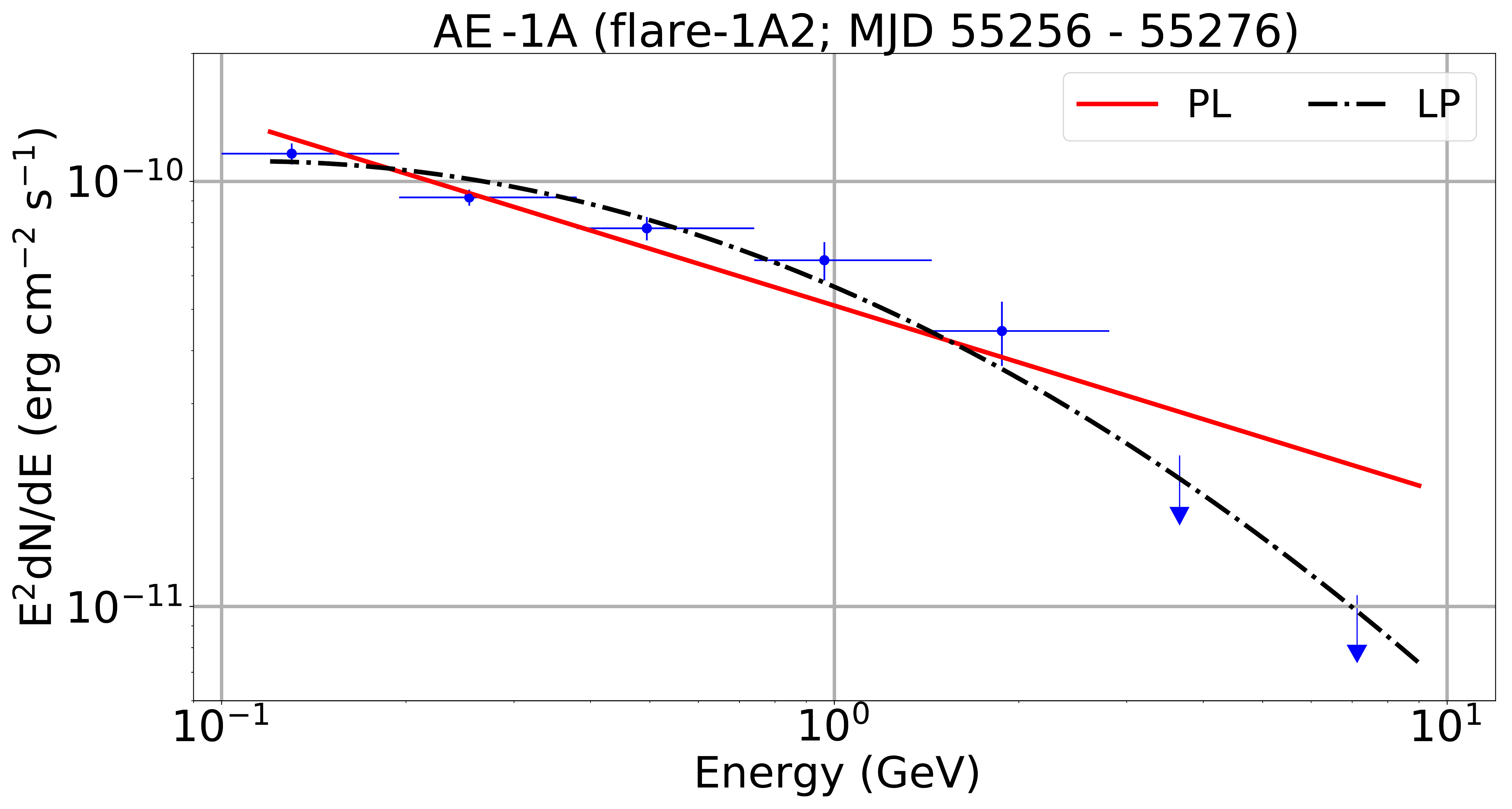}
\includegraphics[height=1.90in,width=2.6in]{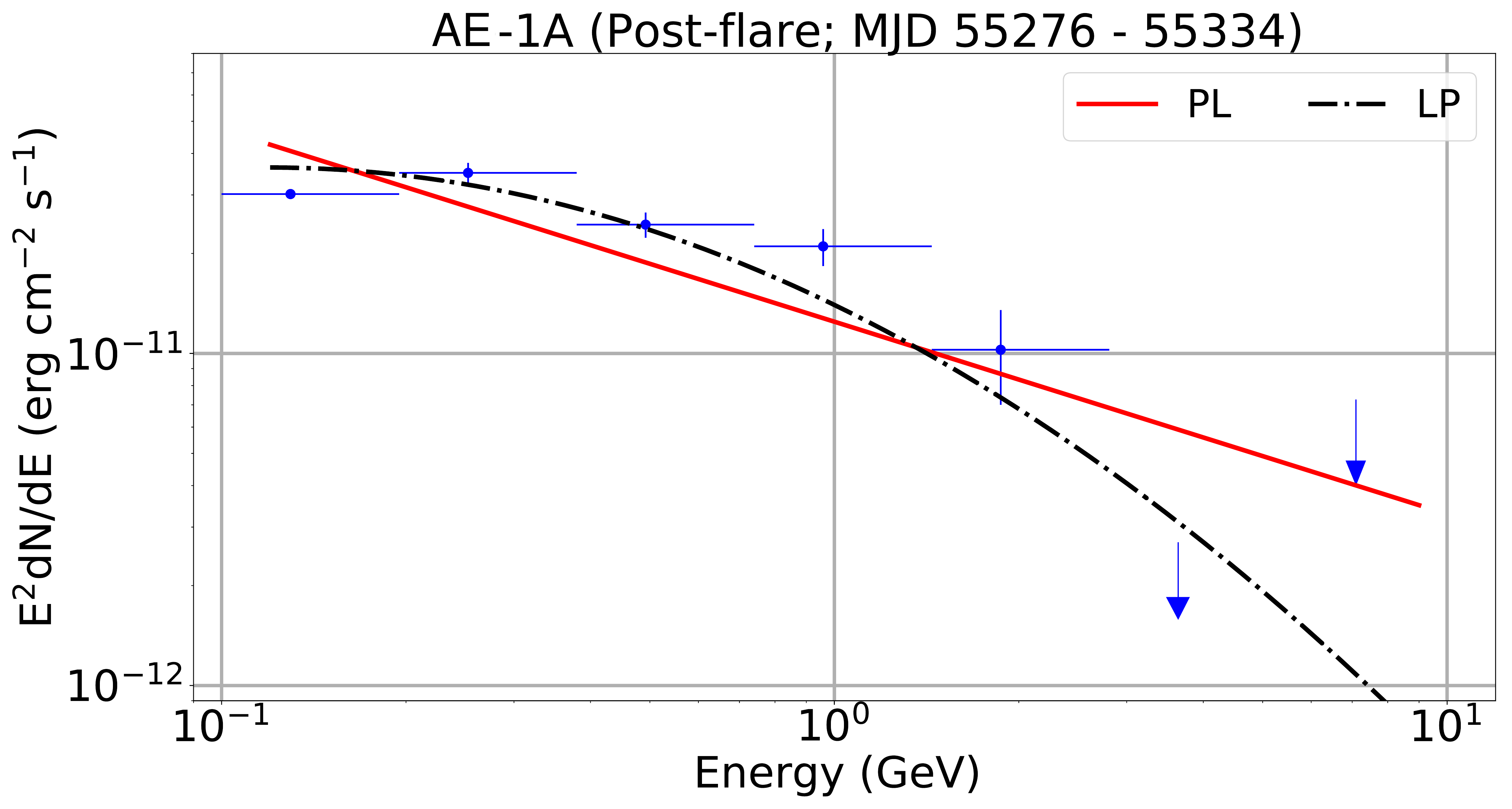}
\caption{Gamma-ray SEDs of different phases (Pre-flare, flare-B1, flare-B2, Post-flare) of AE-1A. PL and LP describe the Power law, and Logparabola model respectively, which are shown by the solid red and dash-dot black lines, respectively.}
\label{fig:3}
\end{figure*}

\subsection{AE-1}
AE-1 epoch (MJD 55176 - 55334) has only one sub-structure, AE-1A (AE-1A is basically the same as AE-1 epoch. hereafter, we call this phase AE-1A). Two-day binning light curve of this phase is shown in upper panel of Figure-\ref{fig:2}. It has four different phases: Pre-flare, flare-1A1, flare-1A2, and Post-flare.

Pre-flare phase has a time duration of 40 days from MJD 55176-55216 with an average flux of 0.25$\pm$0.02 unit. Subsequently, the source entered in flaring state with significant variation in photon flux. This state is divided into two phases, namely, flare-1A1 (MJD 55216 - 55256) and flare-1A2 (MJD 55256 - 55276). The fitted light curves of these phases are shown in lower panel of Figure-\ref{fig:2}. We have recognized three distinct peaks (P1, P2 and P3) in these two phases. The flux value of these peaks are 0.81$\pm$0.11, 0.74$\pm$0.15, and 1.05$\pm$0.13 at MJD 55222.0, 55234.0, and 55266.0 respectively. After Flare-1B, the Flux gradually decreases to a low value (Post-flare phase) from MJD 55276 - 55334 with an average flux of 0.30$\pm$0.11 unit.

The fitted parameters of the light curve have been described in Table-\ref{tab:2}.

\subsection{AE-2}
AE-2 has two different sub-structures, namely, AE-2A (MJD 55795 - 55870) and AE-2B (MJD 55951 - 56011). The one-day binning light curve and different phases of these sub-structures are shown in Figure-\ref{fig:A1} and Figure-\ref{fig:A3}.

AE-2A is the most violent substructure in the entire light curve history. It has three different phases: Pre-flare (MJD 55795 - 55824), flare (MJD 55824 - 55838), and Post-flare (MJD 55838 - 55870). The average flux of the Pre-flare phase is 0.35$\pm$0.03 unit. After that, the source entered in flare phase with a substantial variation in flux. The fitted light curve is shown in Figure-\ref{fig:A2}. In our study, one major peak (P1) is identified with a flux of 6.34$\pm$0.51 unit at MJD 55827.5. Post-flare phase has a time duration of 32 days with an average flux of 0.15$\pm$0.02 unit.

The total time duration of AE-2B is 60 days. Depending on the flux level, this sub-structure is divided into five different phases: Pre-flare (MJD 55951 - 55972), flare-2B1 (MJD 55972 - 55980), Quiescent or Q1 (MJD 55980 - 55987), flare-B2 (MJD 55987 - 55994), and Post-flare (MJD 55994 - 56011) phases. Only one distinctive peak has been identified from each flare phase (flare-2B1 and flare-2B2). The flux values of these peaks (P1 and P2) are 0.95$\pm$0.17 and 1.69$\pm$0.30 units. There is a low flux phase in between two flare phases with a time duration of 7 days. We have defined this phase as the Quiescent or Q1 phase. The fitted light curve of Flare-2B is illustrated in Figure-\ref{fig:A4}.

\begin{table*}
\caption{Result of Gamma-ray SEDs for AE-1A phase. PL and LP models are used to fit the SEDs. Column 1 represents the different phases of the sub-structures. $F_{0}$ is the total flux during each phase. Different model parameters are represented in Columns 3, and column 4. Negative log-likelihood value and the difference in log-likelihood values between two models are given in columns 5, and column 6 respectively (see text for more details).} 
\label{tab:3}
\centering
\begin{tabular}{ccccc rrrr}   
\hline\hline                        
 & & Powerlaw\\
\hline
Activity & $F_0$ & $\Gamma_{PL}$ & &  -log($\mathcal{L}$) \\ [1.5ex]
& [10$^{-6}$ ph cm$^{-2}$ s$^{-1}$] &  \\
\hline
Pre-flare &  0.23$\pm$0.02 & 2.65$\pm$0.07 &  - &  64163.84 & - \\
Flare-1A &  0.62$\pm$0.03 & 2.38$\pm$0.04 &  - &  63921.84 & - \\
Flare-1B &  0.62$\pm$0.03 & 2.45$\pm$0.04 &  - &  67395.00 & - \\
Post-flare &  0.21$\pm$0.02 & 2.58$\pm$0.08 &  - &  74568.59 & - \\
\hline                          
 & & Logparabola\\ 
\hline
Activity & $F_0$ & $\alpha$ & $\beta$ & -log($\mathcal{L}$) & $\Delta$log($\mathcal{L}$) \\ [1.5ex]
& [10$^{-6}$ ph cm$^{-2}$ s$^{-1}$] &  \\
\hline
Pre-flare &  0.24$\pm$0.02 & 2.66$\pm$0.10 &  0.04$\pm$0.07 & 64192.65 & 29.00 \\
Flare-1A &  0.59$\pm$0.03 & 2.24$\pm$0.06 &  0.11$\pm$0.04 & 63916.19 & -5.65 \\
Flare-1B &  0.60$\pm$0.03 & 2.28$\pm$0.06 &  0.14$\pm$0.03 & 67391.22 & -3.78 \\
Post-flare &  0.19$\pm$0.02 & 2.39$\pm$0.12 &  0.21$\pm$0.09 & 74564.93 & -3.66 \\

\hline
\hline 
\end{tabular}
\end{table*}

\subsection{AE-3}
AE-3 is the longest duration (MJD 56812 - 57355) activity phase of the $\sim$ 12.5 years light curve history with moderate variation in flux. There are four different sub-structures in this state. These are, AE-3A (MJD 56834 - 56920), AE-3B (MJD 56992 - 57044), AE-3C (MJD 57036 - 57061) and AE-3D (MJD 57060 - 57168). The light curve of these sub-structures are illustrated in Figure-\ref{fig:A5}, Figure-\ref{fig:A7}, Figure-\ref{fig:A9} and Figure-\ref{fig:A11}.

AE-3A has the most simple kind of structure as Flare-2A. The duration of different phases is 35 days, 17 days and 34 days for Pre-flare, flare and Post-flare phases, respectively. The fitted light curve is shown in Figure-\ref{fig:A6}. The flux values of two major peaks (P1 and P2) are 1.15$\pm$0.12 and 0.74$\pm$0.13 unit at MJD 56874 and MJD 56884, respectively.

There are five different phases in two days binning light curve of AE-3B, namely, Pre-flare (MJD 56992 - 57016), flare-3B1 (MJD 57016 - 57022), Q1 (MJD 57022 - 57029), flare-3B2 (MJD 57029 - 57035) and Post-flare (MJD 57035 - 57044). The fitted light curve of flare-3B1 and flare-3B2 are shown in Figure-\ref{fig:A8}. The Q1 phase is also shown in the grey-shaded region.

Figure-\ref{fig:A9} shows the one binning light curve of AE-3C, which has three phases: Pre-flare, flare and Post-flare. Two significant peaks (P1, and P2) have been identified in the fitted light curve at MJD 57048.5 and MJD 57052.5 (shown in Figure-\ref{fig:A10}). 

AE-3D also shows the complicated structure as AE-2B or AE-3B. This has five different phases: Pre-flare, flare-3D1, Q1, flare-3D2 and Post-flare. The Pre-flare phase lasted only ten days (MJD 57060 - 57070). After that, the source entered into high flux phases, flare-3D1, and flare-3D2. However, there is a quiescent phase (Q1) with an average flux of 0.59$\pm$0.05 unit in between these two flaring phases. The fitted light curve in Figure-\ref{fig:A12} shows two distinct peaks with fluxes of 0.98$\pm$0.12 and 2.36$\pm$0.25 unit at MJD 57072.0 and MJD 57102 respectively. The relatively long duration portion (low flux level) of the light curve
is defined as the Post-flare (MJD 57108 - 57168) phase.

\section{STUDY OF GAMMA-ray SED} \label{sec:4}
We have analyzed the $\gamma$-ray SEDs for different phases (such as Pre-flare, Flare, Post-flare, etc.) of the sub-structures. We have used two different models to fit these SEDs \citep{2010ApJ...710.1271A}:

\begin{enumerate}

\item A power-law model,

\begin{equation}
   \frac{dN}{dE} = N_{0}\Big(\frac{E}{E_0}\Big)^{-\Gamma_{PL}}
\end{equation}

where, $N_{0}$, and $\Gamma_{PL}$ are the normalization factor and spectral index of the model. We have kept free these parameters during likelihood fitting, while the scaling factor ($E_{0}$) is fixed at 100 MeV for all the cases.

\item A log-parabola model,

\begin{equation}
   \frac{dN}{dE} = N_{s}\Big(\frac{E}{E_s}\Big)^{-(\alpha+\beta\log(E/E_s))}
\end{equation}

This model is similar to the power law but with an energy-dependent photon index ($a(E) = \alpha + \beta \log(E/Es)$). $\alpha$ is the photon index at energy $E_s$, which is fixed at 300 MeV, near the low energy part of the spectrum. $\beta$ is known as the curvature index of the model.

\end{enumerate}
In order to compare these two models, we have computed the log-likelihood values for different phases. The values of different model parameters and the negative log-likelihood are given in Table-\ref{tab:3} for AE-1A (Table-\ref{tab:A1} - Table-\ref{tab:A6} for other sub-structures). The differences in log-likelihood ($\Delta log\mathcal{L} = \lvert log\mathcal{L}_{LP} \rvert - \lvert log\mathcal{L}_{PL} \rvert$) values have also been given in the last column of these tables. From $\Delta log\mathcal{L}$, we can see for most of the cases, the $\gamma$-SEDs are best described by the Log-parabola (LP) over Power-aw (PL) model.

The $\gamma$-ray SEDs of different activity phases are shown in Figure-\ref{fig:3} for AE-1A (Figure-\ref{fig:A13} - Figure-\ref{fig:A18} for other sub-structures). Most of the cases, SEDs show `spectral hardening' behavior. This means that the power-law spectral index decreases (increases) when the source transited from a low to high (high to low) flux state. In our study, we have found only one case when the source shows `spectral softening' behavior. This is when the source transited from, Q1 to Flare-2B2 phase ($\Gamma_{PL}$ changes from 2.28$\pm$0.15 to 2.52$\pm$0.12). There are other five cases where the spectral index remains nearly constant, within the uncertainties. These are during the transition of the source from Flare-2B1 to Q1, Flare-2B2 to Post-flare, Flare-3D1 to Q1, Q1 to Flare-3D2, and Flare-3D2 to Post-flare phase.

\begin{figure*}
\includegraphics[height=4.7in,width=8.0in]{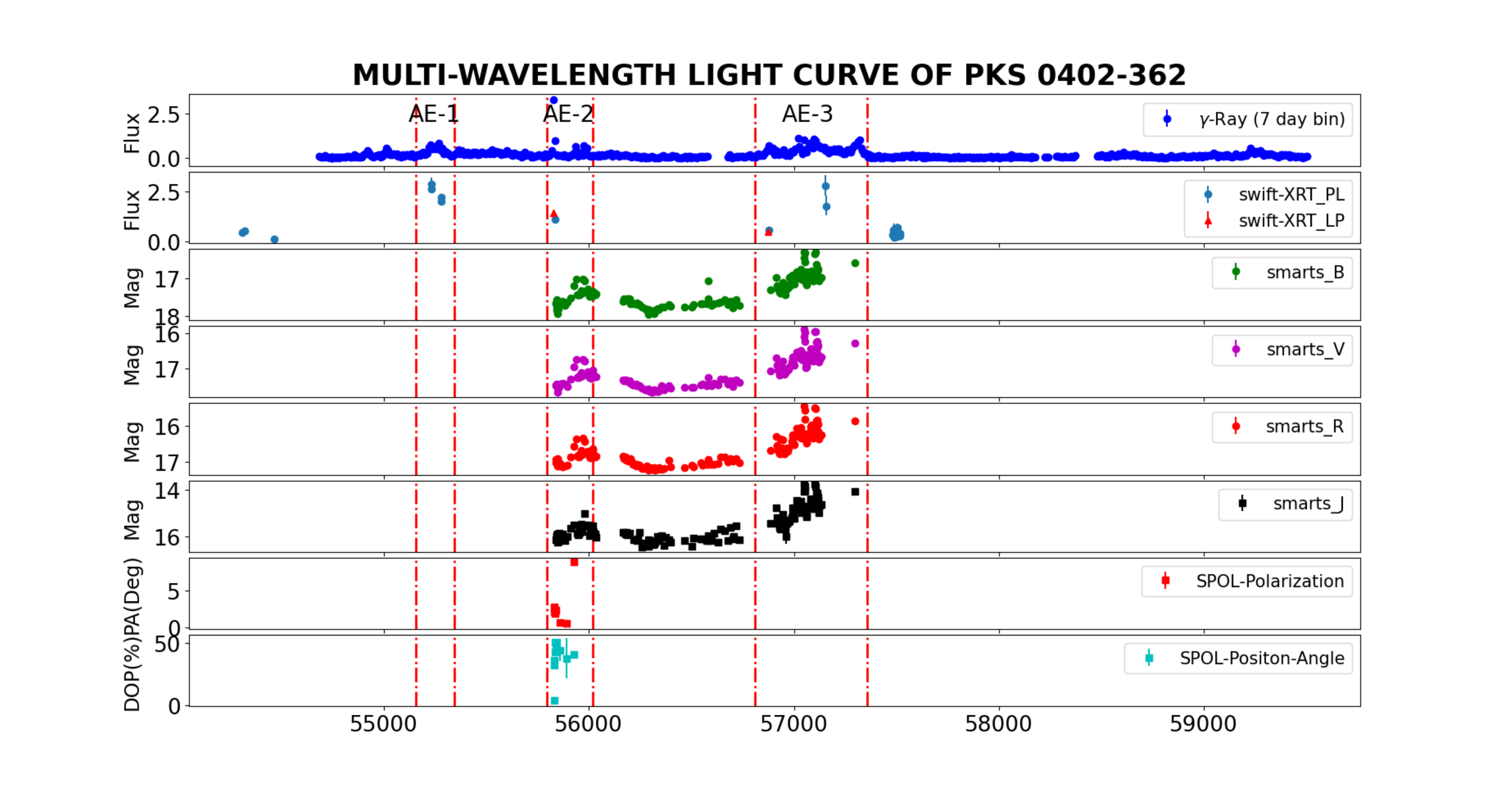}
\caption{Multi-wavelength light curve of PKS 0402-362. The first panel shows the Fermi-LAT data in the unit of $10^{-6}$ ph cm$^{-2}$ s$^{-1}$. Swift-XRT data are shown in the second panel in units of $10^{-11}$ erg cm$^{-2}$ s$^{-1}$. Light blue-circle and Red-triangle in the second panel represent the flux value computed using power law and log parabola model, respectively. SMART data in the BVRJ band are shown from the third to the sixth panel. Polarimetric data from the steward observatory are also shown in the seven and eighth panels.}
\label{fig:4}

\end{figure*}

\begin{figure*}
\centering

\includegraphics[height=2.4in,width=3.3in]{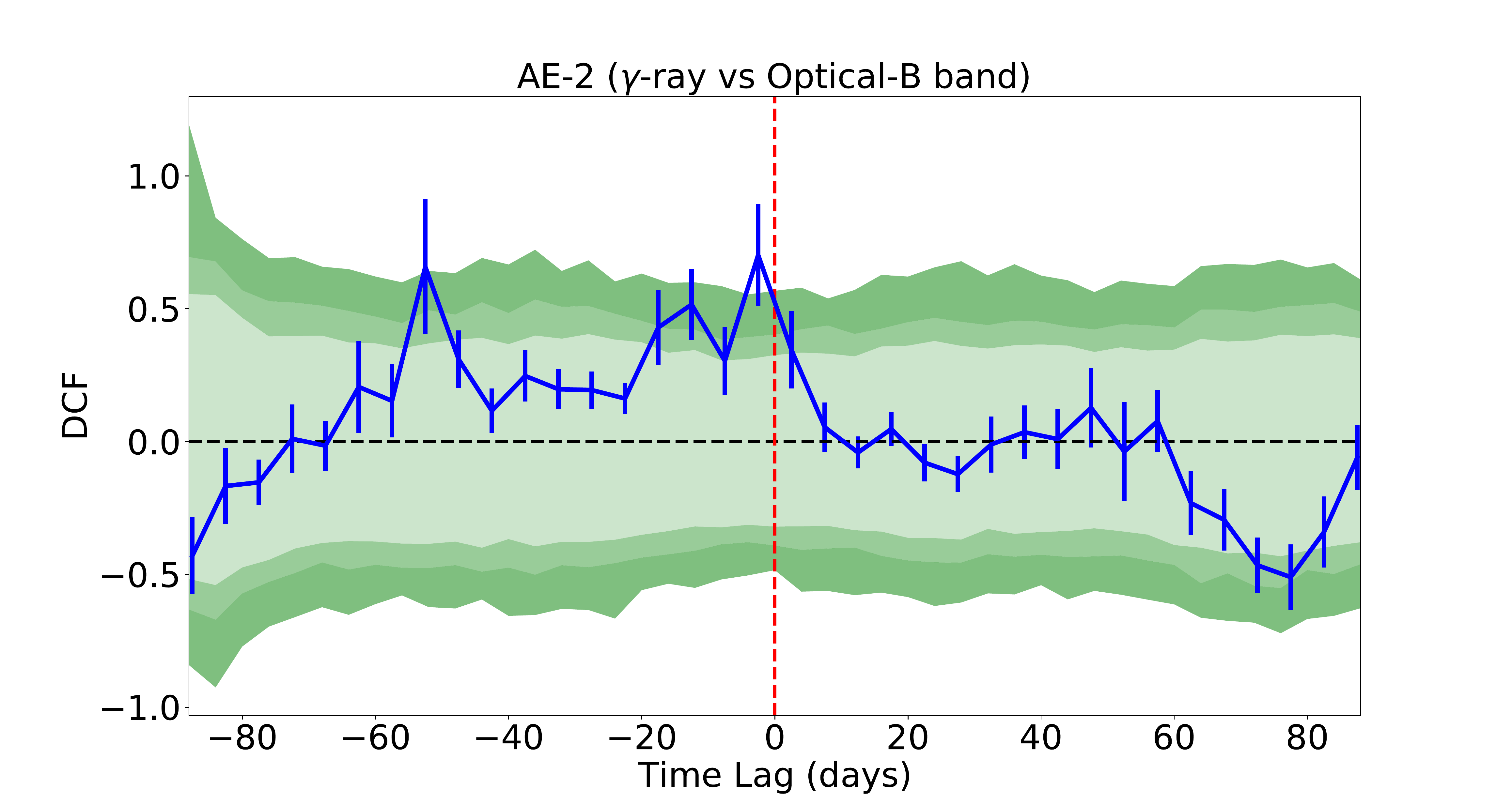}
\includegraphics[height=2.4in,width=3.3in]{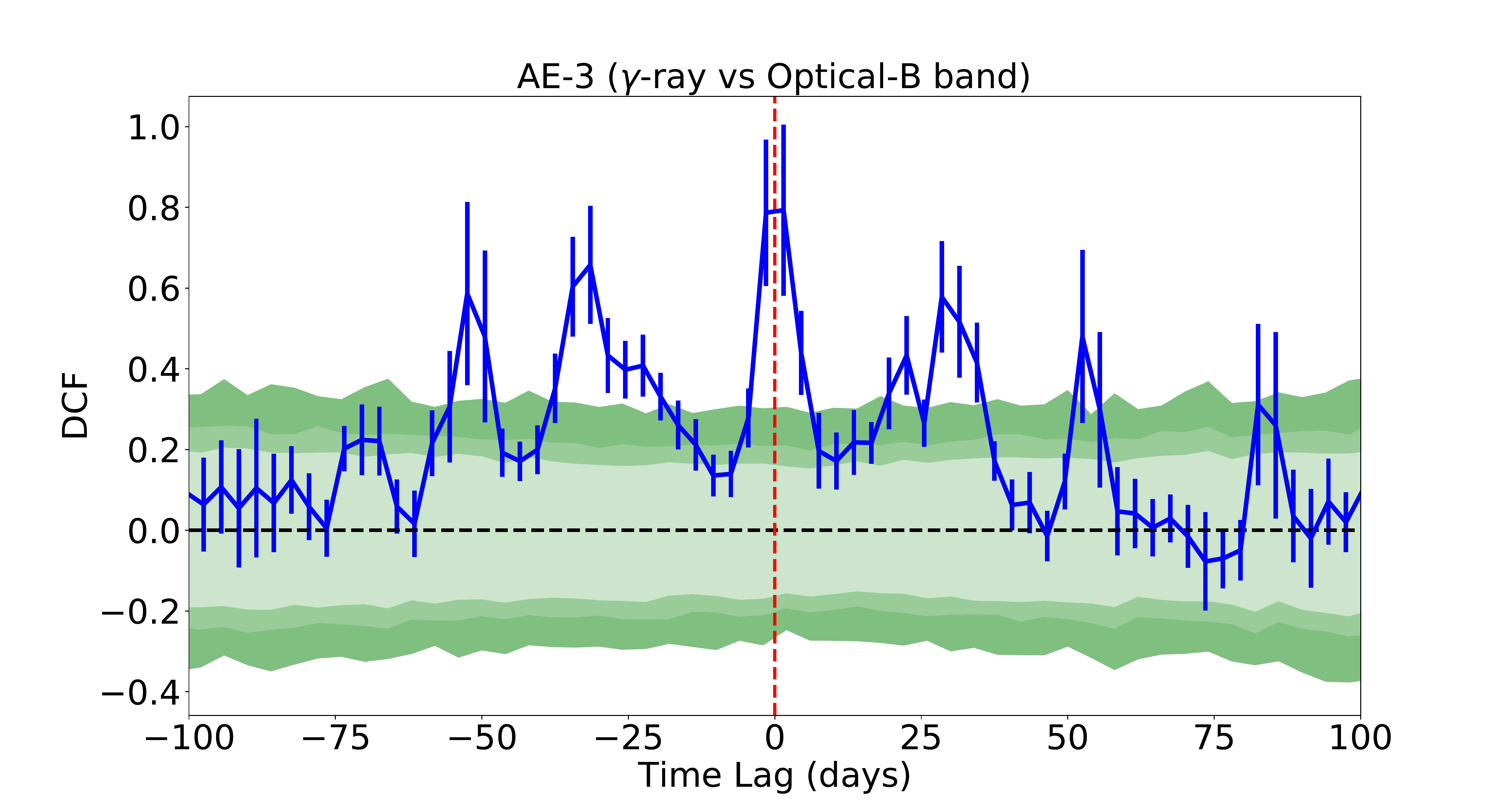}

\includegraphics[height=2.4in,width=3.3in]{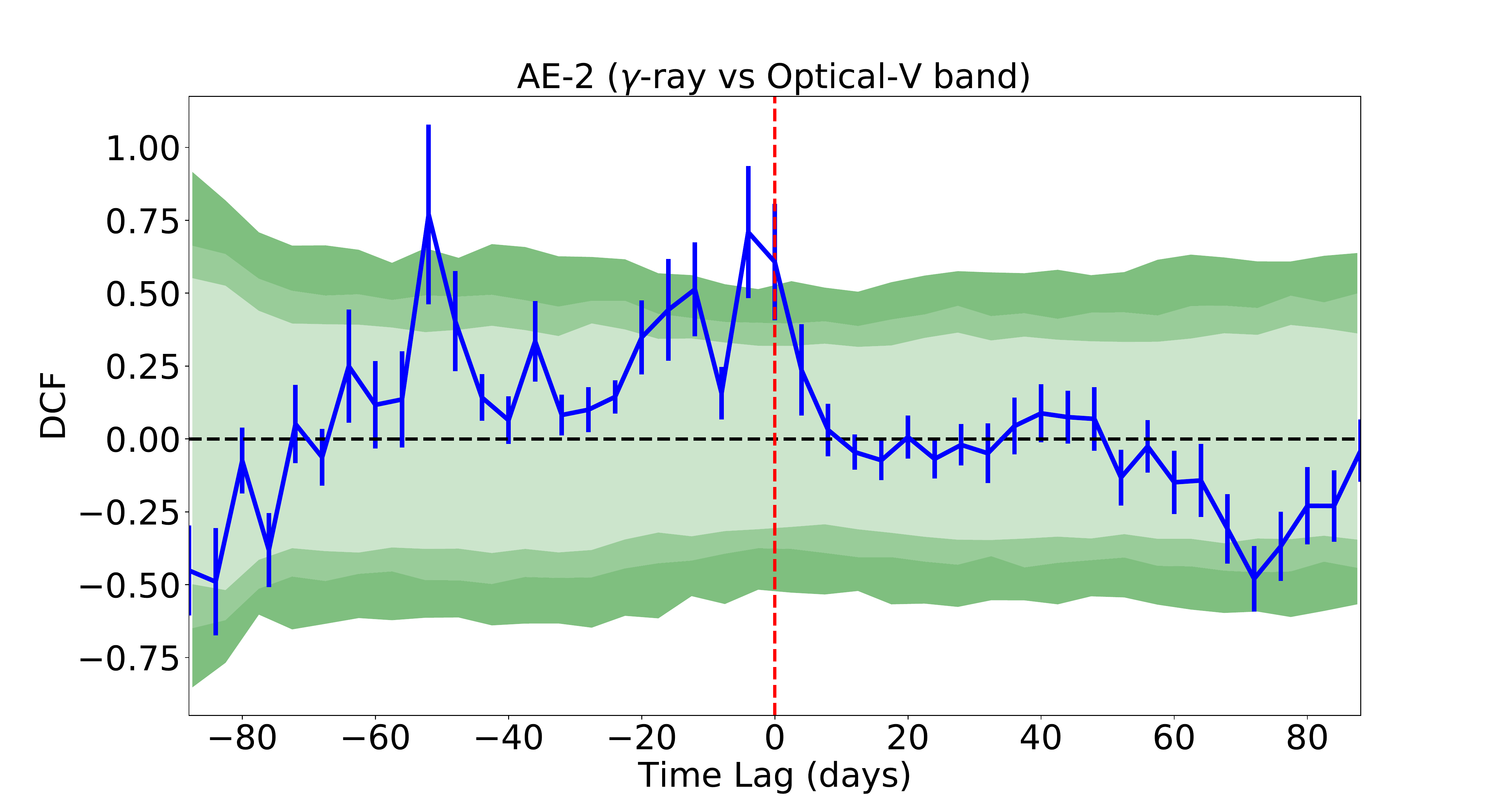}
\includegraphics[height=2.4in,width=3.3in]{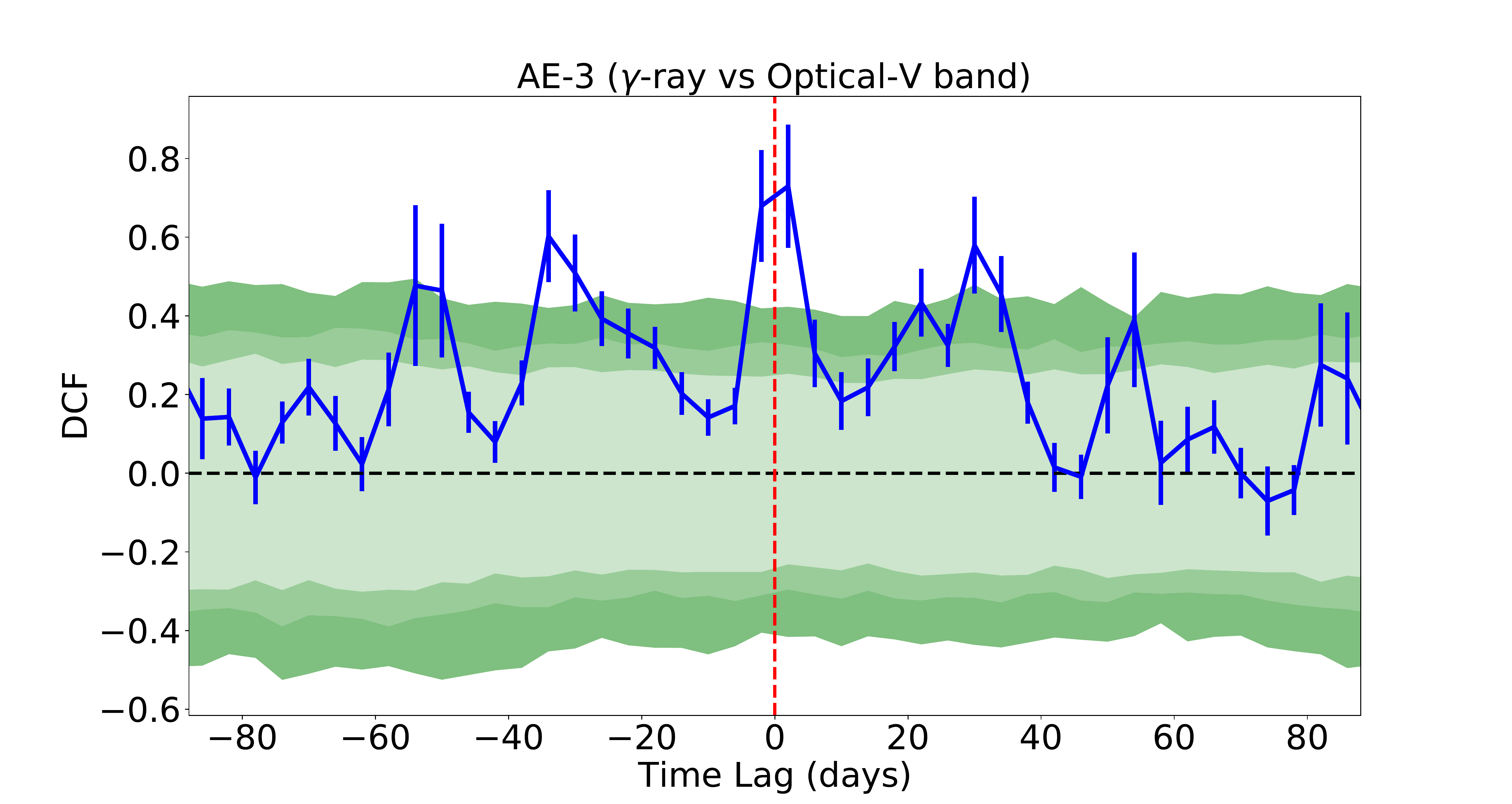}

\includegraphics[height=2.4in,width=3.3in]{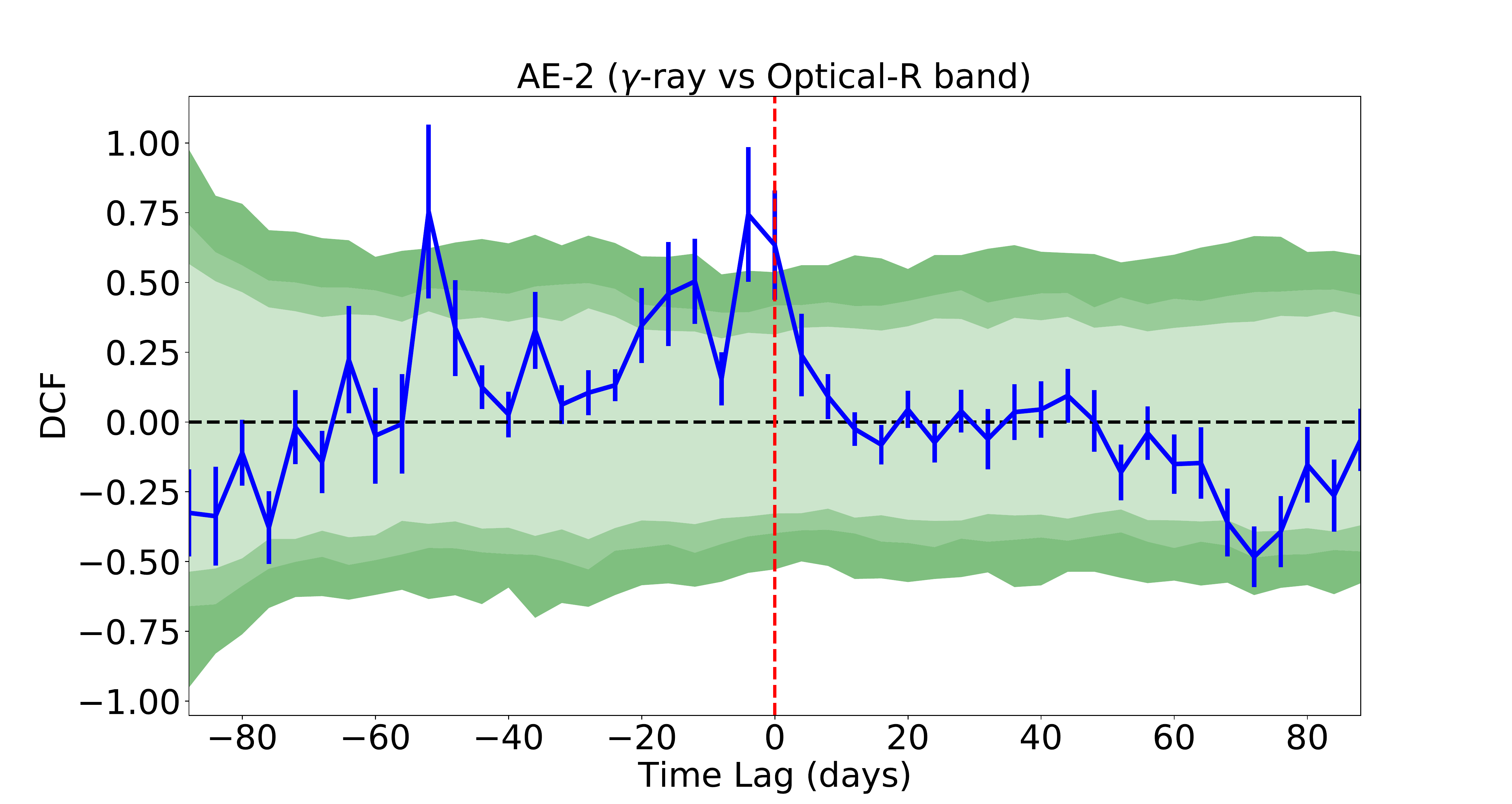}
\includegraphics[height=2.4in,width=3.3in]{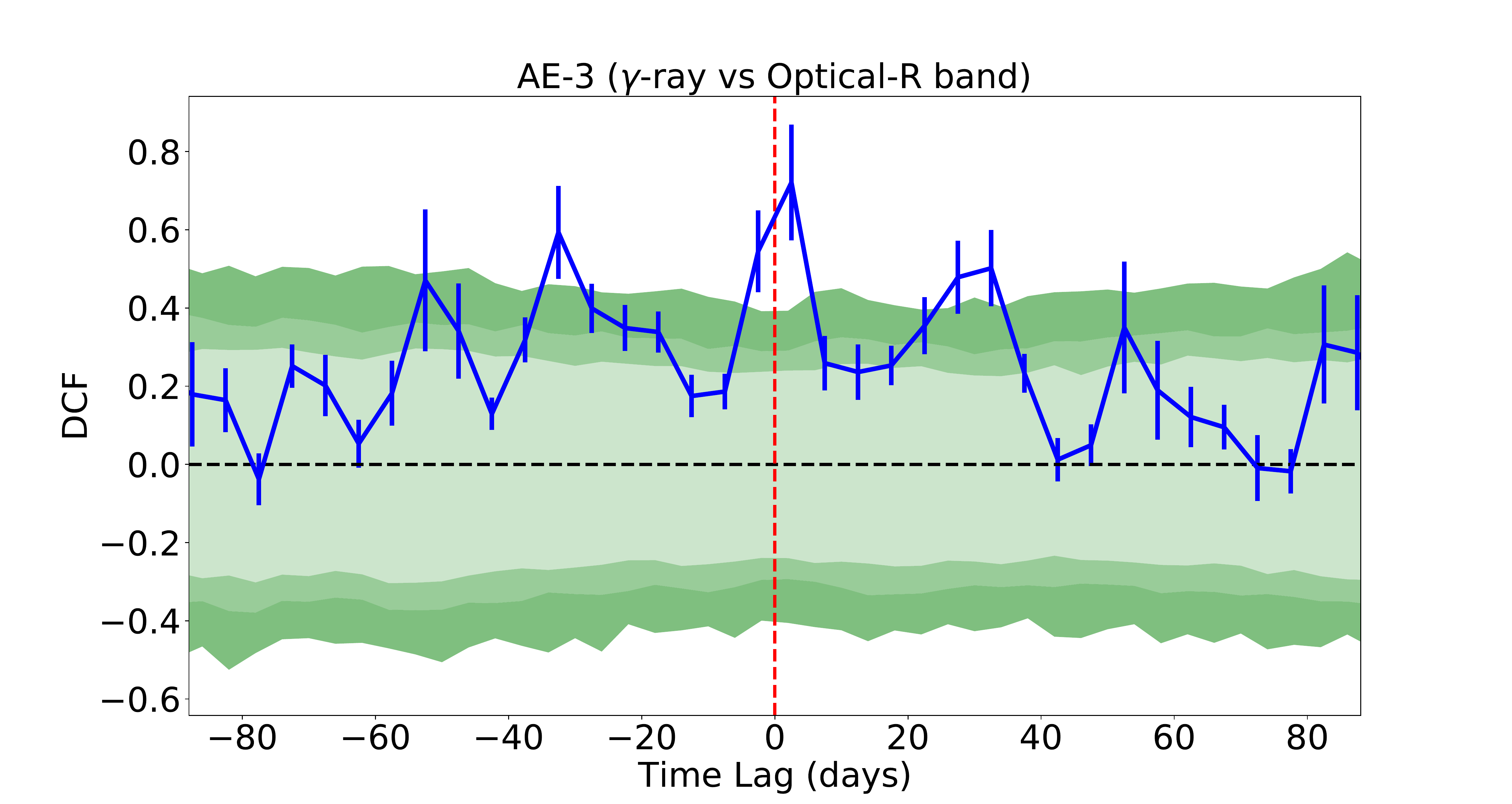}

\includegraphics[height=2.4in,width=3.3in]{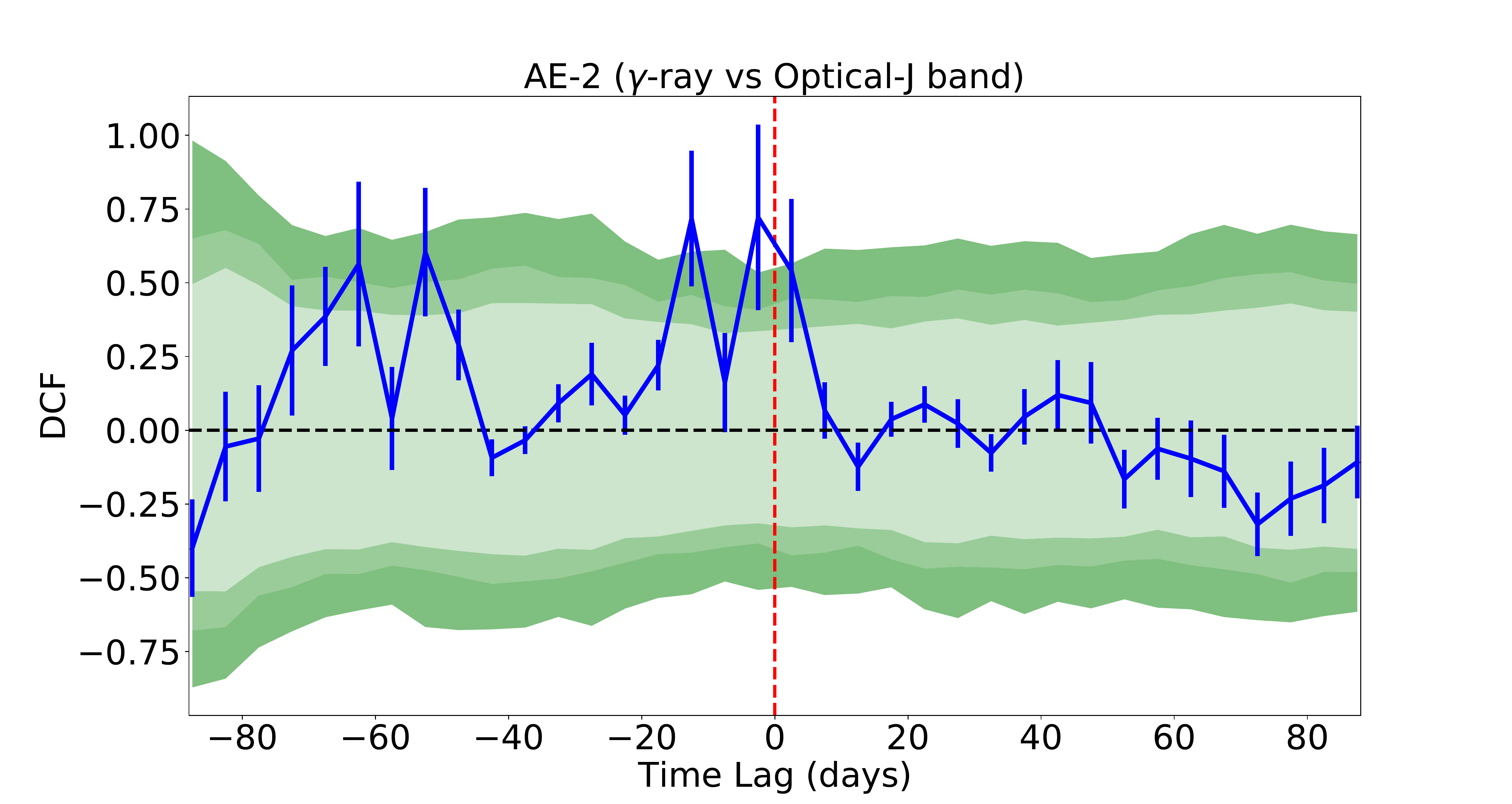}
\includegraphics[height=2.4in,width=3.3in]{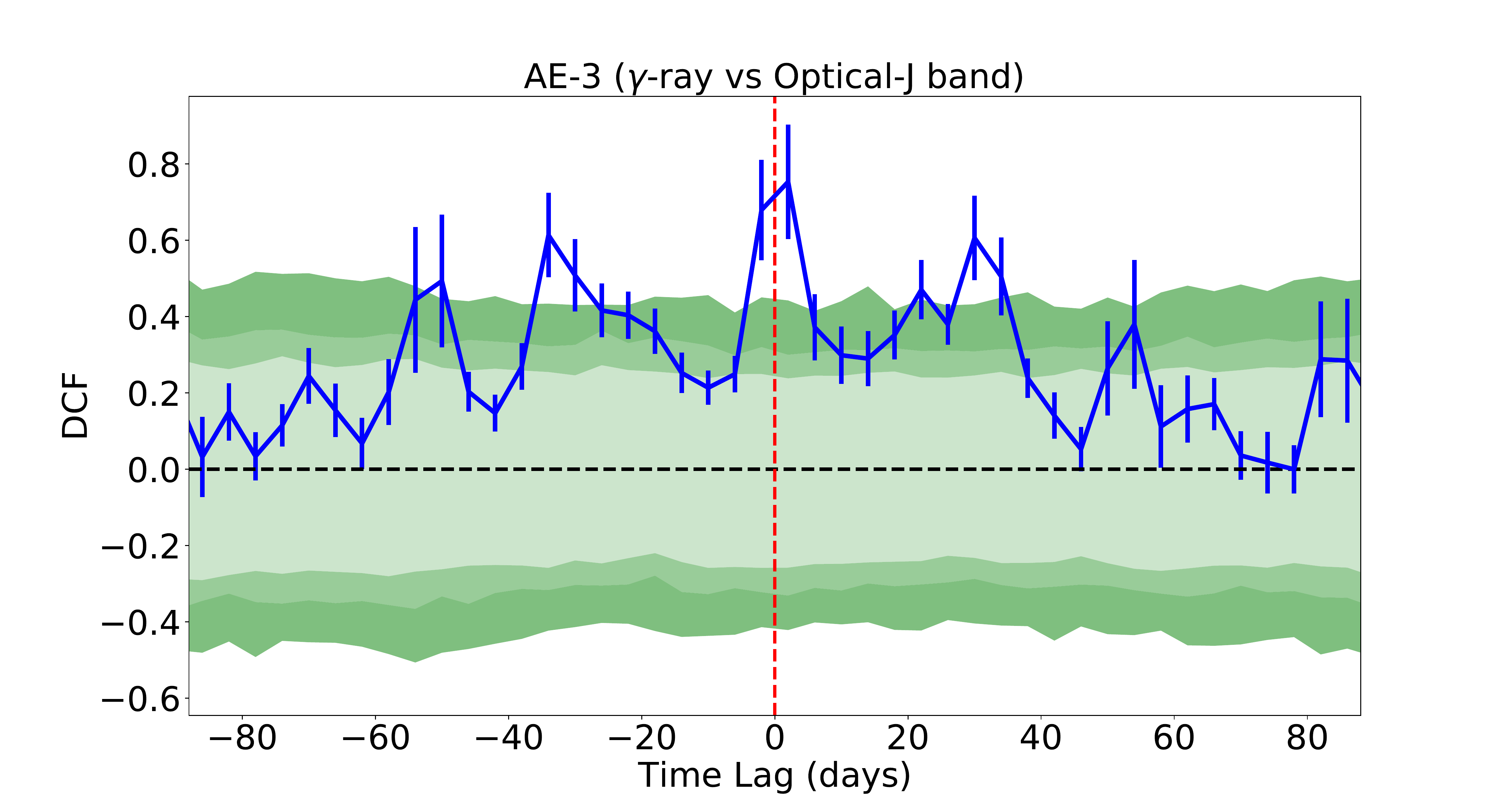}

\caption{Correlation plots (DCF) between Gamma-ray and different Optical band light curves (see text for more details). 90\%, 95\%, and 99\% contours (from light green to deep green) are shown in the plots.}
\label{fig:5}

\end{figure*}

\section{CORRELATION STUDY} \label{sec:5}

We have done a correlation study between $\gamma$-ray and different optical bands (B, V, R, and J band) for AE-2 and AE-3, respectively. Due to the low number of data points in the X-ray band, we can not study the correlation between $\gamma$-ray and X-ray data. The multi-wavelength light curve of this source is shown in Figure-\ref{fig:4}. Different Flares and simultaneous data in other bands are shown by broken red lines.

Unbinned Discrete Correlation Function (UDCF) is given by \citep{1988ApJ...333..646E}:
\begin{equation}\label{eq:6}
    UDCF_{ij} = \frac{(a_i - <a>)(b_j - <b>)}{\sqrt{ (\sigma_a^2 - e_a^2)(\sigma_b^2 - e_b^2)}}
\end{equation}

where, $a_i$, and $b_j$ are the two discrete time series. $\sigma_a$ and $\sigma_b$ are the standard deviations of the two-time series, respectively. Here, each value of $UDCF_{ij}$ is associated with the time delay, $\Delta t_{ij} = t_{j} -  t_{i}$.
If we average the Equation-\ref{eq:6} over M number of pairs for which $\tau - \Delta \tau/2 \leq \Delta t_{ij}  (= t_j - t_i) \leq \tau + \Delta \tau/2$, we get,

\begin{equation}
    DCF(\tau) = \frac{1}{M} \sum UDCF_{ij} \pm  \sigma_{DCF}(\tau)  
\end{equation}

where,

\begin{equation}
    \sigma_{DCF}(\tau) = \frac{1}{M-1} \sqrt {\sum [UDCF_{ij} - DCF(\tau)]^2}
\end{equation}

$DCF(\tau)$, and $\sigma_{DCF}(\tau)$ are the discrete correlation function and the associated error at the lag $\tau$.

The results of the correlation study are shown in Figure-\ref{fig:5}. All the figures on the left and right columns are for AE-2 and AE-3 epochs, respectively. 

The correlation between $\gamma$-ray and all the four optical bands (B, V, R, J) for AE-2 show minimal negative time lag of -2.50, -4.00, -4.00 and -2.50 days with DCF values of 0.70$\pm$0.19, 0.71$\pm$0.23, 0.73$\pm$0.24, 0.71$\pm$0.31 respectively (Left column of Figure-\ref{fig:5}). This suggests that light curves of $\gamma$-ray and optical bands are positively correlated. Here, Negative lag implies that the $\gamma$-ray light curve lags the optical emission. 

For epoch AE-3, the DCF plots between $\gamma$-ray and optical bands show very small positive time lag (1.50, 2.00, 2.50, and 2.00 days) with high DCF values of 0.79$\pm$0.21, 0.73$\pm$0.16, 0.71$\pm$0.15, and 0.75$\pm$0.15, respectively (Right column of Figure-\ref{fig:5}). 

To estimate the significance level of the DCF peaks, we have followed the procedure given in \citet{2014MNRAS.445..428M}. We have first modeled the observed $\gamma$-ray and optical periodograms with the power-law function ($P(f) \propto f^{-\beta}$). Next, we have computed the values of $\beta$ by the `PSRESP' method \citep{Uttley2002May}. After modeling the observed periodograms, we have simulated 1000 light curve pairs by \citet{Timmer1995Aug} algorithm for each case. Finally, we cross-correlate the simulated light curve pair to estimate the significance level for each time lag. In Figure-\ref{fig:5}, 90\%, 95\%, and 99\% (from light green to deep green) significance levels are shown for different cases. We noticed that all the combinations show a strong correlation above 99$\%$.

However, for both epochs, it is important to note that the time lags seen here in all the combinations are shorter than the average time gap of the optical light curve ($\sim$5 days for AE-2 and $\sim$3 days for AE-3). The observed small time lags could be caused by the gap in the light curves and, in that case, can not be considered reliable. We also checked the correlation study results with ICCF method\footnote{\url{http://ascl.net/code/v/1868}} and computed uncertainty on time lags by Monte Carlo (MC) analysis which involves Flux Randomization (FR) and Random Subset Selection (RSS) techniques \citep{Peterson1998Jun}. For the AE-2 epoch, we found centroid time lags of -3.1$^{+1.9}_{-8.9}$, -2.9$^{+2.9}_{-6.8}$, -3.0$^{+1.9}_{-2.0}$, and -11.9$^{+13.9}_{-1.0}$ days for B-gamma, V-gamma, R-gamma, and J-gamma correlation respectively. Whereas, for AE-3 epoch, the value of centroid lags are: 0.05$^{+0.99}_{-0.95}$,  0.04$^{+0.18}_{-0.98}$,  0.04$^{+0.11}_{-0.96}$, and  0.06$^{+0.12}_{-1.67}$ days. Lags obtained by ICCF are consistent with DCF results within the error bar.

The $\gamma$-ray and optical light curve of AE-3 (MJD 56886 - 57132) show Quasi-Periodic Oscillation (QPO) with periods of $\sim$ 27 days for five and seven cycles, respectively. We have also checked this periodic nature by the Lomb-Scargle Periodogram (LSP) and Weighted-Wavelet Transform (WWZ) methods. Due to this reason, there are nearly equally spaced multiple peaks observed in all the DCF plots of AE-3.

\section{MULTI-WAVELENGTH MODELLING} \label{sec:6}
 We use `GAMERA', an open-source code to model the multi-wavelength SEDs of the source PKS 0402-362. This code is publicly available on GitHub\footnote{\url{https://github.com/libgamera/GAMERA}}. This code solves the time-dependent transport equation; it estimates the propagated particle spectrum N(E,t) for an input injected particle spectrum, and further it uses the propagated spectrum to calculate the Synchrotron and Inverse-Compton (IC) emissions. GAMERA solves the following transport equation:
 \begin{equation}
\frac{\partial N(E,t)}{\partial t}= Q(E,t)-\frac{\partial}{\partial E}(b(E,t)N(E,t))-\frac{N(E,t)}{\tau\textsubscript{esc}(E,t)}
\end{equation}
where, Q(E,t) is the input particle spectrum and b(E,t) corresponds to the energy loss rate by Synchrotron and IC and can be defined as
\big($\frac{dE}{dt}$\big). In the last term, $\tau\textsubscript{esc}$(E,t) denotes the energy-dependent escape time of particles from the emission region.\\
Here, all the multi-wavelength SEDs can be explained by one-zone leptonic modeling; so, we consider electrons as injected particles. As LogParabola (LP) gives the best-fit parameters for $\gamma$-ray spectrum for all the sub-structures, we have used LP as a particle distribution to model the multi-wavelength SEDs. Following \cite{2004A&A...413..489M}, an LP photon spectrum can be produced by the radiative losses of an LP electron spectrum. The functional form of the injected electron is the following:
\begin{equation}
Q(E)=L\textsubscript{o}\Bigg(\frac{E}{E_o}\Bigg)^{-\big(\alpha+\beta log\big(\frac{E}{E_o}\big)\big)}
\end{equation}
where L\textsubscript{o} is the normalization constant and E\textsubscript{o} is the scaling factor. This code uses the `Klein-Nishina' cross-section to compute Inverse Compton emission \citep{Blumenthal1970Jan}.

\begin{figure*}
\centering

\includegraphics[height=2.5in,width=3.2in]{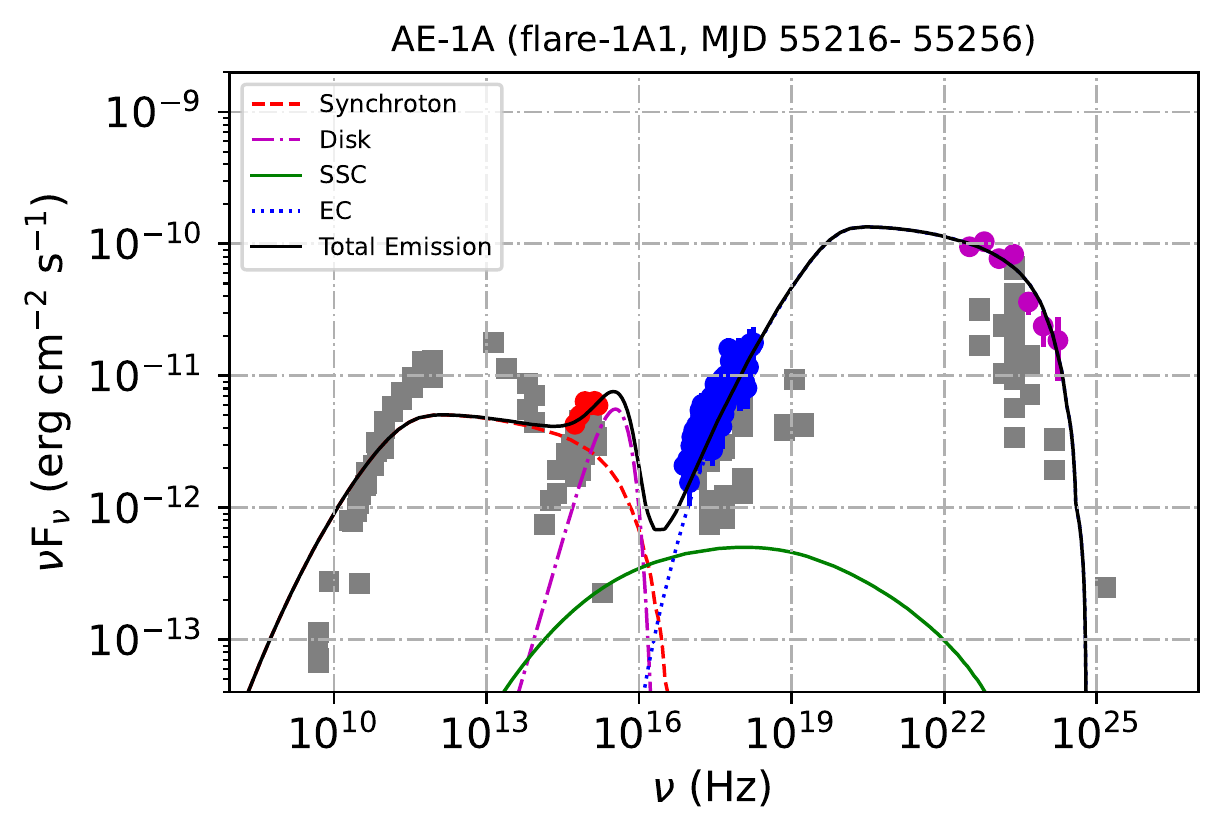}
\includegraphics[height=2.5in,width=3.2in]{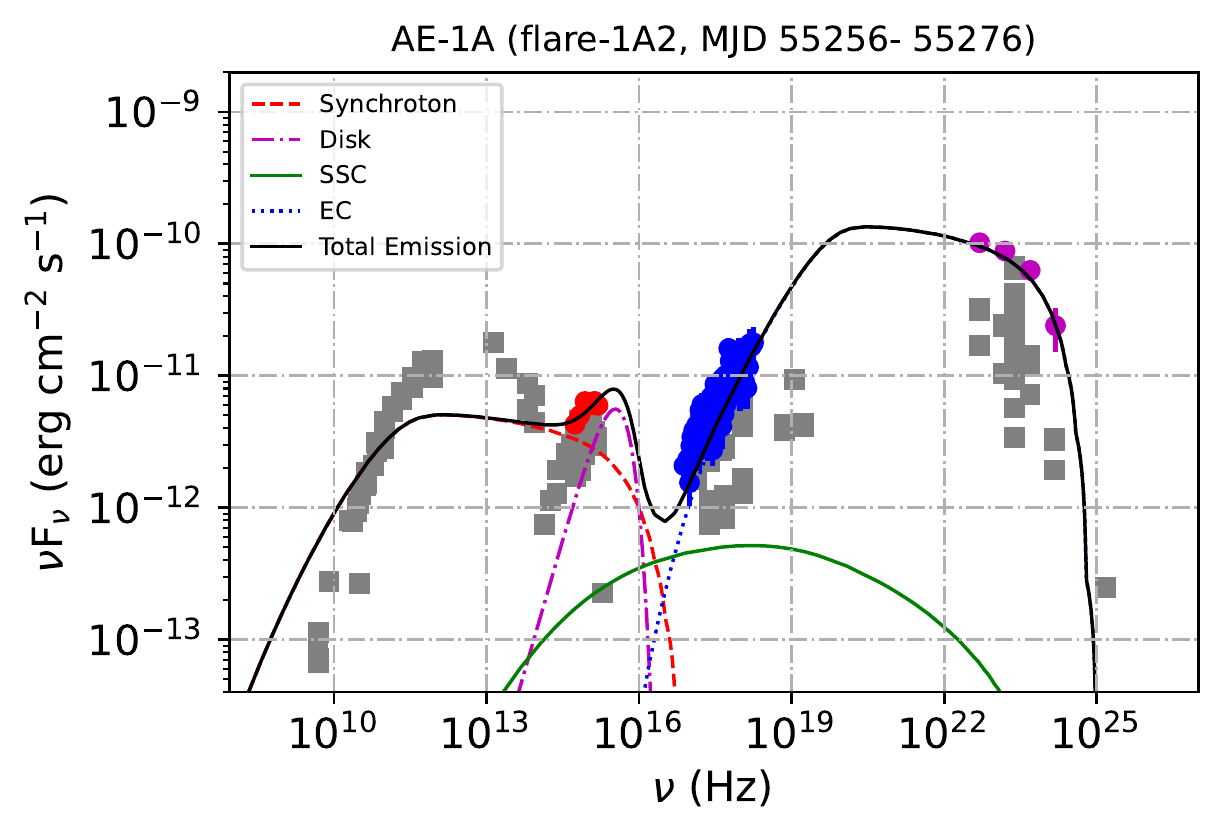}
\includegraphics[height=2.5in,width=3.2in]{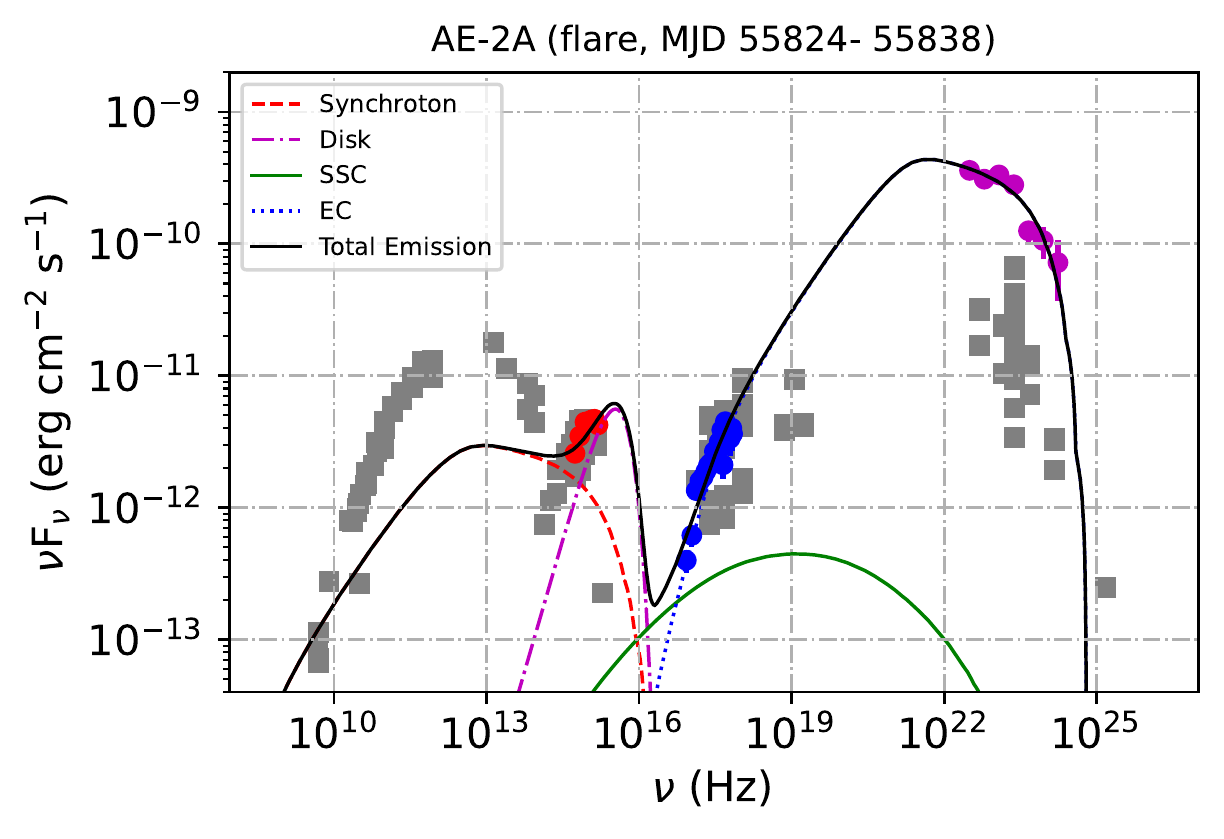}
\includegraphics[height=2.5in,width=3.2in]{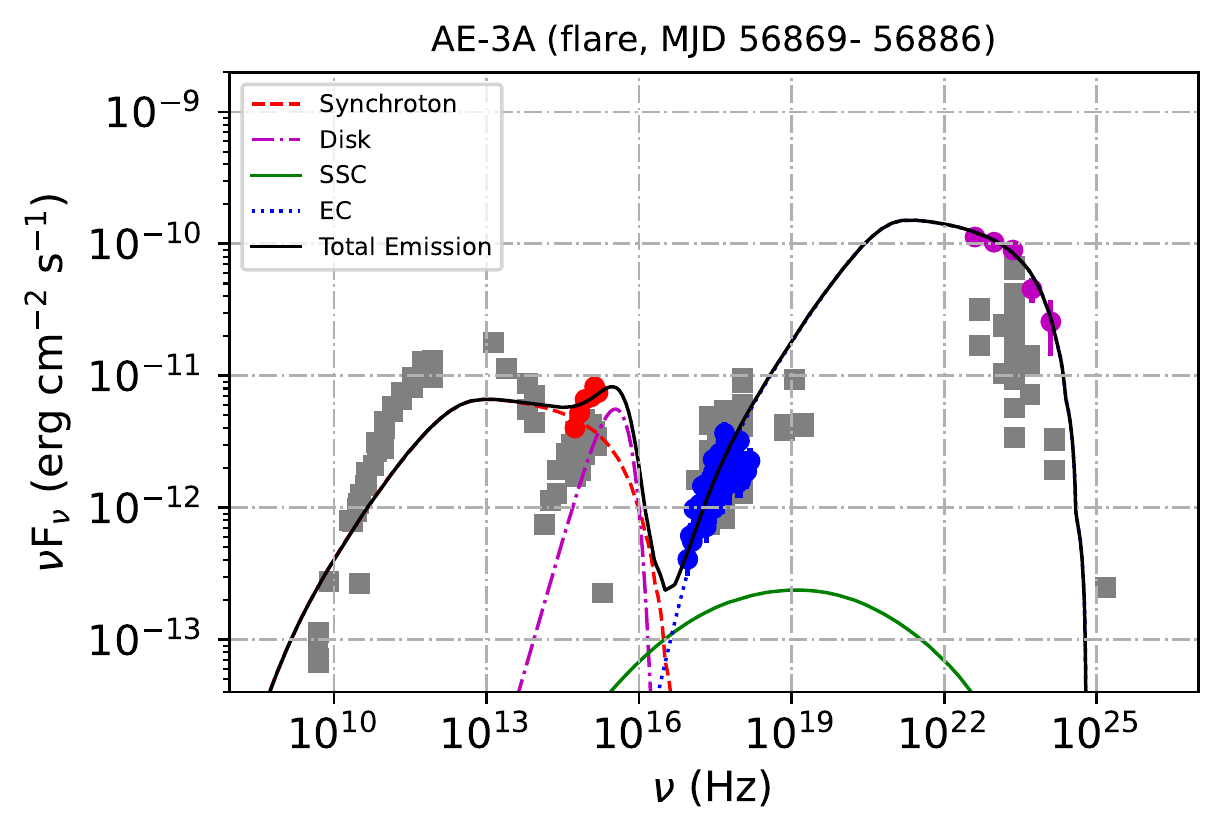}
\includegraphics[height=2.5in,width=3.2in]{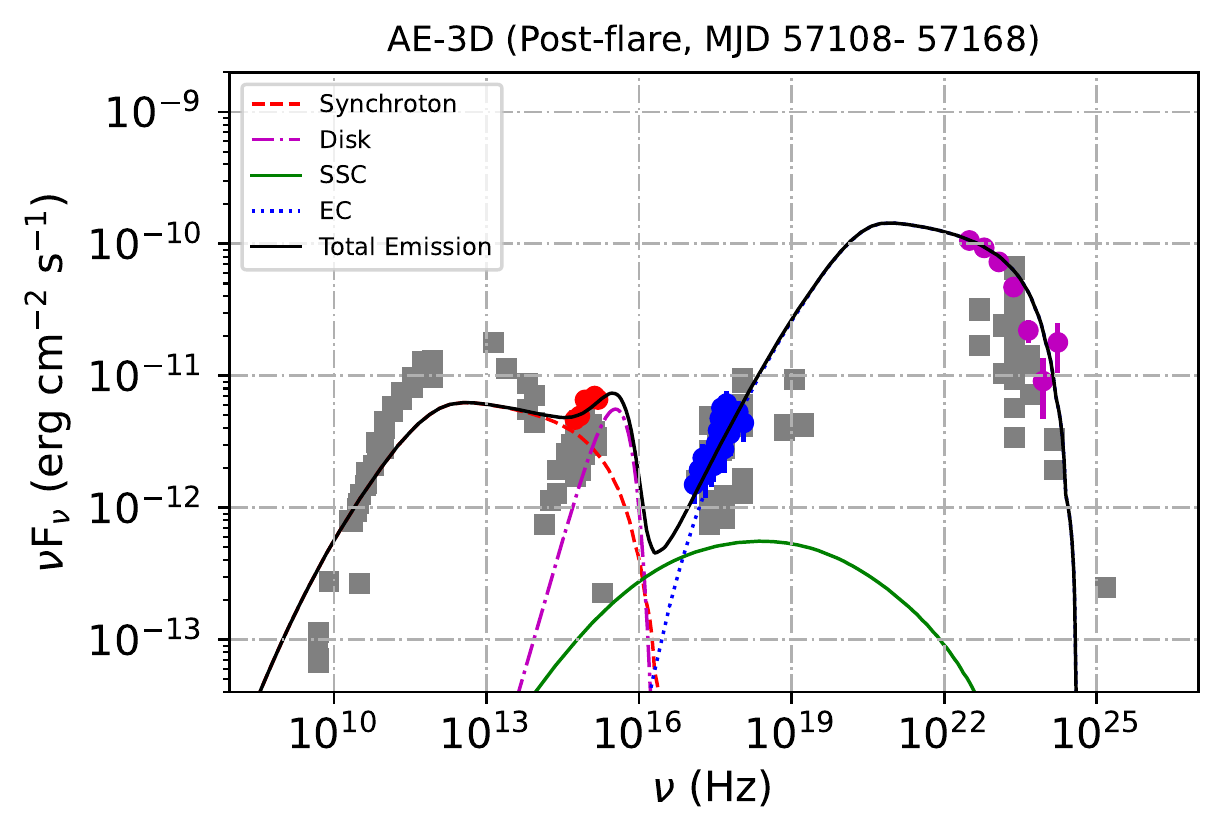}
\caption{Simple one-zone model fits the multi-wavelength SEDs of different phases. Swift-UVOT and XRT data are shown in red and blue colors respectively. Magenta color points represent the $\gamma$-ray data. All the non-simultaneous archival data from the different observatories are shown in square grey colors.}

\label{fig:6}
\end{figure*}

\subsection{Physical Constraint for Multi-Wavelength SED Modeling}
PKS 0402-362 is an FSRQ-type object. During the IC process, besides SSC (Synchrotron self-Compton) we consider External Compton (EC) process in which the seed photons are coming from Broad Line Region (BLR) and accretion disk. We have used Synchrotron and SSC+ EC emission to model the SEDs. Here we have discussed the physical constraints on the model parameters that we have used in our modeling:

    \begin{enumerate}
        \item To calculate the EC emission by the relativistic electrons, Broad Line Region (BLR) photons are taken into account as target photons. The energy density of the BLR photons has been calculated (in the co-moving frame) with the following equation:
        \begin{equation}
            U_{BLR}^{\prime}=\frac{\Gamma^2 \zeta_{BLR}L_{Disk}}{4\pi cR^2_{BLR}}
        \end{equation}
        where $\Gamma$ is the bulk Lorentz factor of the emitting blob whose value is assumed to be 16.24 \citep{Hovatta2009Feb}. The BLR photon energy density is only a fraction of 10\% ($\zeta_{BLR}\sim$ 0.1) of the accretion disk photon energy density. The assumed value of the disk luminosity $L_{Disk}$= 5$\times$10$^{46}$ erg/sec, which is a typical value of FSRQ-type blazars. The radius of the BLR region is derived by the scaling relation, $R_{BLR} \sim (\frac{L_{Disk}}{10^{45}})^{\frac{1}{2}} \times 10^{17}$ cm \citep{Ghisellini2009Aug}.
        
        \item We also include the effect of EC emission for accretion disk photons. We computed the disk energy density in the co-moving jet frame by the following equation \citep{Dermer:1225453}:
        \begin{equation}
            U_{Disk}^{\prime}=\frac{0.207R_g L_{Disk}}{\pi c Z^3 \Gamma^2}
        \end{equation}
        The mass of the central engine or Black Hole (M$_{BH}$) is 10$^9$M$_{\odot}$ \citep{Decarli2011May}. So, the gravitational radius R$_g$=3$\times$10$^{14}$ cm (gravitational radius of sun= 3$\times$10$^{5}$ cm.). The distance of the emission region from the black hole is represented by `Z'. The lower limit of this quantity is estimated by the following equation:
        \begin{equation} \label{eq:13}
            Z\le \frac{2 \Gamma^2 c t_{var}}{1+z}
        \end{equation}
        where z is the redshift of the source. The variability time is estimated as $\sim$ 1.18 days by the procedure described in \citep{2020ApJS..248....8D}. This has been used to estimate `Z'. The value of `Z' is 6.7$\times$10$^{17}$ cm.
        
        \item In the modeling, we consider a spherical emission region or blob of radius R. The size of the emission region (R) can be constrained from the following equation:
        \begin{equation} \label{eq:14}
            R\le\frac{c t_{var} \delta}{1+z}
        \end{equation}
        where t$_{var}$ is the observed variability time, $\delta$ is the Doppler factor of the emission region, which is assumed as 17.0 \citep{Hovatta2009Feb}. z is the redshift of the source. This gives just an approximate constraint on the size of the emission region, as there are several other factors that may affect this estimate  \citep{2002PASA...19..486P}.

        \item The Optical-UV part of the broadband SED is explained by the multi-temperature accretion disk model with the temperature profile given as:
        \begin{equation}
            T(R) =  \Big[\frac{3 R_{s} L_{Disk}}{16 \pi \eta \sigma R^{3}} \Big(1 - \sqrt{\frac{3 R_{s}}{R}} \Big) \Big]^{\frac{1}{4}}
        \end{equation}
        Here, $R_{s}$ (= $\frac{2 G M}{c^{2}}$) is Schwarzchild radius. The maximum flux of the accretion disk spectrum occurs at $\sim 5 R_{s}$ \citep{Ghisellini2009Nov}. $\sigma$ and $\eta$ are the Stefan-Boltzman constant and efficiency ($L_{Disk} = \eta \dot{M} c^{2}$) of the accretion disk, respectively.
    \end{enumerate}

\subsection{Modelling the SEDs}
After constraining the above model parameters, we model the multi-wavelength SEDs of PKS 0402-362 by varying the free parameters in the code `GAMERA'. We consider the constant escape of the leptons from the emission region with escape time $\tau_{esc}\sim$ R/c, where R is the radius of the spherical emission region and c is the speed of light.\\

We have modelled five activity phases: flare phases of AE-1A (flare-1A1, and flare-1A2), AE-2A, AE-3A, and Post-flare phase of AE-3D, shown in the figure Figure-\ref{fig:6}. We have modeled considering a one-zone emission region. During modeling, we have adjusted the values of different parameters e.g. magnetic field (B), the radius of the emission region (R), minimum and maximum electron energy (e$_{min}$ and e$_{max}$, spectral index ($\alpha$) and curvature index ($\beta$). All the values of the fitted parameters for various phases are given in Table-\ref{tab:4}.\\
We also checked the correction due to EBL absorption \cite{Gilmore:2011ks} and did not find a significant effect on the detected gamma-ray band for the source redshift 1.4228. \\

We have calculated the total jet power using the following equation:
    \begin{equation}
        P_{tot}=\pi R^2 \Gamma^2 c (U^\prime_e+U^\prime_B+U^\prime_p)
    \end{equation}
where P$_{tot}$ is the total jet power, $\Gamma$ is the bulk Lorentz factor; U$^\prime_e$, U$^\prime_B$, and U$^\prime_p$ are the energy density of the electrons(and positrons), magnetic field, and cold protons respectively in the co-moving jet frame (prime denotes `co-moving jet frame'; unprime denotes `observer frame').\\
 The power carried by the leptons is given by:
    \begin{equation}
        P_e=\frac{3 \Gamma^2 c}{4 R} \int_{e_{min}}^{e_{max}} EQ(E)dE 
    \end{equation}
    where Q(E) is the injected particle spectrum.\\
    The power due to the magnetic field is calculated by, 
    \begin{equation}
        P_B=R^2 \Gamma^2 c \frac{B^2}{8}
    \end{equation}
where B is the magnetic field, used to model the SED.\\
The energy density in cold protons U$_p^\prime$ is calculated assuming the ratio of electron-positron pair to proton is 10:1. We have maintained the charge neutrality condition in the jet. The jet power of protons is computed using the energy density of cold protons. The Eddington luminosity of this source is $\sim$ 1.26$\times$10$^{47} erg/sec$, which is greater than the total required jet power ($P_{tot}$) to model the SEDs in each case. The results of our modeling for different activity states are shown in Table-\ref{tab:4}. \\

\begin{table*}
\caption{Results of multi-wavelength SED modeling (One-zone). 1st column represents the study of different phases. The time duration of different phases is given in the last column. Second to tenth columns represent the value of various parameters used in modeling. Here, $\alpha$, $\beta$ = Spectral and Curvature index of injected electron spectrum,  $e_{min}$, $e_{max}$ = Minimum and Maximum energy of injected electrons, R = Size of the emission region, $P_e$ = Power in the injected electrons, $P_B$ = Power in the magnetic field, $P_{tot}$ = Total required jet power. Several parameters are kept fixed at a specific value during modelling/fitting: $T^\prime_{BLR}$ = 2.0$\times10^{4}$ K, $T^\prime_{Disk}$ = 1.0$\times10^{6}$ K, $U^\prime_{BLR}$ = 7.14 erg/$cm^3$, $U^\prime_{Disk}$ = 2.08$\times10^{-7}$ erg/$cm^3$, $\delta$ = 17.0, $\Gamma$ = 16.24 (see text for more details).}
\label{tab:4}
\centering
\begin{tabular}{ccccccc rrrr}   
\hline\hline
Activity & $\alpha$ & $\beta$ & $e_{min}$ & $e_{max}$ & B & R & $P_e$ & $P_B$ & $P_{tot}$ & Time duration  \\
& & & (MeV) & (MeV) & (G) &(cm.) & (erg/sec) & (erg/sec) & (erg/sec) & (days) \\
\hline\hline
AE-1A & 2.00 & 0.09 & 24.53  & 5110.00 & 2.6 & 4.0$\times10^{16}$ & 2.37$\times10^{46}$ & 1.07$\times10^{46}$  & 4.35$\times10^{46}$ & 40 \\
(flare-1A1) & \\
\hline
AE-1A & 2.00 & 0.09 & 24.53 & 6132.00 & 2.4 & 4.0$\times10^{16}$ & 2.45$\times10^{46}$  & 9.11$\times10^{45}$ & 4.23$\times10^{46}$ & 20 \\
(flare-1A2) & \\
\hline
AE-2A  & 2.00 & 0.09 & 132.86 & 4854.50 & 1.1 & 5.0$\times10^{16}$ & 4.34$\times10^{46}$  & 2.99$\times10^{46}$ & 5.09$\times10^{46}$ & 14  \\
(flare phase) & \\
\hline
AE-3A & 2.00 & 0.09 & 76.65 & 4854.00 & 2.8 & 6.5$\times10^{16}$ & 1.36$\times10^{46}$  & 3.28$\times10^{45}$  & 4.84$\times10^{46}$ & 17 \\
(flare phase) & \\
\hline
AE-3D & 2.00 & 0.09 & 48.54 & 3832.00 & 2.8 & 4.0$\times10^{16}$ & 2.18$\times10^{46}$  & 1.24$\times10^{46}$   & 3.93$\times10^{46}$  & 60  \\
(Post-flare phase) & \\
\hline\hline
\end{tabular}
\label{tab:MWSED_Param}
\end{table*}

\section{Discussion} \label{sec:7}
Based on the $\sim$ 12.5 years of $\gamma$-ray light curve the source has been seen three times in a major activity phases identified as AE-1, AE-2, and AE-3. Among these phases, AE-2 happens to be the brightest one with flux more than 20 times the average $\gamma$-ray flux (see Figure-\ref{fig:1}). To establish the size and location of the emission region we calculated the flux variability time as 1.18 days using the $\gamma$-ray light curve. Using this variability time we estimated the size of the emission region as 2.1$\times$10$^{16}$cm (ref Equation-\ref{eq:14}) and it is located at 6.7$\times$10$^{17}$cm or 0.2 pc (computed using Equation-\ref{eq:13}) from the base of the jet. This is an FSRQ type source and using the typical value of disk luminosity (5$\times$10$^{46}$ erg/s) we can derive the size of the broad-line region (BLR) which is found to be 7.1$\times$10$^{17}$cm. Comparing this size with the location of the emission region we can say that the emission region is located just at the boundary of the BLR. This finding can be confirmed by producing the $\gamma$-ray spectra. It is known since 1992 (CGRO-EGRET mission) that BLR is opaque to $\gamma$-ray photons in the FSRQ due to $\gamma$-$\gamma$ pair production (\citealt{Becker1995Nov}, \citealt{Donea2003Jan}). The BLR starts to be opaque above $\sim$20/1+z GeV (e.g., \citet{2006ApJ...653.1089L}). Indeed, we do not observe any high energy photons above $\sim$10 GeV for this source with redshift, z = 1.4228  (see Figure-\ref{fig:3} and Figure-\ref{fig:A13} - Figure-\ref{fig:A18}). However, the absence of higher than 10 GeV photons may also be due to the short integration times used in the spectra. \\
We used the Bayesian Block method to identify the flares and further these flares were explored with days scale finer binned light curves. Flare phases were found to be a combination of many peaks which were fitted with a standard sum of exponentials to estimate the rise and decay time of the flares (see Table-\ref{tab:2}). To compare the rise and decay timescales (T$_{r}$ and T$_{d}$), we define a asymmetry parameter K = $\frac{T_d - T_r}{T_d + T_r}$. A peak is symmetric if the value of $\lvert K \rvert \leq$ 0.3 \citep{Das_2021}. The value of K for all the peaks has been given in Table-\ref{tab:2}. In our study, most of the time we found asymmetric peaks either with fast rise slow decay or slow rise fast decay except peak P2, P3 of AE-1A and P2 of AE-3B where it is more symmetric within the error bar. The variability in the flare can be caused by the interaction of the blob (emission region) with a standing shock (\citealt{1979ApJ...232...34B, 1985ApJ...298..114M}). However, there could be other possibilities, such as the interaction of internal shocks with moving blob or magnetic reconnection (\citealt{Spada2001Aug, Giannios2013May}). The symmetric peaks are expected to occur when the particles radiate all their energy within the light-crossing time ($\sim$ R/c), and observation of such peaks can be compatible with the scenario of a blob passing through a standing shock. A similar explanation is also given by \citet{10.1093/mnras/sty2748}. Considering the scenario where the blob interacts with shocks, once the blob is moved out or the shock is moving we expect to see asymmetric peaks. This can also be seen as a scenario where the radiative cooling time is longer than the light travel time and as a result, a long-decay flare is observed. This means the radiative cooling time scale which includes synchrotron and IC cooling time scales is longer than the particle injection which is more or less instantaneous. Symmetric and asymmetric peaks are commonly seen in many blazars one of the examples is PKS 1510-089 (\citealt{Prince_2017}).

\begin{figure}
\includegraphics[height=2.5in,width=3.7in]{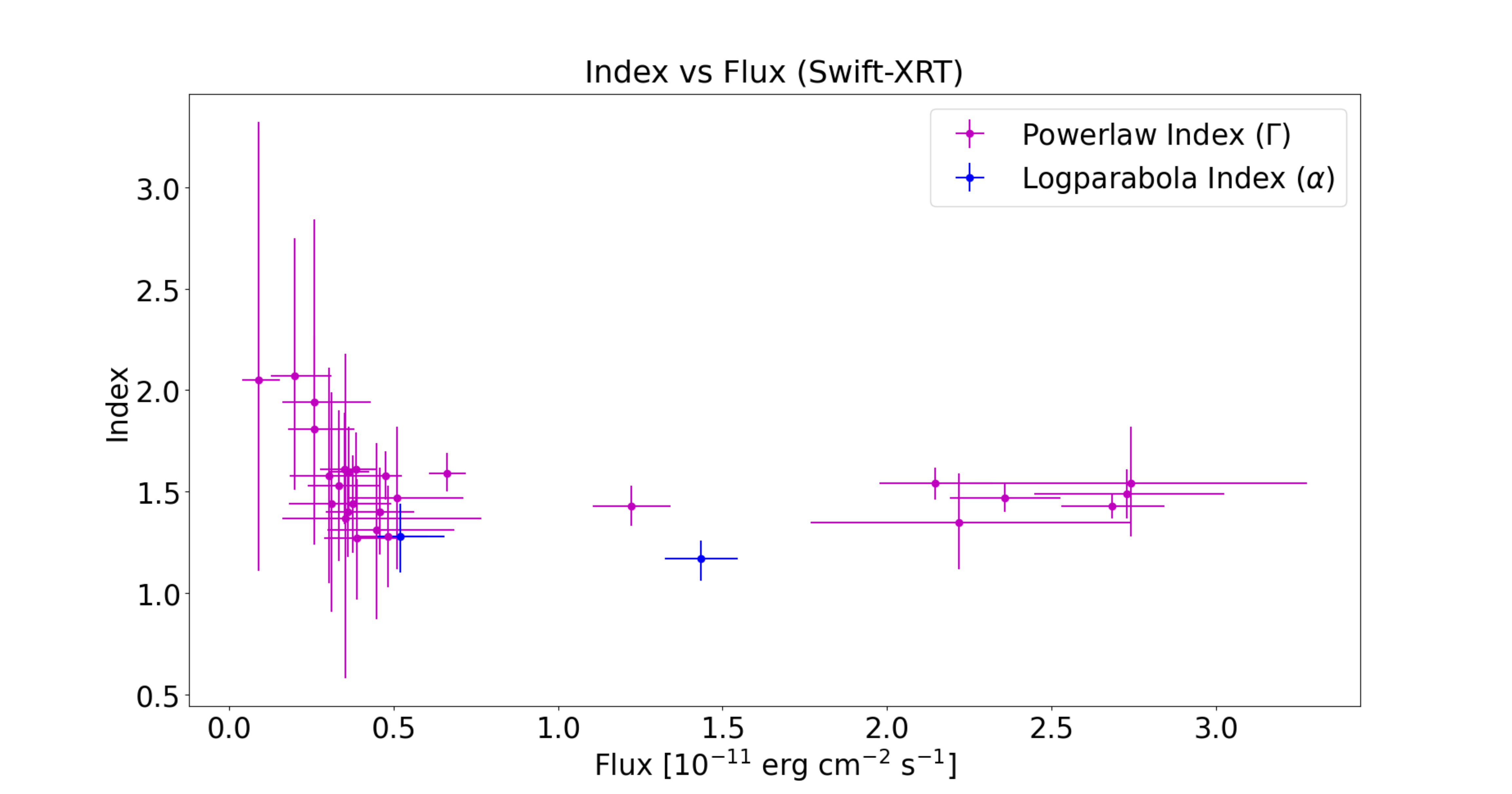}
\caption{Flux-index diagram for Swift-XRT light curve.}
\label{fig:flux-index}
\end{figure}

The X-ray photon index plotted against the 0.3-10 keV flux shows a weak correlation (correlation coefficient, r = 0.30 and p-value = 0.50) at flux value higher than $> 1.0 \times 10^{-11}$ erg cm$^{-2}$ s$^{-1}$ with a mild softer-when-brighter (Figure-\ref{fig:flux-index}) trend which is generally seen in FSRQ. However, in BL Lac type sources, the trend is opposite to this, as shown in \citet{10.1093/mnras/stac1866} for TeV blazar 1ES 1727+502.

The broadband light curves shown in Figure-\ref{fig:4} clearly show the simultaneous flaring in optical and infrared. In X-ray, we do not see much activity because of the very bad sampling of the data. Nonetheless, we performed the correlation study among the $\gamma$-ray and BVRJ band light curves and the results are shown in Figure-\ref{fig:5}.
Apparently, the optical-IR light curves were available only for the AE-2 and AE-3 phase. A nearly zero time lag was found between $\gamma$-ray and optical-IR emission suggesting the co-spatiality of their origin and demanding a single emission zone or blob in the broadband SED modeling. In the leptonic scenario, the IR-optical-UV emission is governed by the synchrotron process and the $\gamma$-ray emission is most suitably produced by the Inverse-Compton scattering of internal or external photons.  Single-zone emission model is one of the most well-established models under the leptonic scenario.  In our study, we found that one-zone emission is sufficient to explain the broadband simultaneous SED in FSRQ PKS 0402-362. This is an FSRQ-type source and hence the disk emission is expected to be higher than the BL Lac type of sources. A dominant accretion disk is observed for this source which is well fitted by the combination of synchrotron and the thermal disk emission (see Figure-\ref{fig:6}). This source has a strong accretion disk as it was also observed by \citet{2018A&A...610A...1M} (see there Figure A.2) which makes this source a good candidate to explore the disk-jet connection, which is out of scope for this paper. As we discussed earlier that the emission region is located at the boundary of the BLR, therefore, it can provide seed photons for the external Compton (EC). As a result, we found that the high-energy peak of the SED is well described by the EC from the BLR. It is shown for SSC, the contribution is much below the observed $\gamma$-ray data points. The broadband SED of all the phases is shown in Figure-\ref{fig:6} along with the archival data in grey color. In our modeling, we have a few free parameters such as magnetic field inside the blob, particle minimum and maximum energy, particle distribution index, and size of the emission region, which we optimized to get the best fit to the data. The slope and curvature index of the particle injection spectrum were found as $\alpha$=2.00 and $\beta$ = 0.09, respectively, from the fit. The value of the power-law index ($\alpha$), what we get from the modeling, is similar to what is expected in First-order diffusive shock acceleration, which confirms that the shock acceleration is the dominant process behind the particle acceleration to a very high energy such as $e_{max}$ = 4000-6000 MeV (ref Table-\ref{tab:4}). The magnetic field was found to be between 1.1 to 2.8 Gauss which is expected in blazar jets. The fitted size of the emission region is in close agreement with the observation value estimated from the variability time. The calculated jet power in the electron and magnetic field are comparable but dominated by the electron power. The total required jet power is well within the Eddington luminosity of the source. Having the disk domination in this source, it is highly likely that the disturbance in the jet (and so flaring event) is a propagation effect of perturbation produced in the disk. However, other explanations are also possible. To explore the main reason behind the flaring event inside the jet more rigorous disk-jet connection study is required.

\section{Conclusions} \label{sec:8}
The long-term study reveals multiple Activity Epochs (AE) of different durations spread over 12 years of the timeline. The long-term correlation study suggests that broadband emissions are co-spatial, and the variability study implies that they are produced within the BLR. The broadband SED modeling of all the flare phases is consistent in terms of parameters value suggesting similar conditions for high flux phase production.  As expected in FSRQs, the accretion disk dominates the optical-UV emission which is visible across the flaring states and that makes this object
a good source for variability studies of the disk-jet connection.

\section*{Acknowledgements}
We thank the anonymous referee for their comments and suggestions which have helped us to improve the manuscript and the scientific gain of the project.
R.P. acknowledges the support by the Polish Funding Agency National Science Centre, project 2017/26/A/ST9/00756 (MAESTRO 9), and European Research Council (ERC) under the European Union’s Horizon 2020 research and innovation program (grant agreement No. [951549]).

\section*{Data Availability}

This work has made use of publicly available Fermi-LAT data obtained from FSSC's website data server and provided by NASA Goddard Space Flight Center. This work has also made use of data, software/tools obtained from NASA High Energy Astrophysics Science Archive Research Center (HEASARC) developed by the Smithsonian Astrophysical Observatory (SAO) and the XRT Data Analysis Software (XRTDAS) developed by ASI Science Data Center, Italy. The archival data from the Submillimeter Array observatory has also been used in this study \citep{2007ASPC..375..234G}. This paper has made use of up-to-date SMARTS optical/near-infrared light curves that are available at \url{www.astro.yale.edu/smarts/glast/home.php}. The Submillimeter Array is a joint project between the Smithsonian Astrophysical Observatory and the Academia Sinica Institute of Astronomy and Astrophysics and is funded by the Smithsonian Institution and the Academia Sinica.

\bibliographystyle{mnras}
\bibliography{example}
\appendix
\section{Figures and Tables}

\begin{figure*}
\centering
\includegraphics[height=2.6in,width=5.3in]{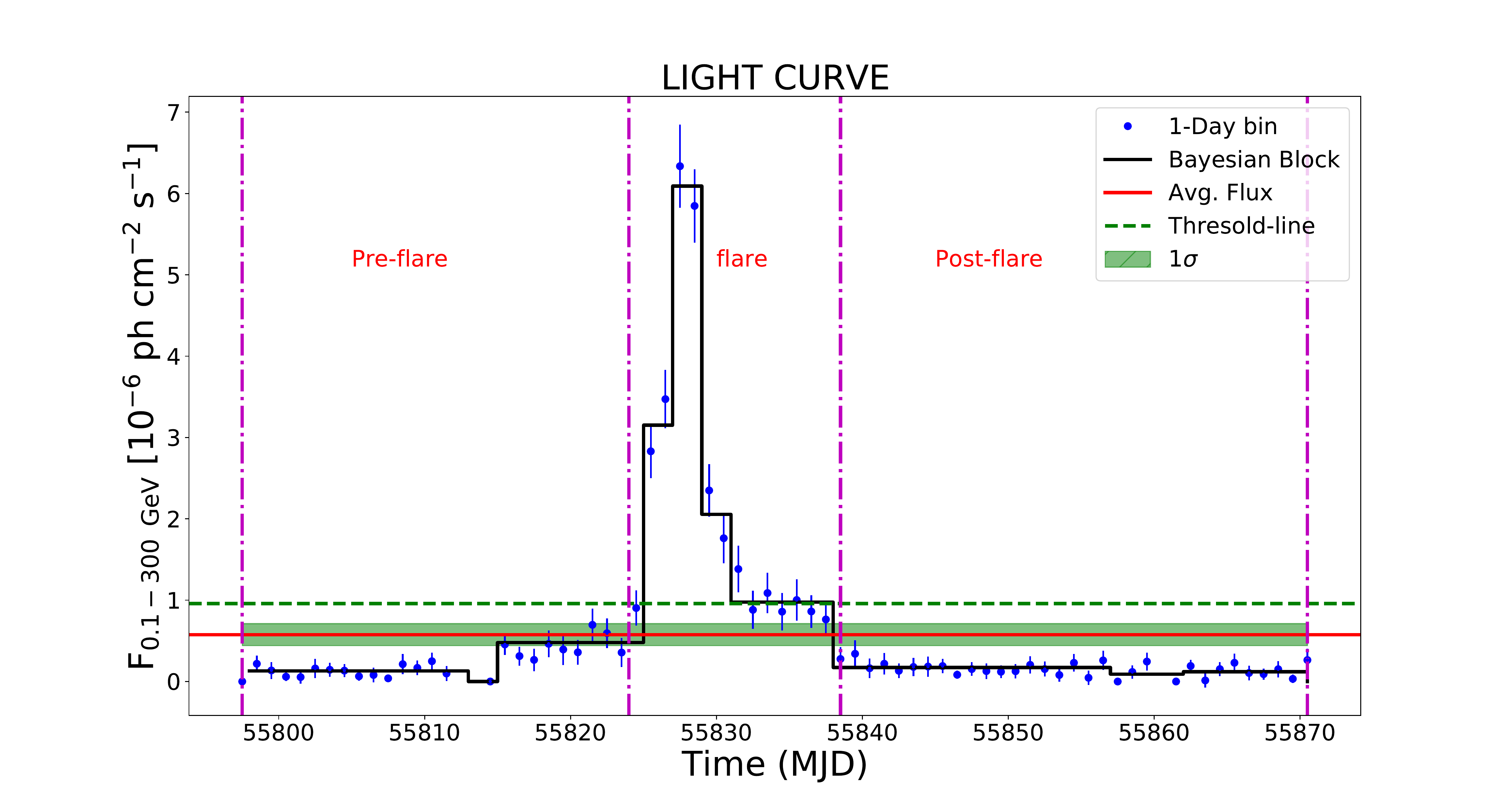}
\caption{One day binning light curve of \textbf{AE-2A}. The time duration of the different phases are MJD 55797 - 55824 (Pre-flare), MJD 55824 - 55838 (flare), MJD 55838 - 55870 (Post-flare).}
\label{fig:A1}

\end{figure*}

\begin{figure*}
\centering

\includegraphics[height=2.6in,width=5.3in]{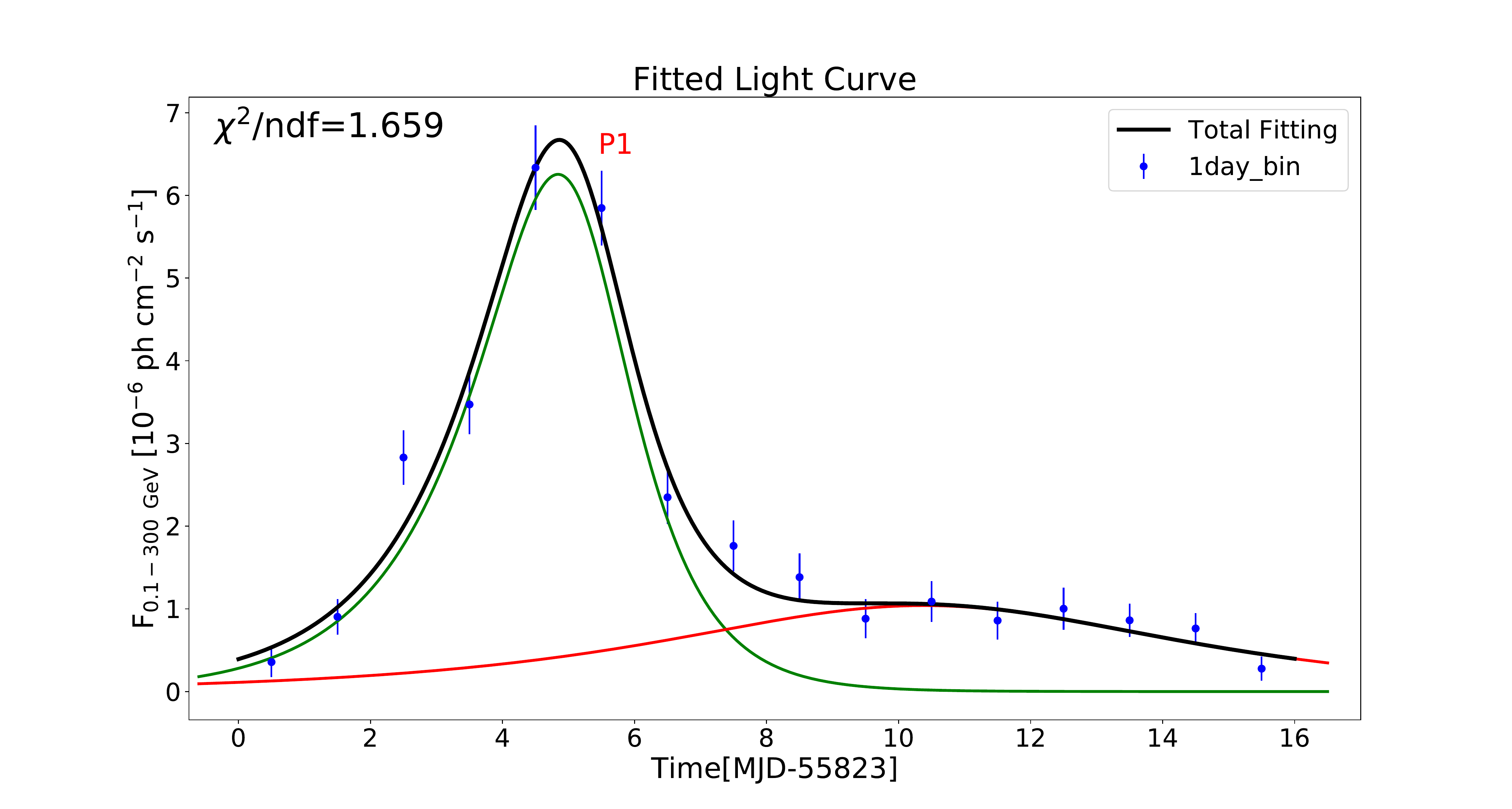}
\caption{Fitted light curve of flare phase of AE-2A with time span of 14 days (MJD 55824 - 55838).}
\label{fig:A2}

\includegraphics[height=2.8in,width=5.3in]{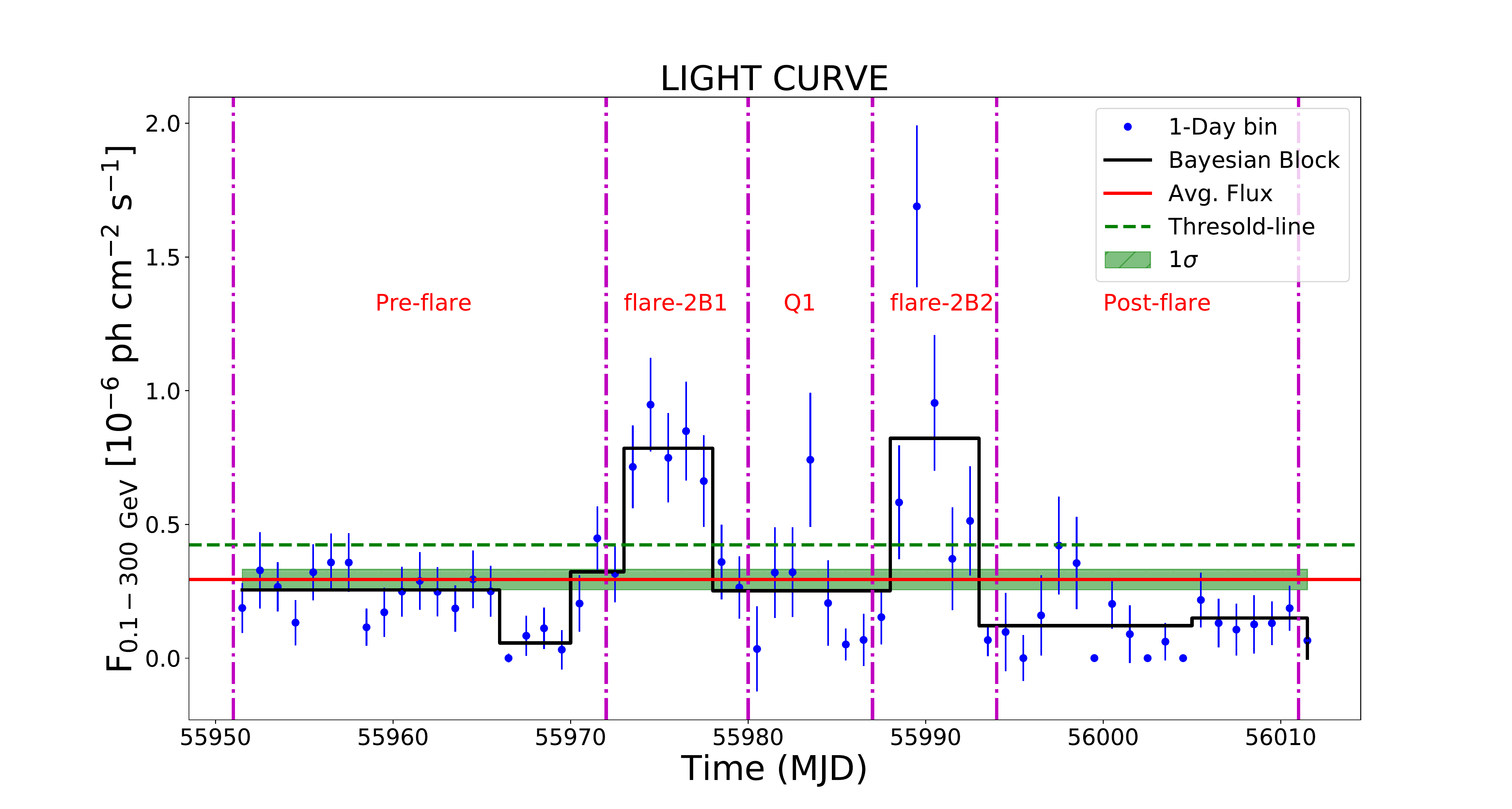}
\caption{One day binning light curve of AE-2B. The time duration of the different phases are  MJD 55951 - 55972 (Pre-flare), MJD 55972 - 55980 (flare-2B1), MJD 55980 - 55987 (Q1), MJD 55987 - 55994 (flare-2B2), MJD 55994 - 56011 (Post-flare).}
\label{fig:A3}

\includegraphics[height=2.6in,width=5.3in]{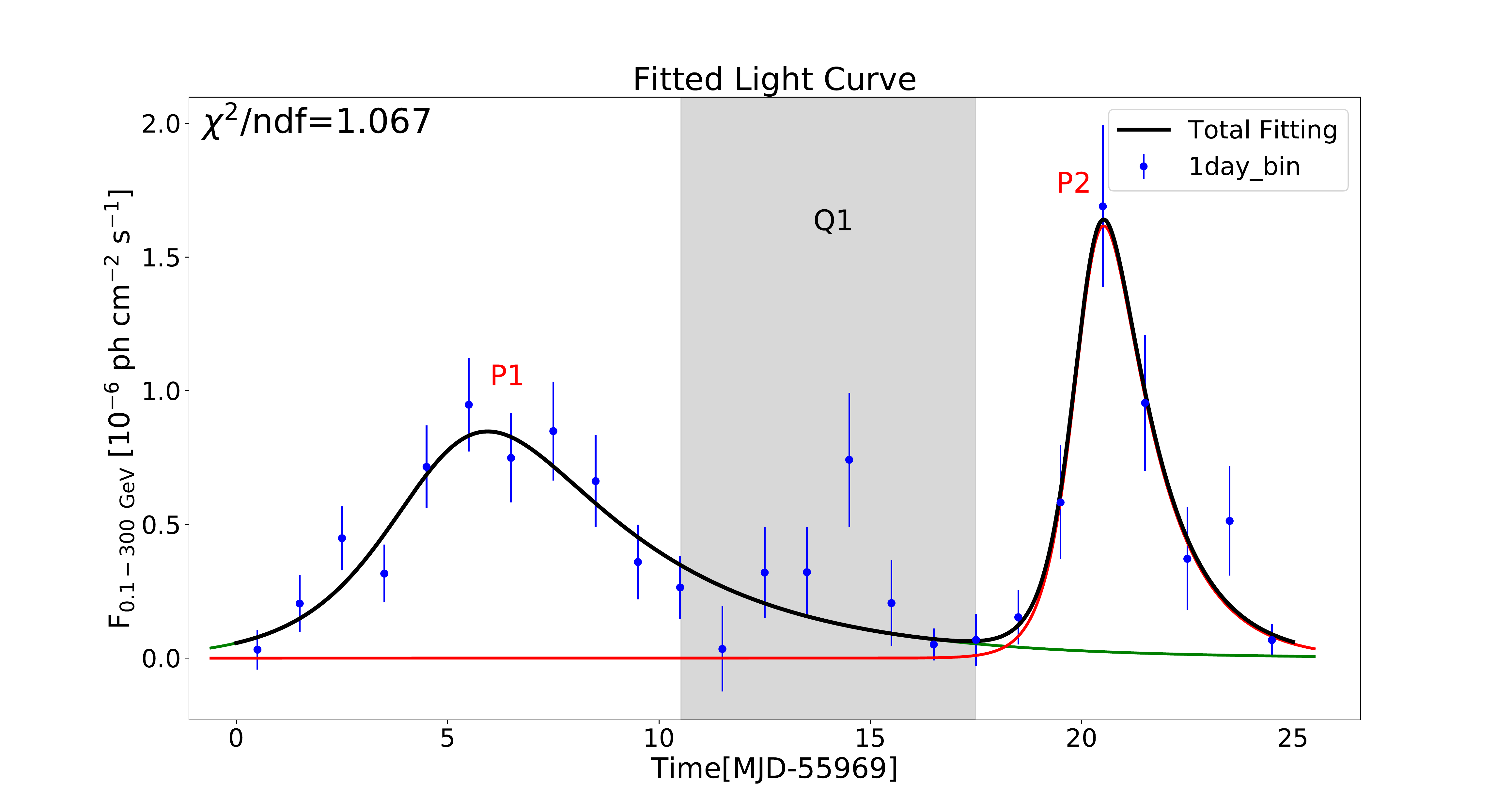}
\caption{Fitted light curve of flare-2B1 and flare-2B2 phases with time span of 8 days (MJD 55972 - 55980) and 7 days (MJD 55987 - 55994), respectively. Q1 phase (MJD 55980 - 55987) is shown in grey shaded region.}
\label{fig:A4}

\end{figure*}

\begin{figure*}
\centering
\includegraphics[height=2.8in,width=5.3in]{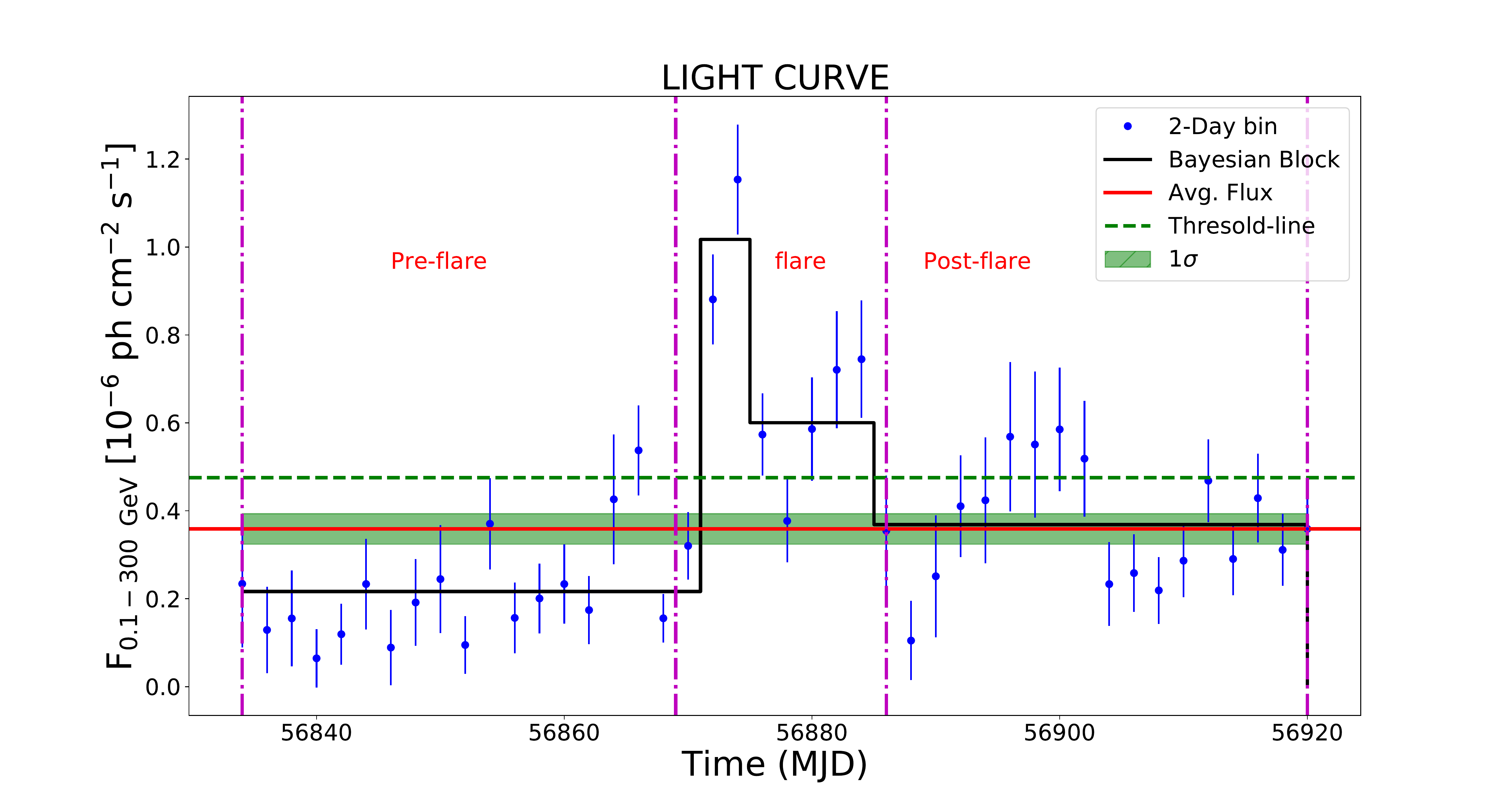}
\caption{Two day binning light curve of AE-3A. Time duration of the different phases, MJD 56834 - 56869 (Pre-flare), MJD 56869 - 56886 (flare), MJD 56886 - 56920 (Post-flare).}
\label{fig:A5}

\includegraphics[height=2.8in,width=5.3in]{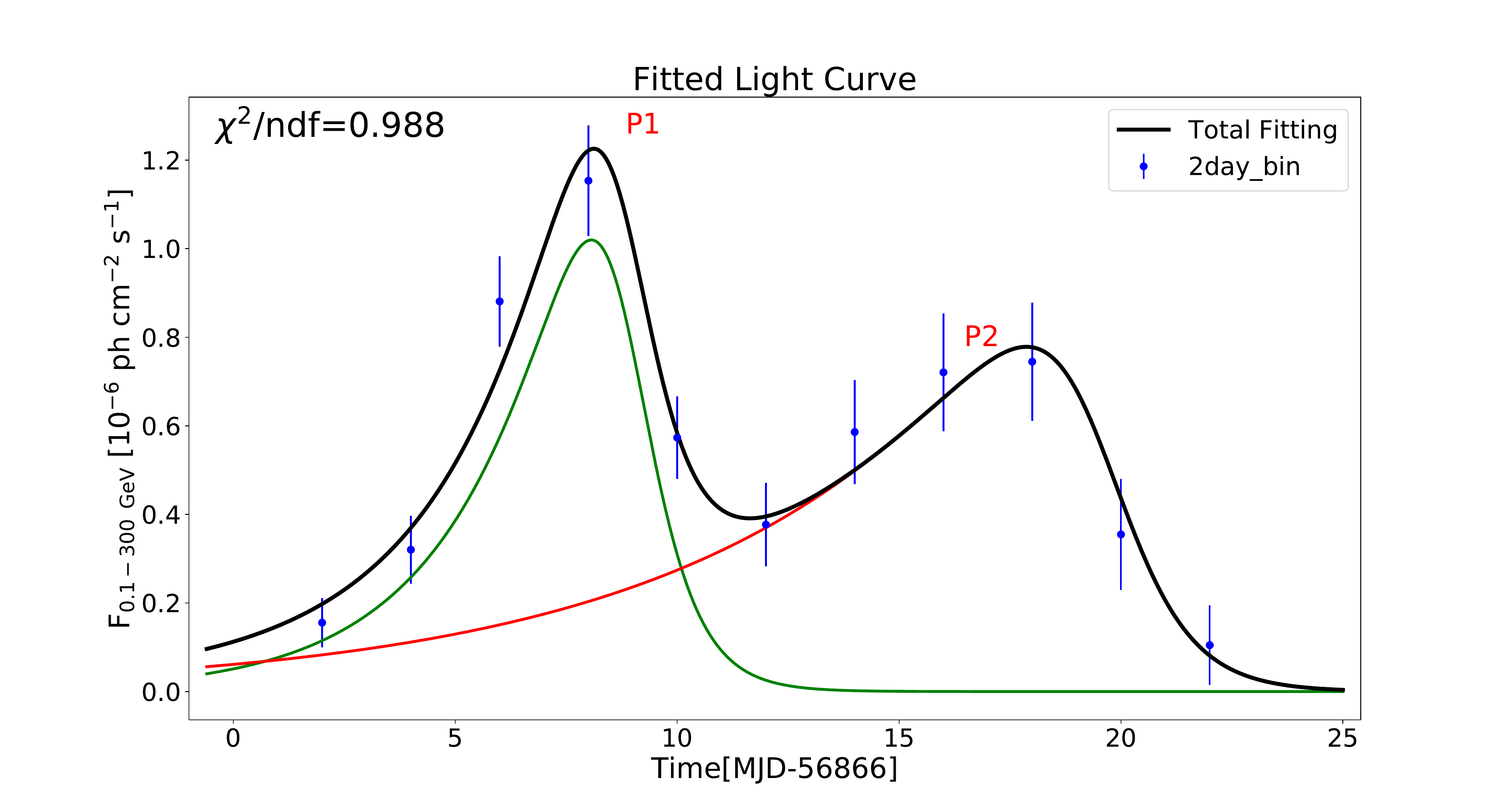}
\caption{Fitted light curve of flare phase of AE-3A with time span of 17 days (MJD 56869-56886).}
\label{fig:A6}

\includegraphics[height=2.8in,width=5.3in]{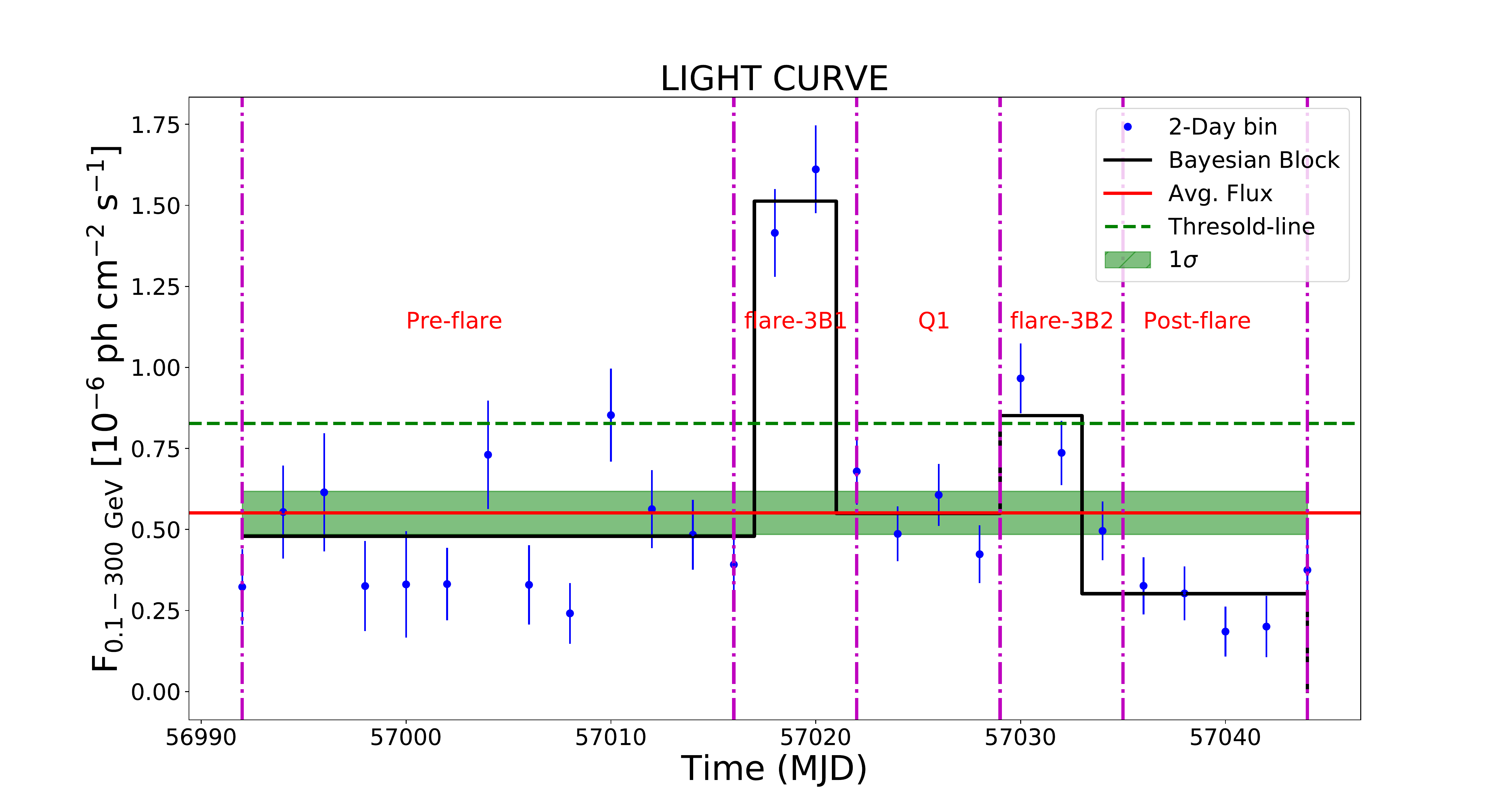}
\caption{Two day binning light curve of AE-3B. The time duration of the different phases are MJD 56992 - 57016 (Pre-flare), MJD 57016 - 57022 (flare-3B1), MJD 57022 - 57029 (Q1), MJD 57029 - 57035 (flare-3B2), MJD 57035 - 57044 (Post-flare).}
\label{fig:A7}
\end{figure*}

\begin{figure*}
\centering
\includegraphics[height=2.8in,width=5.3in]{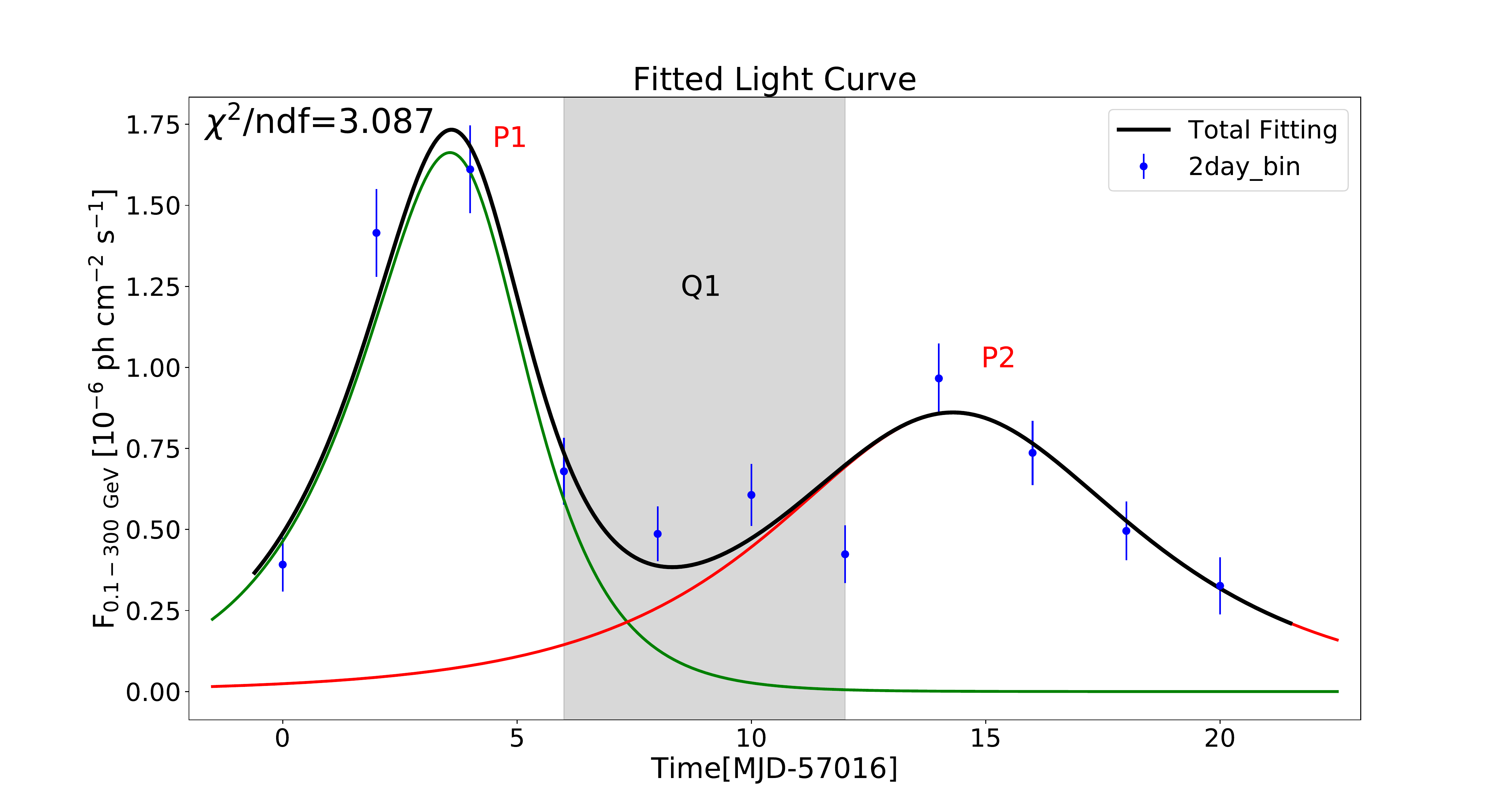}
\caption{Fitted light curve of flare-3B1 and flare-3B2 phases with time span of 8 days (MJD 57016 - 57022) and 6 days (MJD 57029 - 57035) respectively. Q1 phase (MJD 57022 - 57029) is shown in grey shaded region.}
\label{fig:A8}

\includegraphics[height=2.8in,width=5.3in]{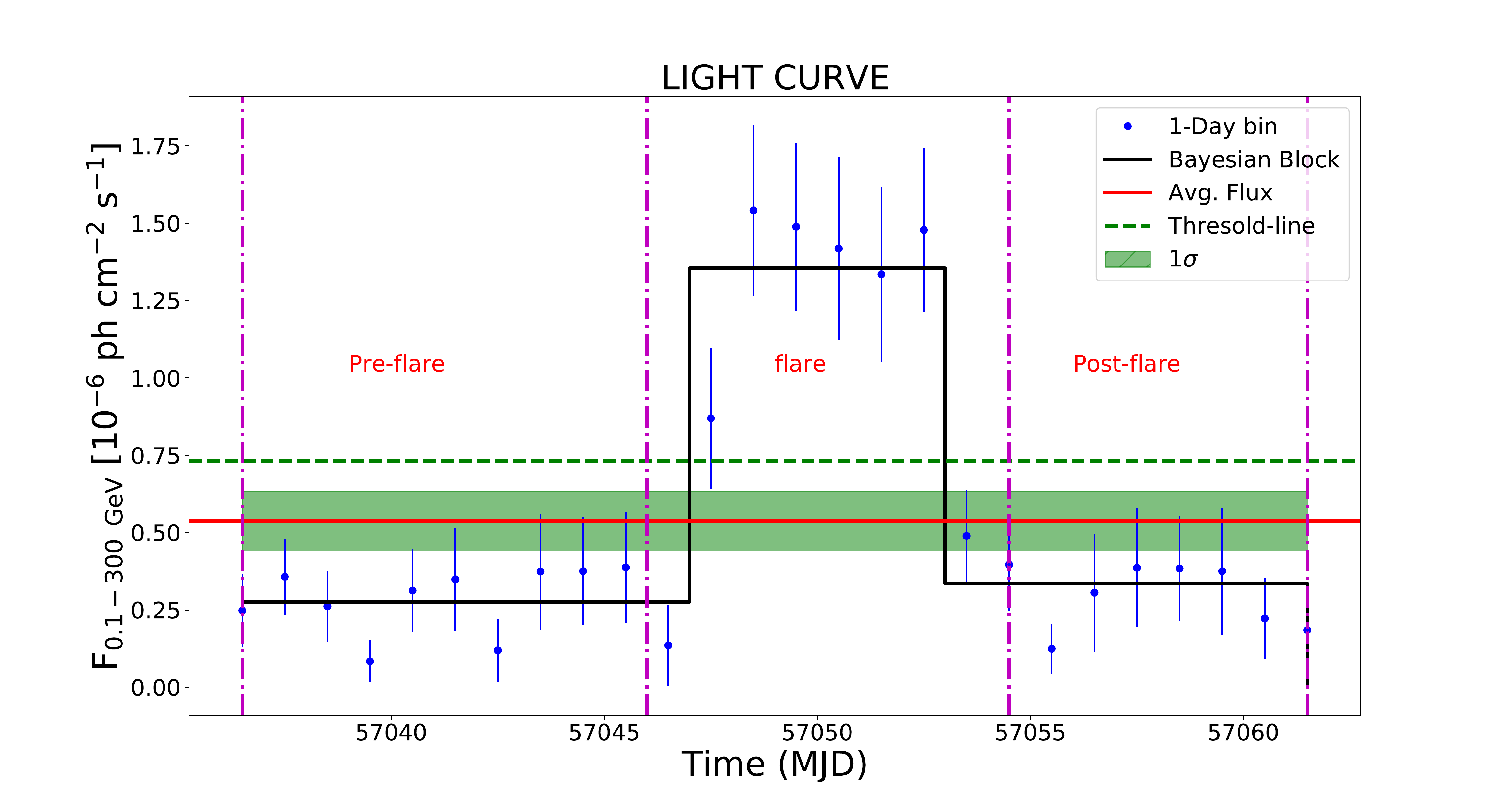}
\caption{One day binning light curve of AE-3C. The time duration of the different phases are MJD 57036 - 57046 (Pre-flare), MJD 57046 - 57054 (flare), MJD 57054 - 57061 (Post-flare).}
\label{fig:A9}

\includegraphics[height=2.8in,width=5.3in]{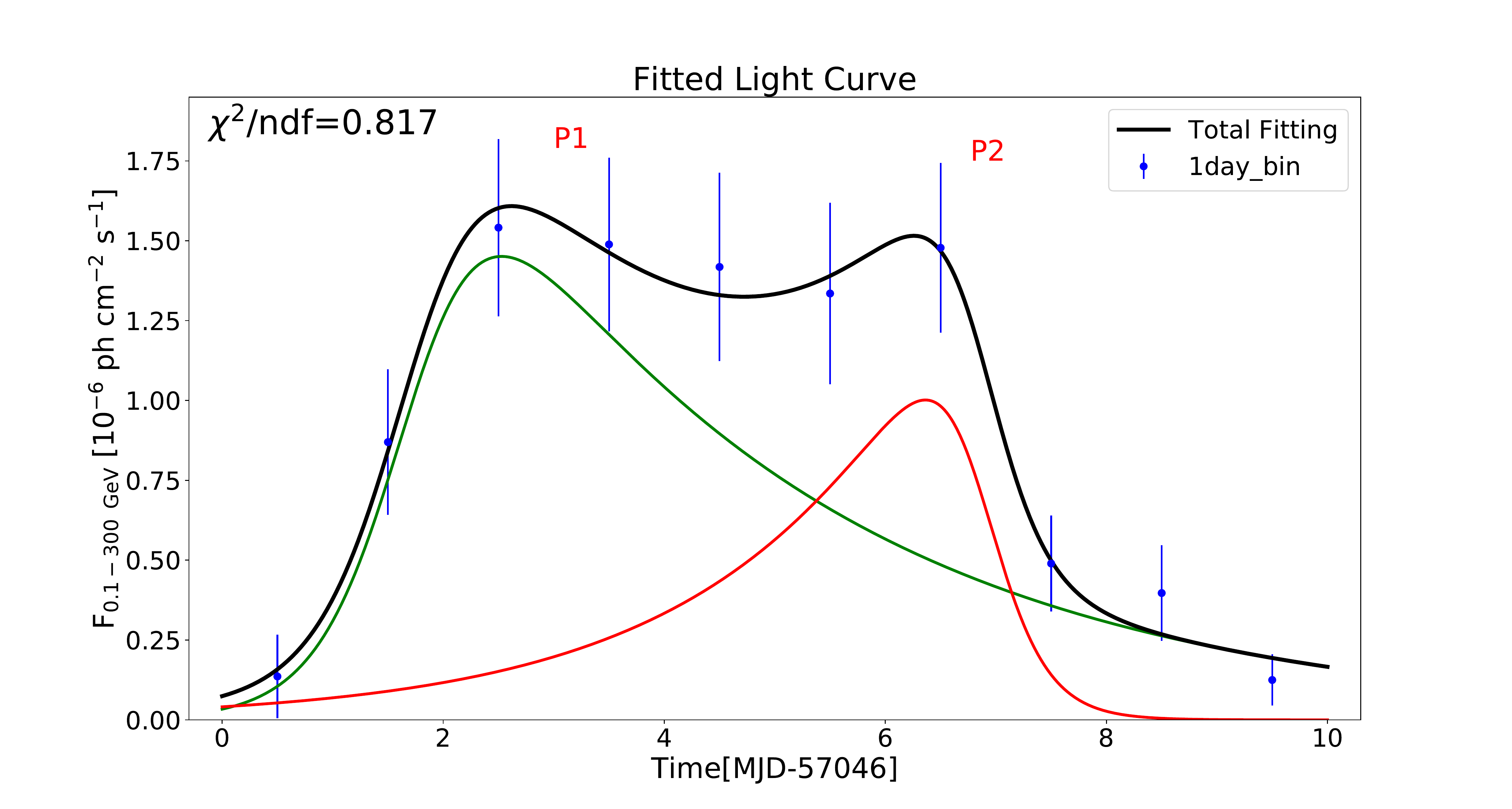}
\caption{Fitted light curve of flare phase of AE-3C with time span of 8 days (MJD 57046 - 57054).}
\label{fig:A10}
\end{figure*}

\begin{figure*}
\centering
\includegraphics[height=2.8in,width=5.3in]{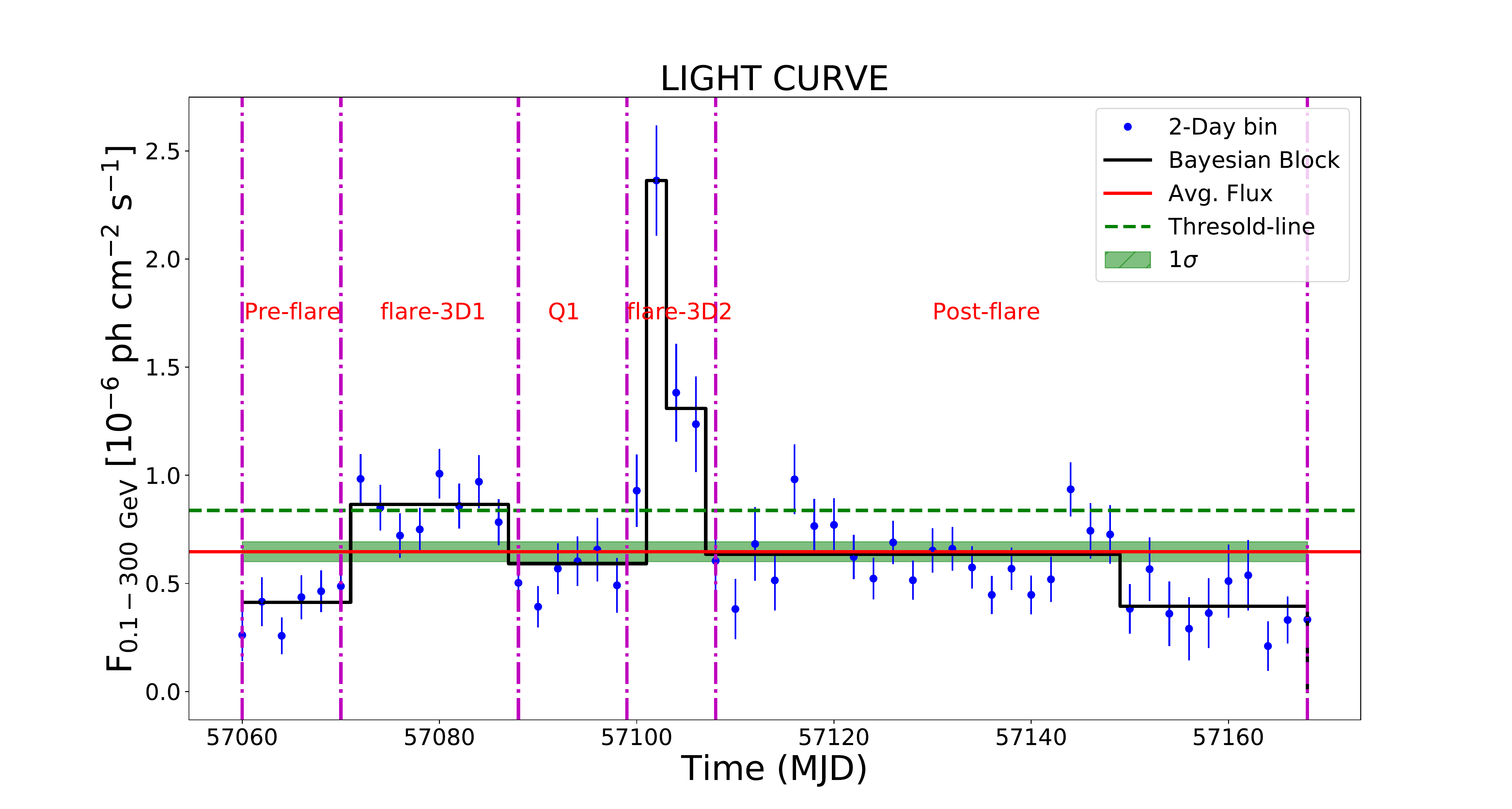}
\caption{Two day binning light curve of AE-3D. The time duration of the different phases are MJD 57060 - 57070 (Pre-flare), MJD 57070 - 57088 (flare-3D1), MJD 57088 - 57099 (Q1), MJD 57099 - 57108 (flare-3D2), MJD 57108 - 57168 (Post-flare).}
\label{fig:A11}

\includegraphics[height=2.8in,width=5.3in]{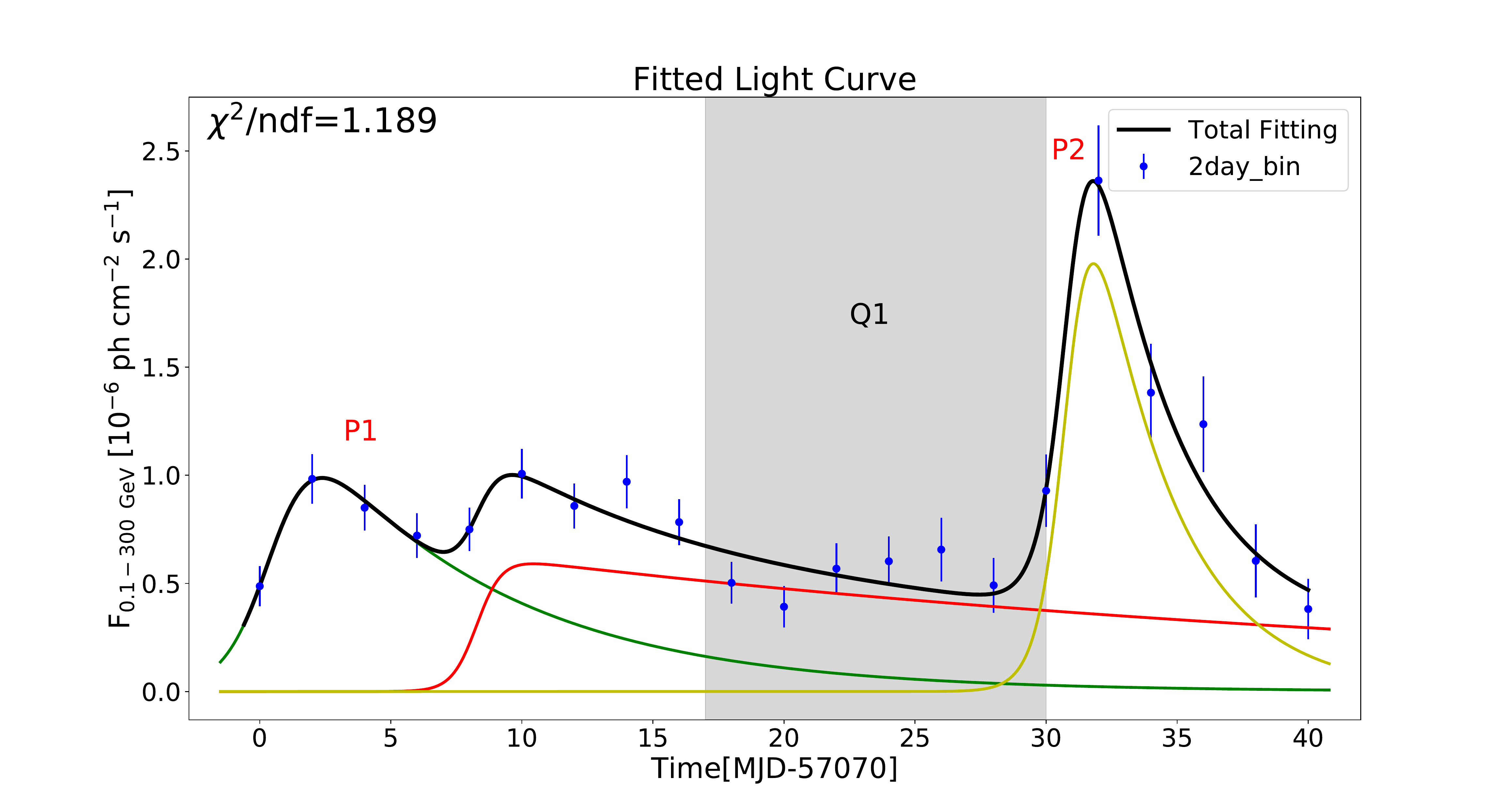}
\caption{Fitted light curve of flare-3D1 and flare-3D2 phases with time span of 18 days (MJD 57070 - 57088) and 9 days (MJD 57099 - 57108), respectively. Q1 phase (MJD 57088 - 57099) is shown in grey shaded region.}
\label{fig:A12}

\end{figure*}

\begin{figure*}
\centering
\includegraphics[height=1.90in,width=2.6in]{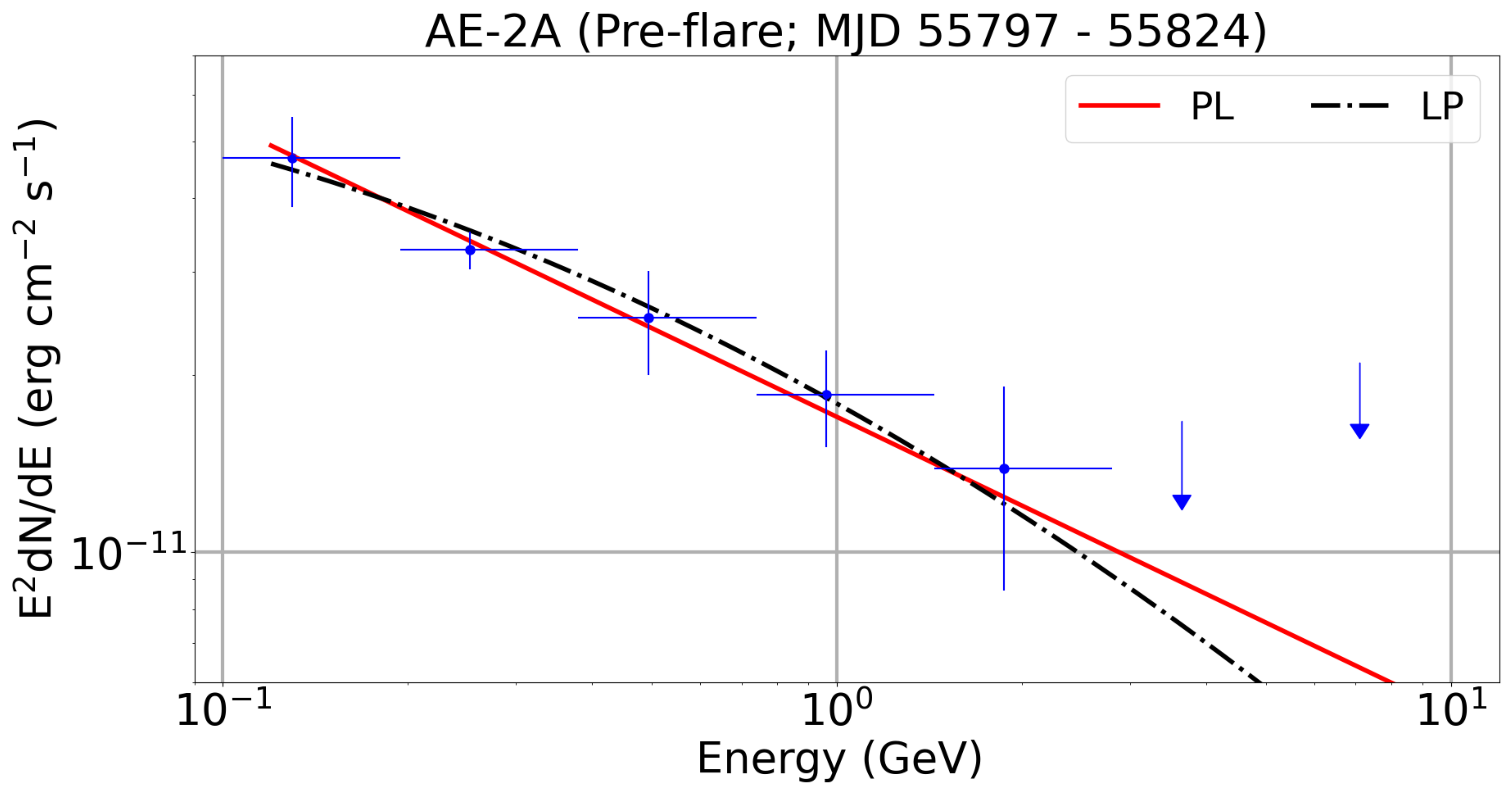}
\includegraphics[height=1.90in,width=2.6in]{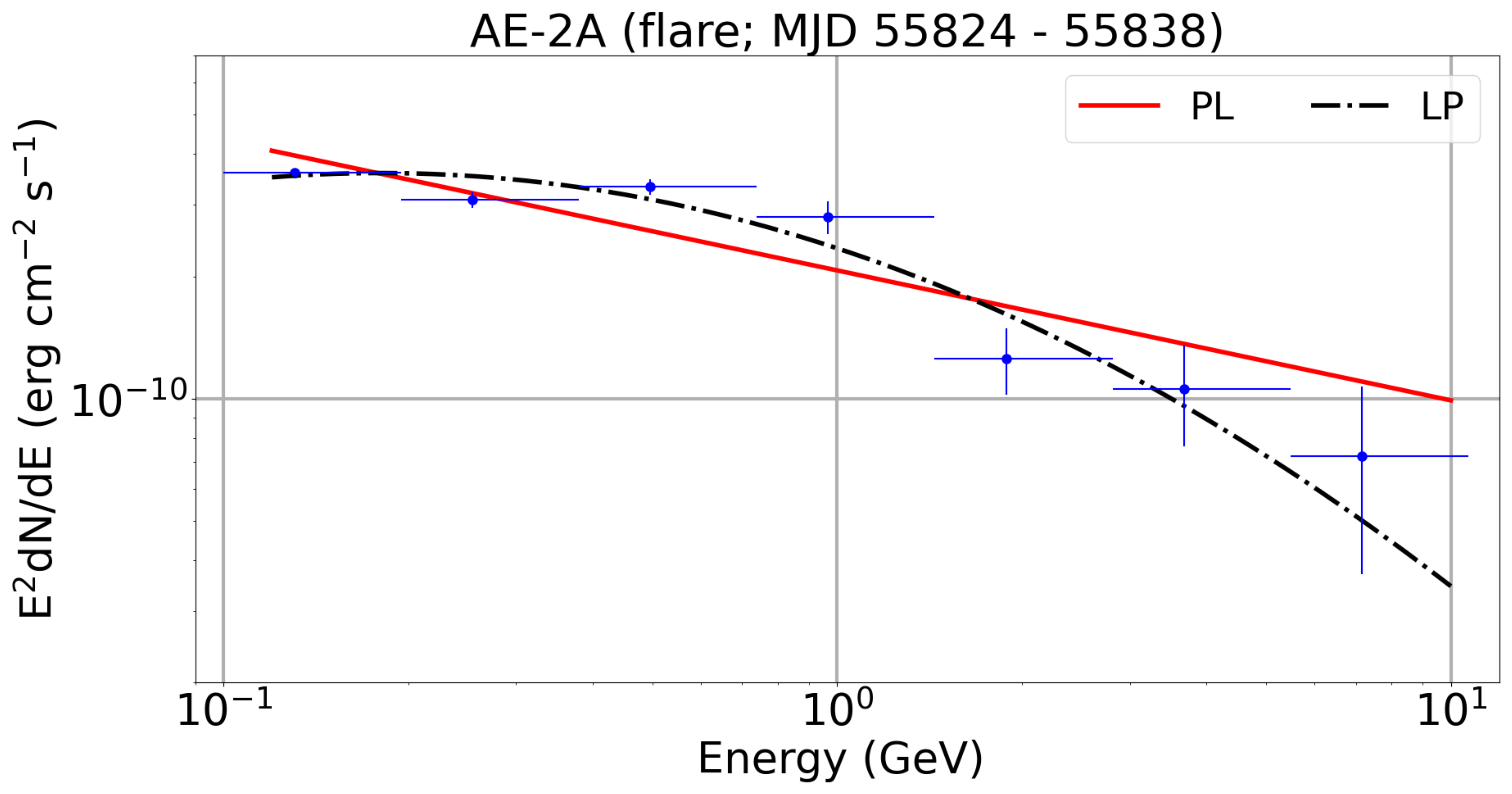}
\includegraphics[height=1.90in,width=2.6in]{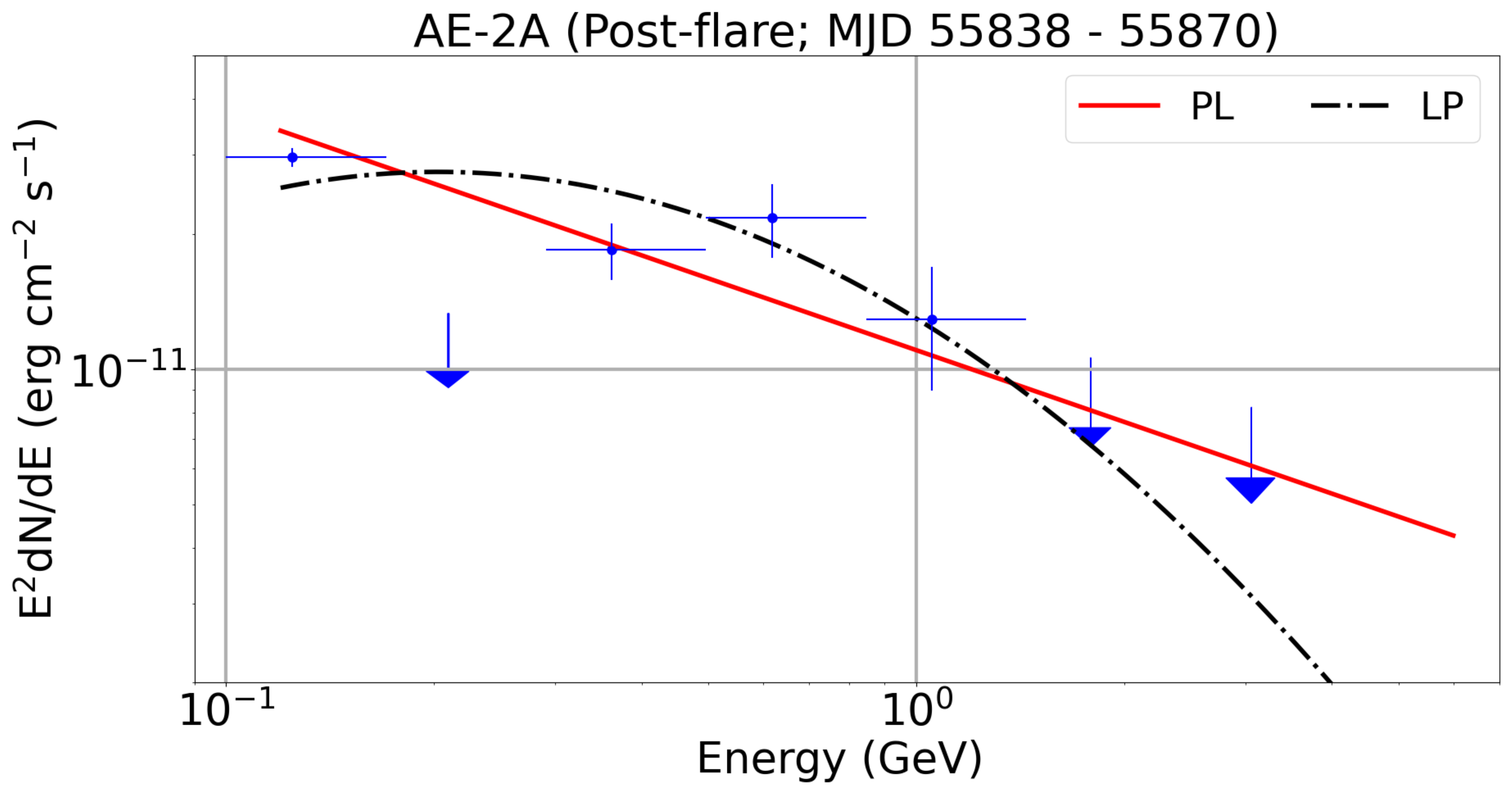}
\caption{Gamma-ray SEDs of different phases (Pre-flare, flare, Post-flare) of AE-2A. PL, LP describe the Powerlaw, Logparabola model respectively, which are shown by solid red and dash-dot black line, respectively.}
\label{fig:A13}

\end{figure*}

\begin{figure*}
\centering
\includegraphics[height=1.90in,width=2.6in]{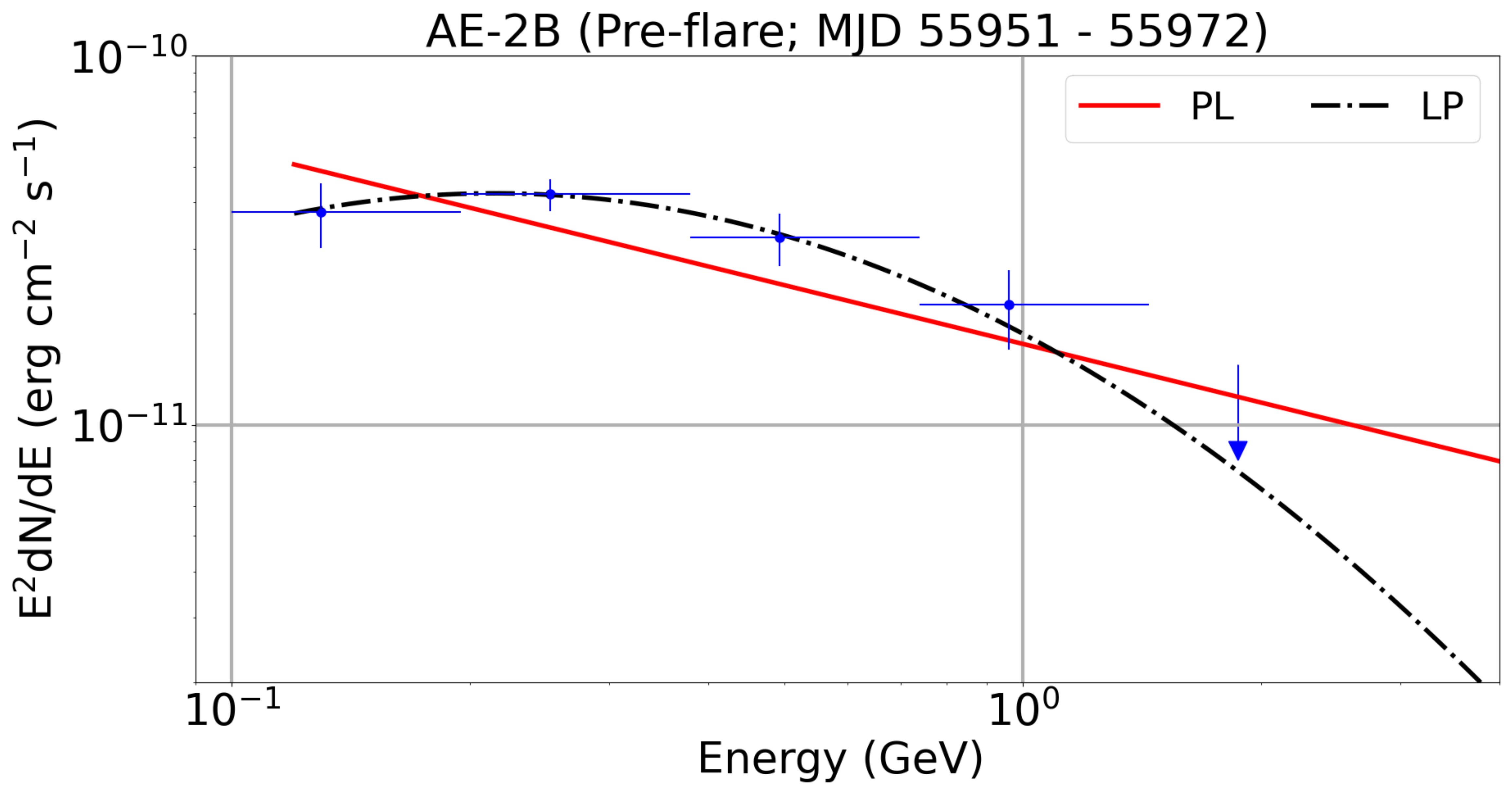}
\includegraphics[height=1.90in,width=2.6in]{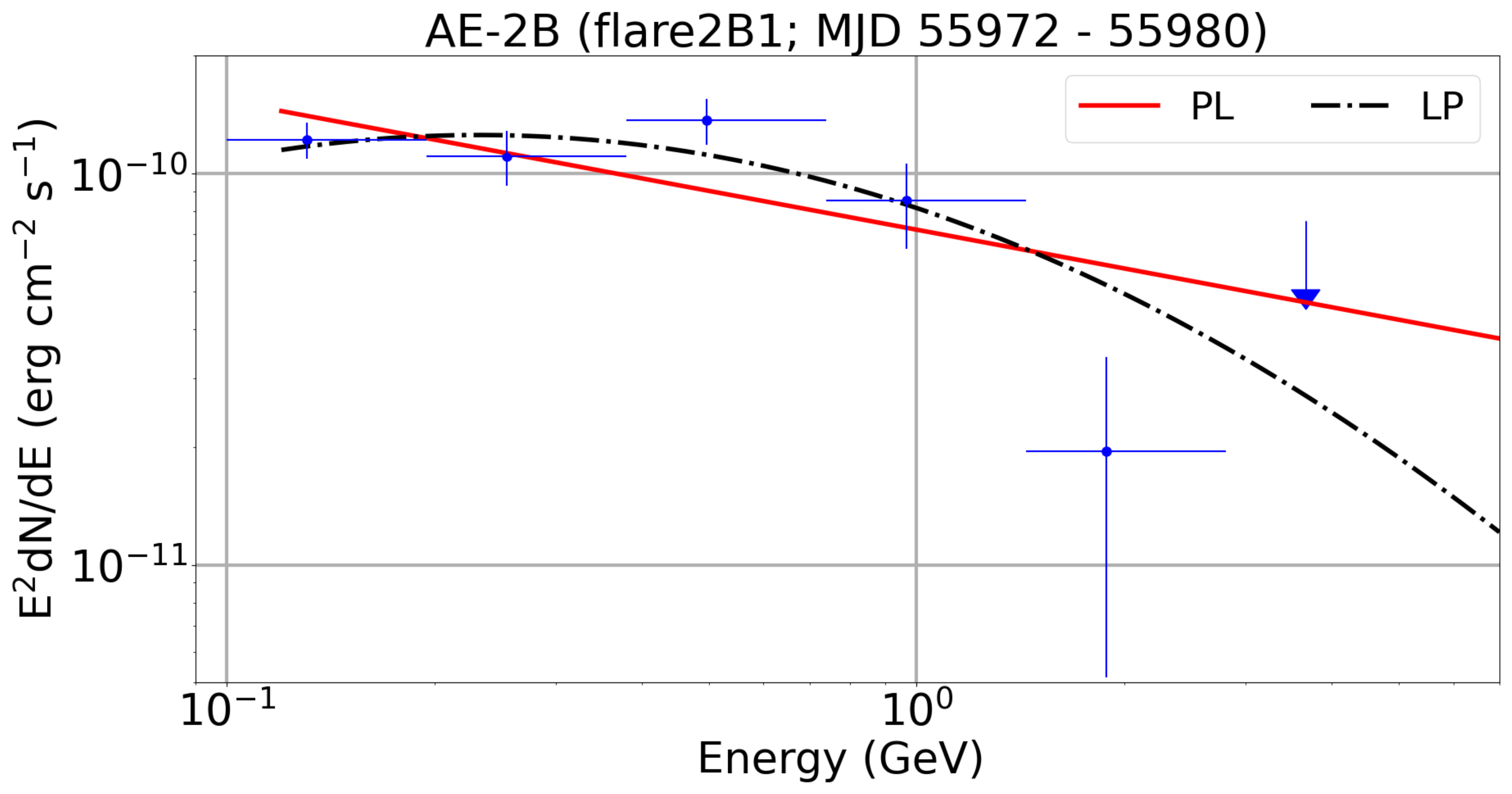}
\includegraphics[height=1.90in,width=2.6in]{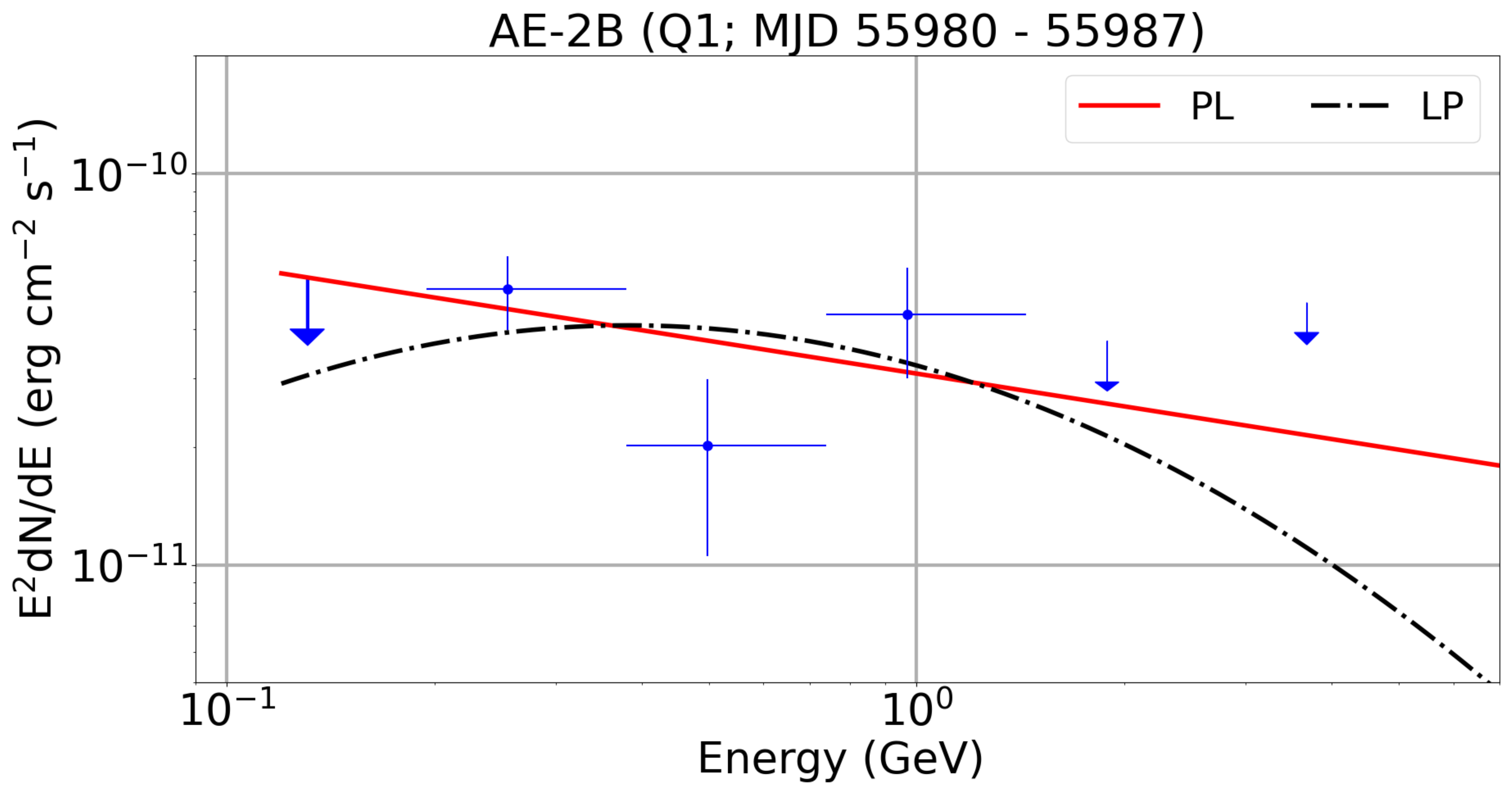}

\includegraphics[height=1.90in,width=2.6in]{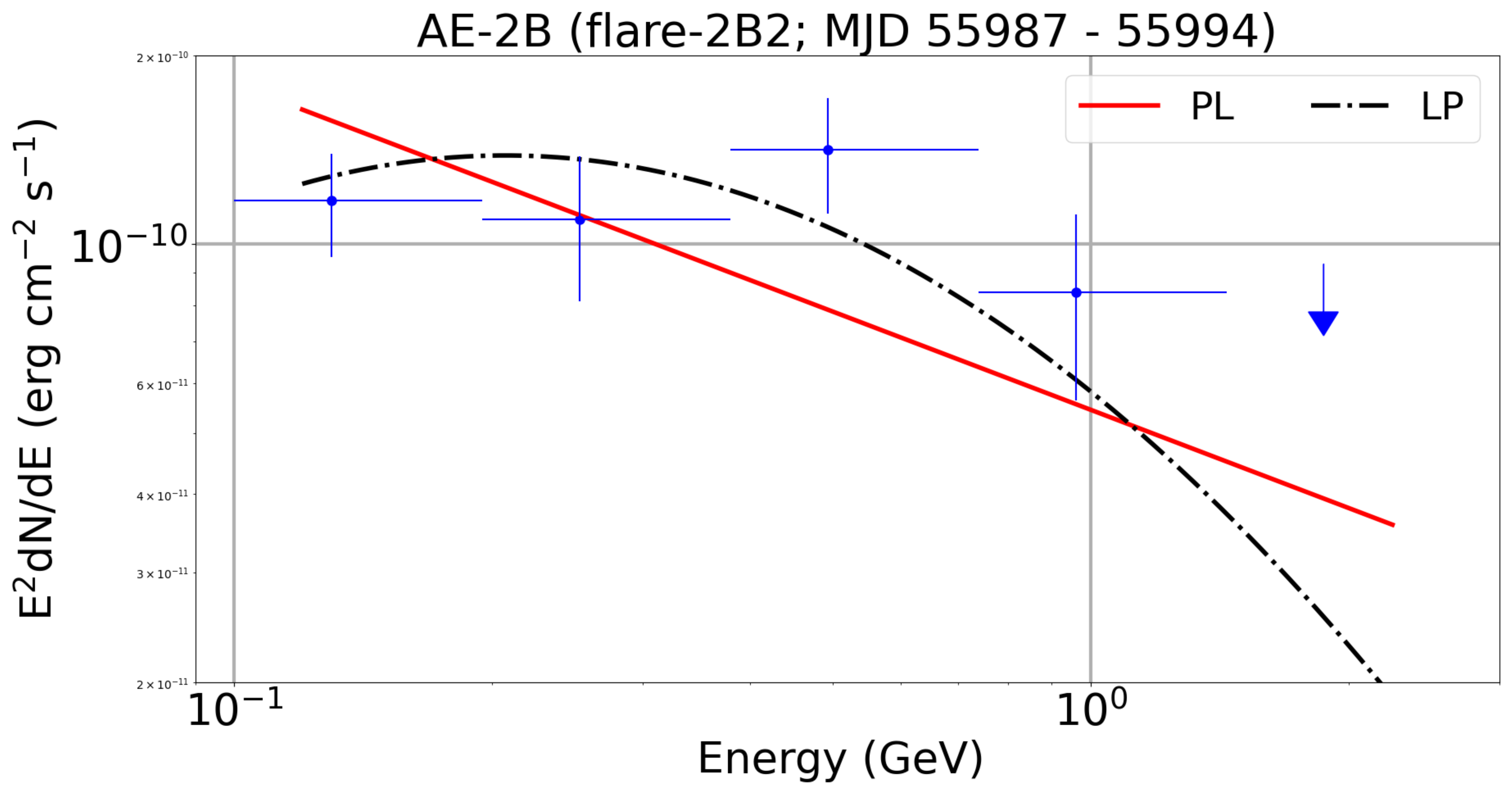}

\includegraphics[height=1.90in,width=2.6in]{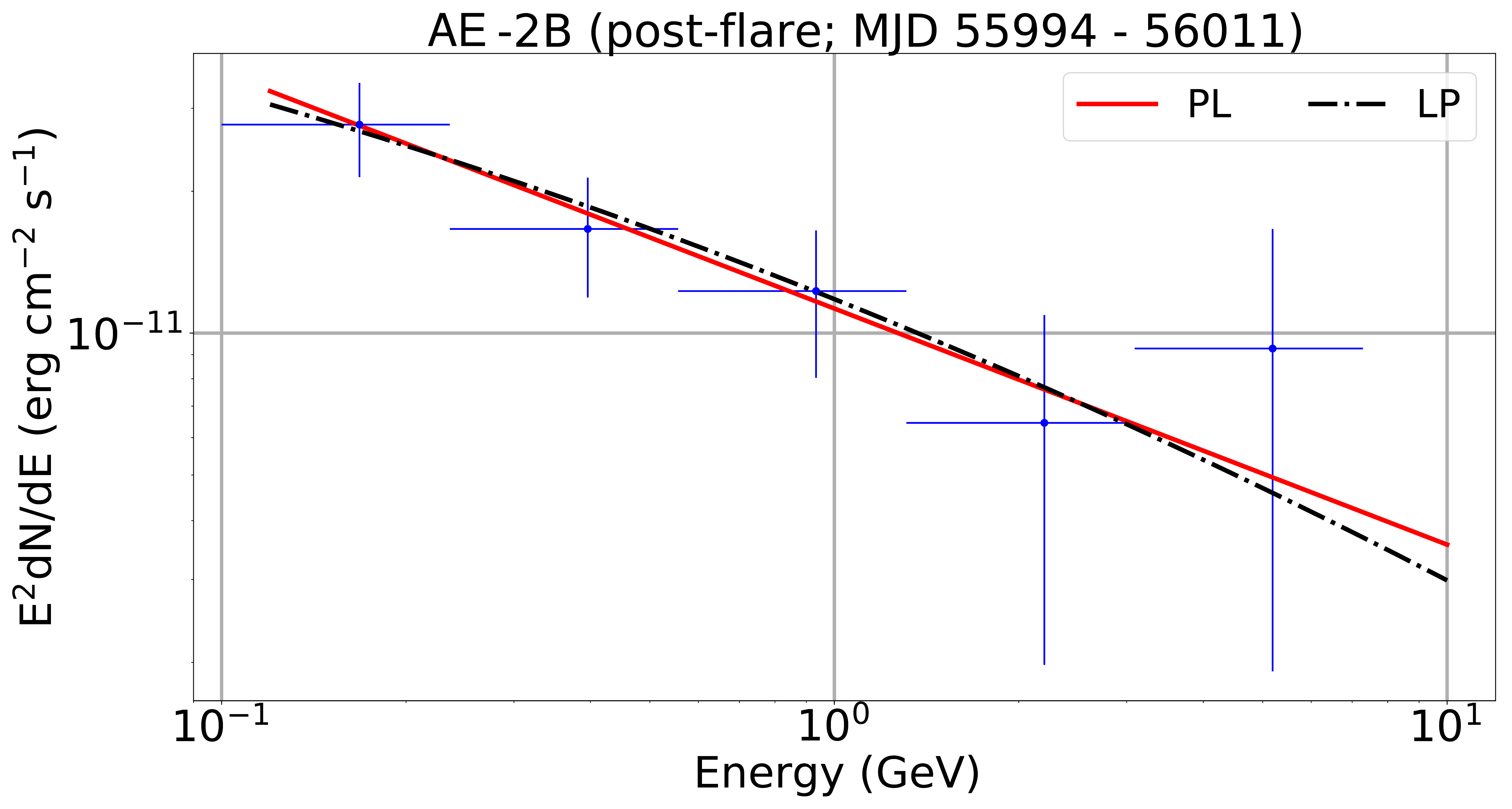}

\caption{Gamma-ray SEDs of different phases (Pre-flare, flare-2B1, Q1, flare-2B2, and Post-flare) of AE-2B. PL, LP describe the Powerlaw, Logparabola model which are shown by solid red and dash-dot black line, respectively.}
\label{fig:A14}

\end{figure*}

\begin{figure*}
\centering
\includegraphics[height=1.90in,width=2.6in]{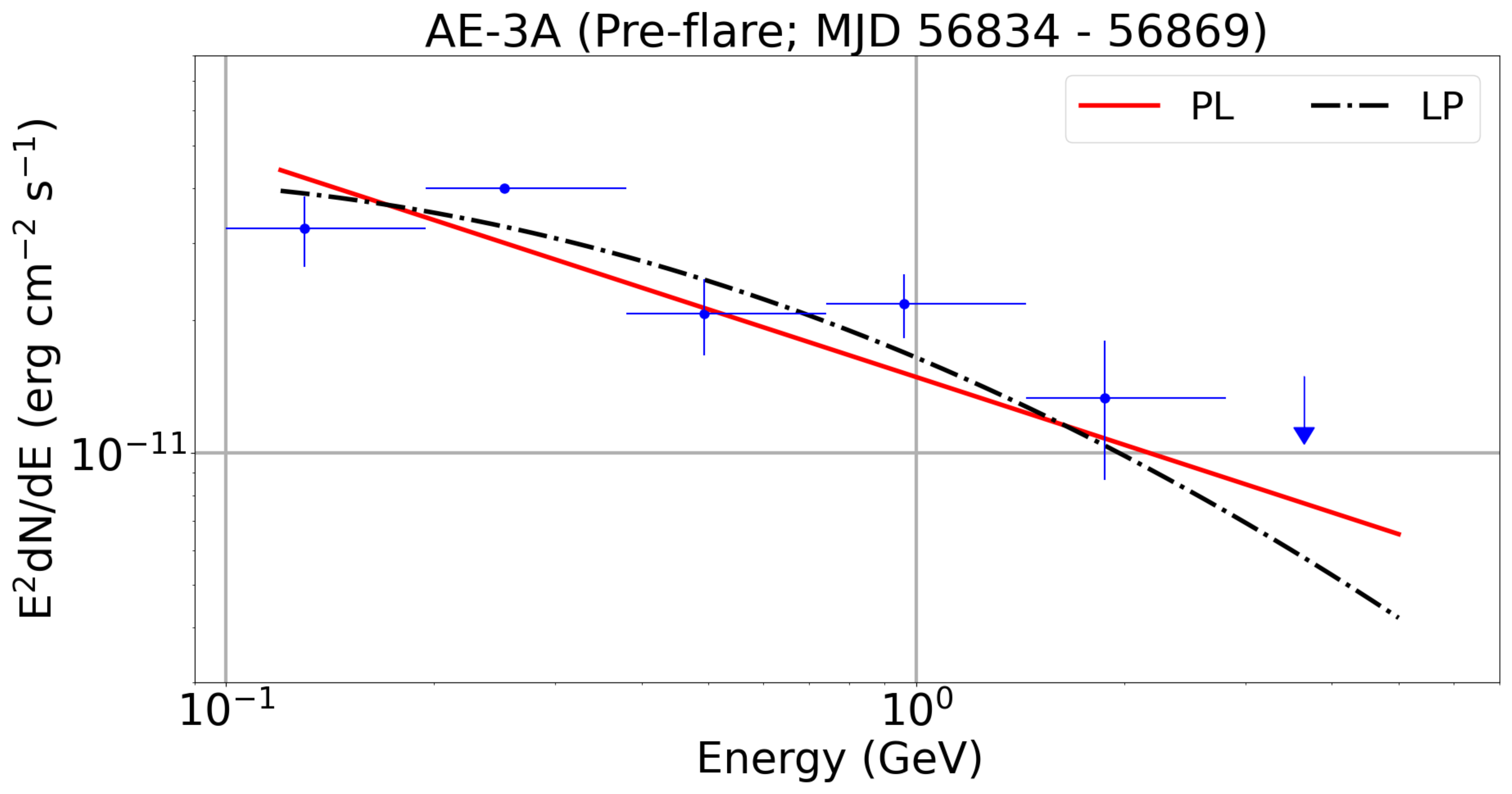}
\includegraphics[height=1.90in,width=2.6in]{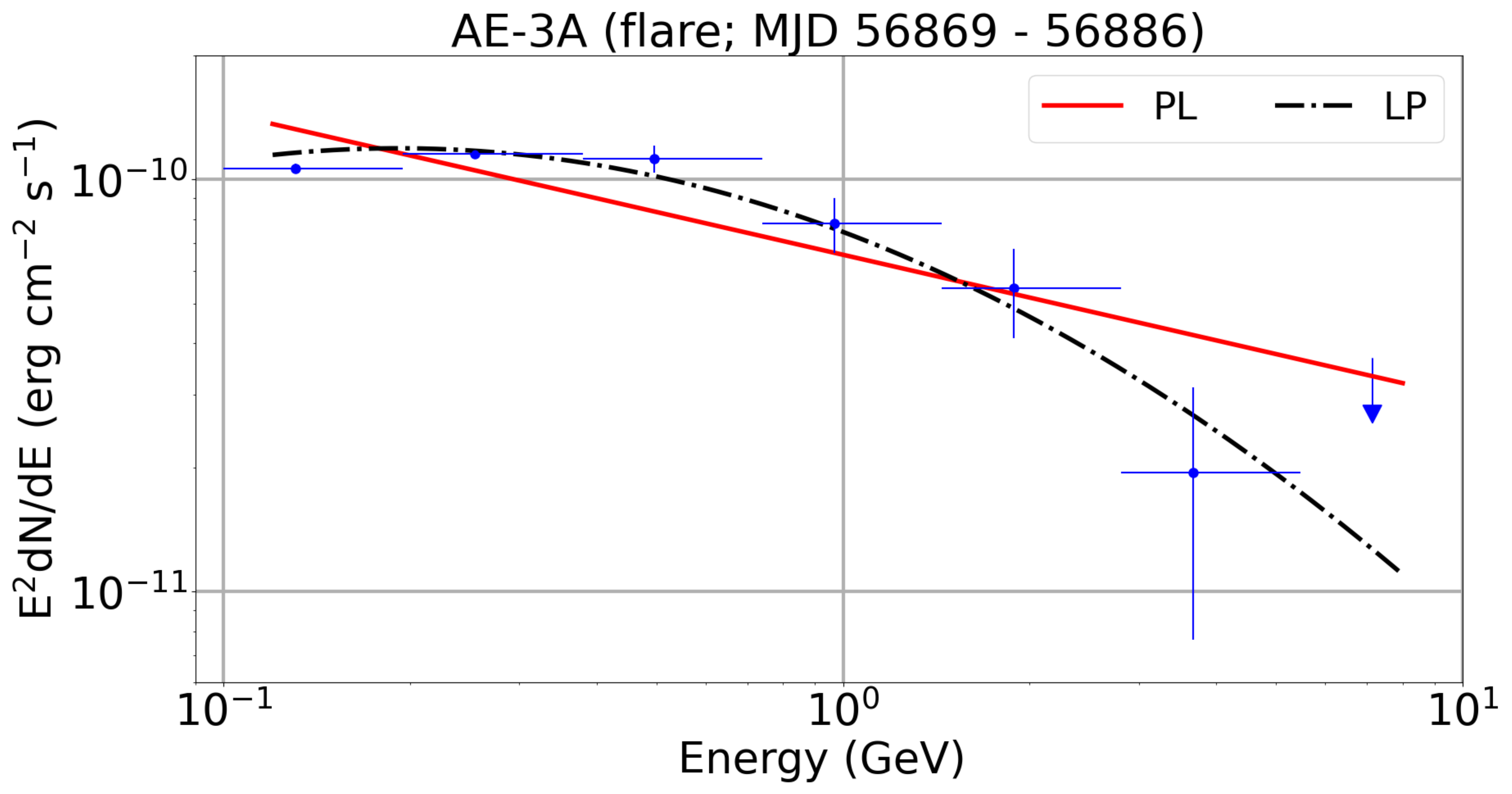}
\includegraphics[height=1.90in,width=2.6in]{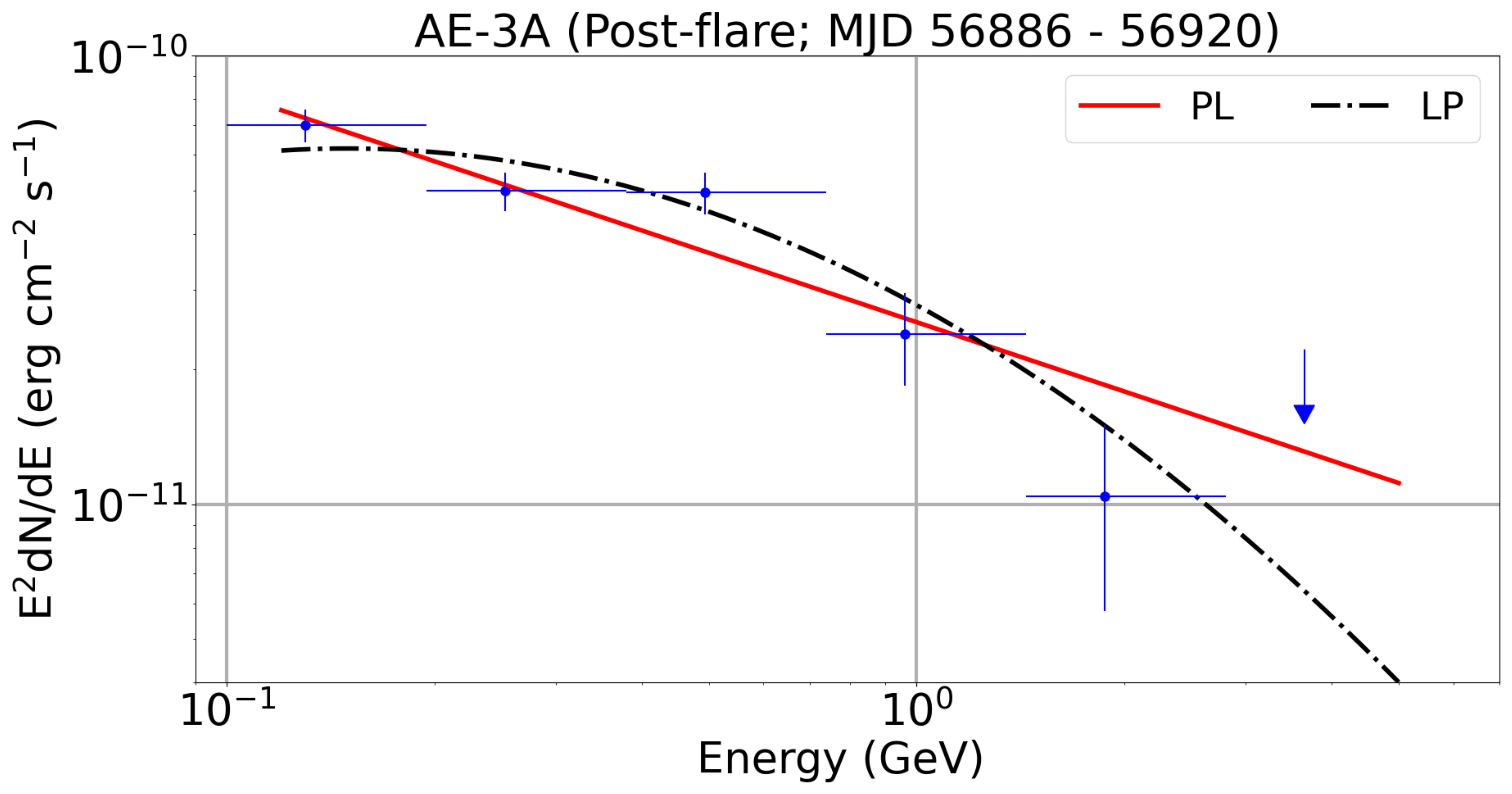}

\caption{Gamma-ray SEDs of different phases (Pre-flare, flare, and Post-flare) of AE-3A. PL, LP describe the Powerlaw, Logparabola model which are shown by solid red and dash-dot black line, respectively.}
\label{fig:A15}

\end{figure*}

\begin{figure*}
\centering
\includegraphics[height=1.90in,width=2.6in]{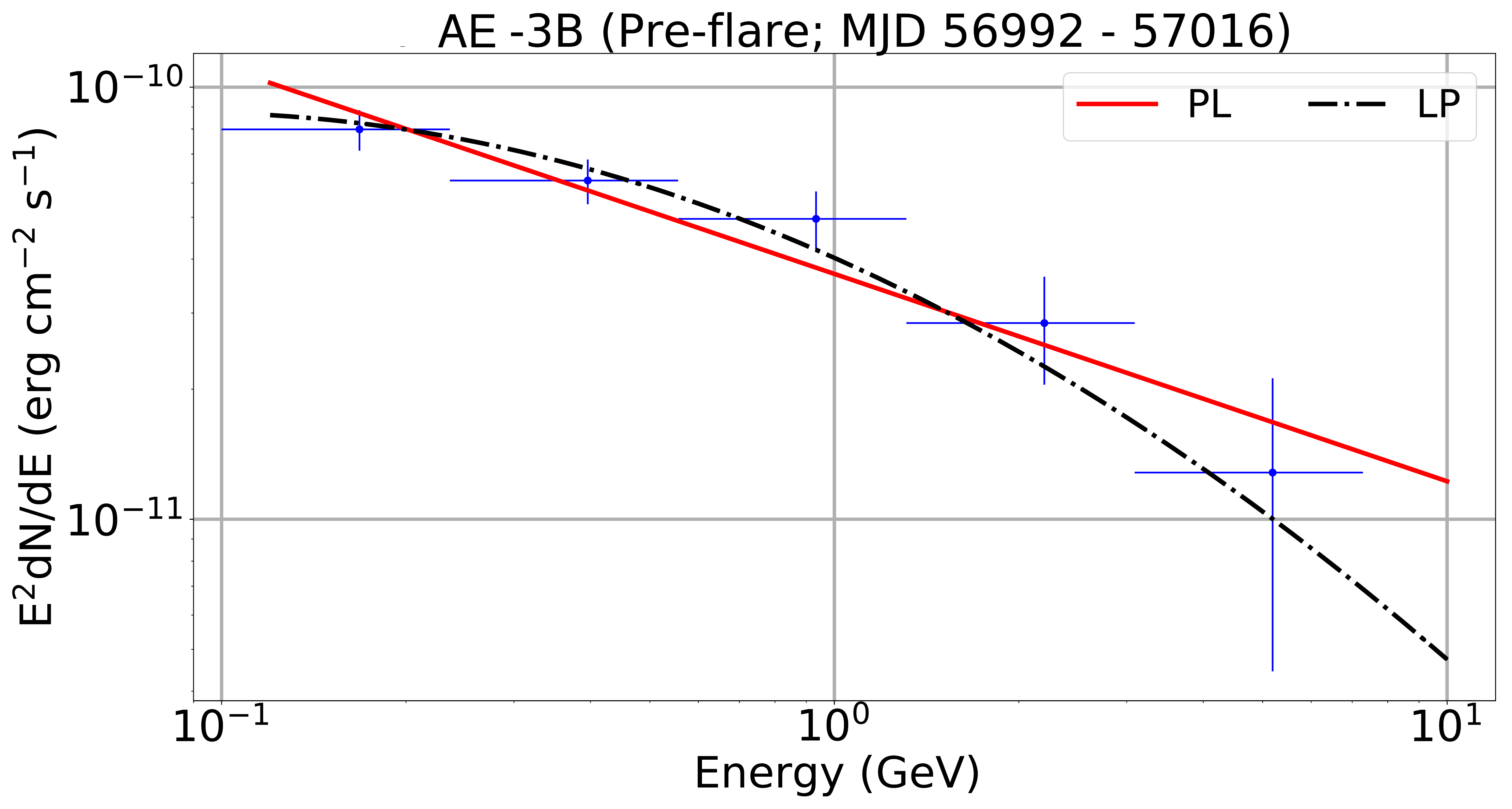}
\includegraphics[height=1.90in,width=2.6in]{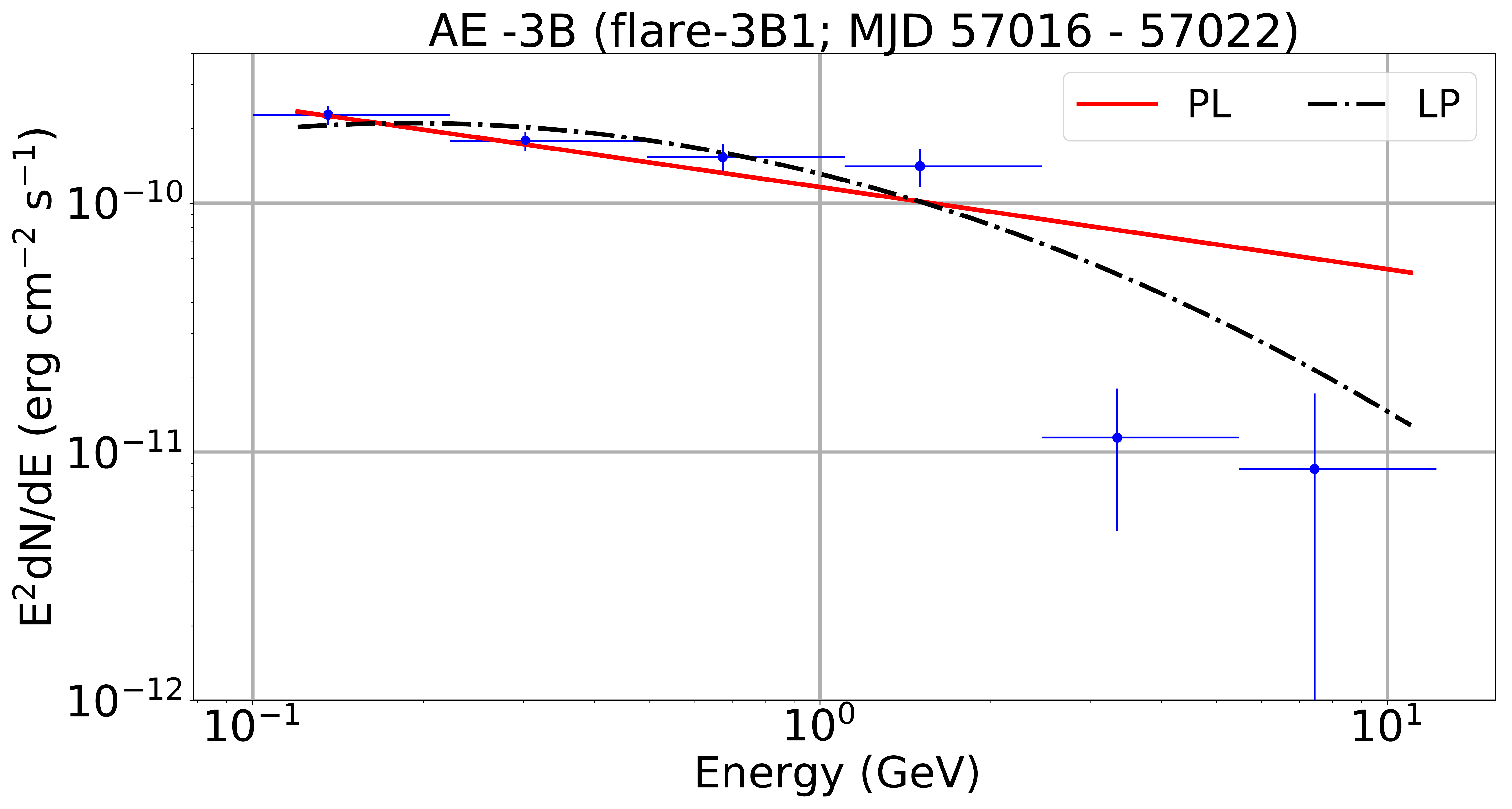}
\includegraphics[height=1.90in,width=2.6in]{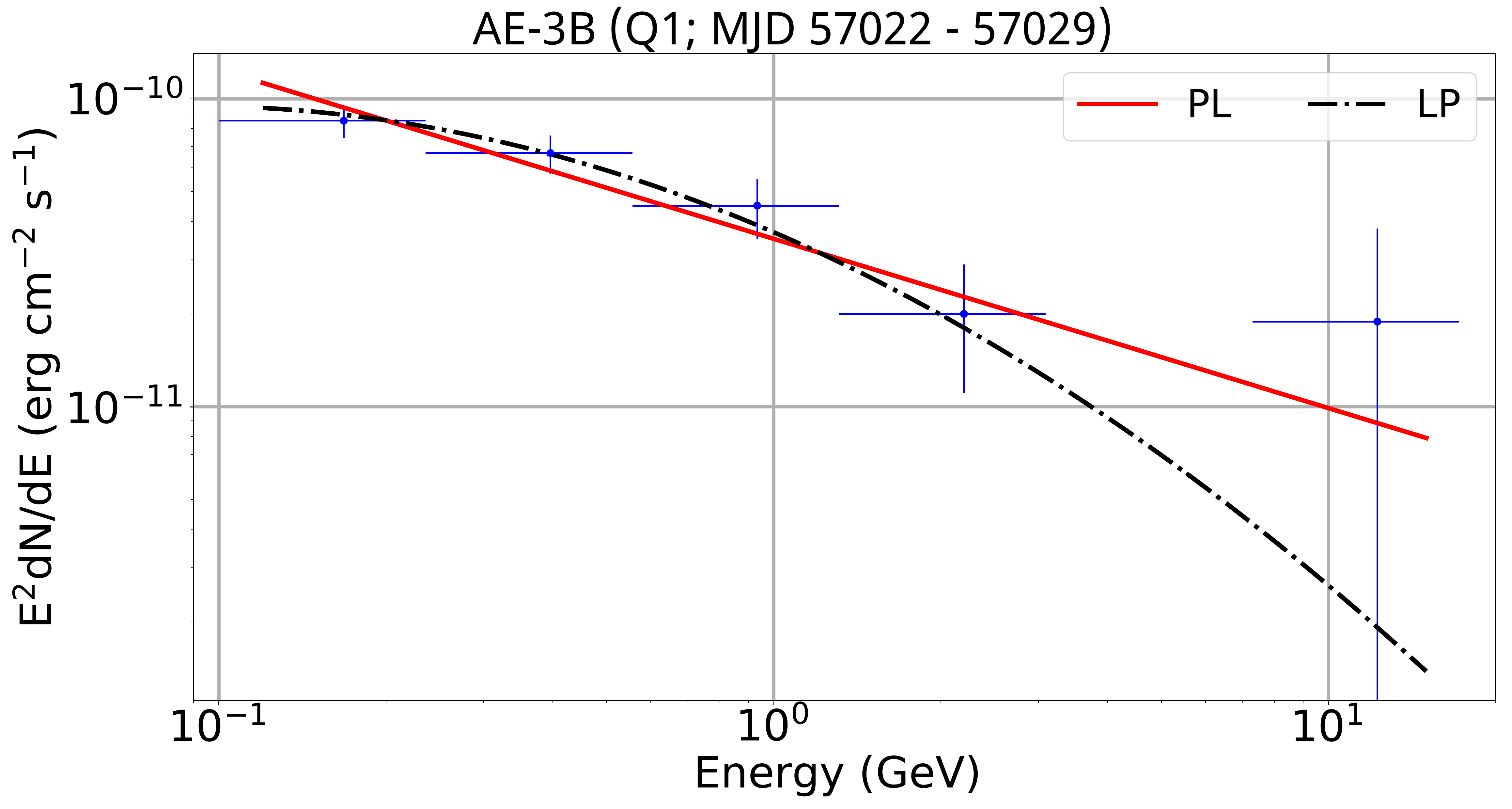}
\includegraphics[height=1.90in,width=2.6in]{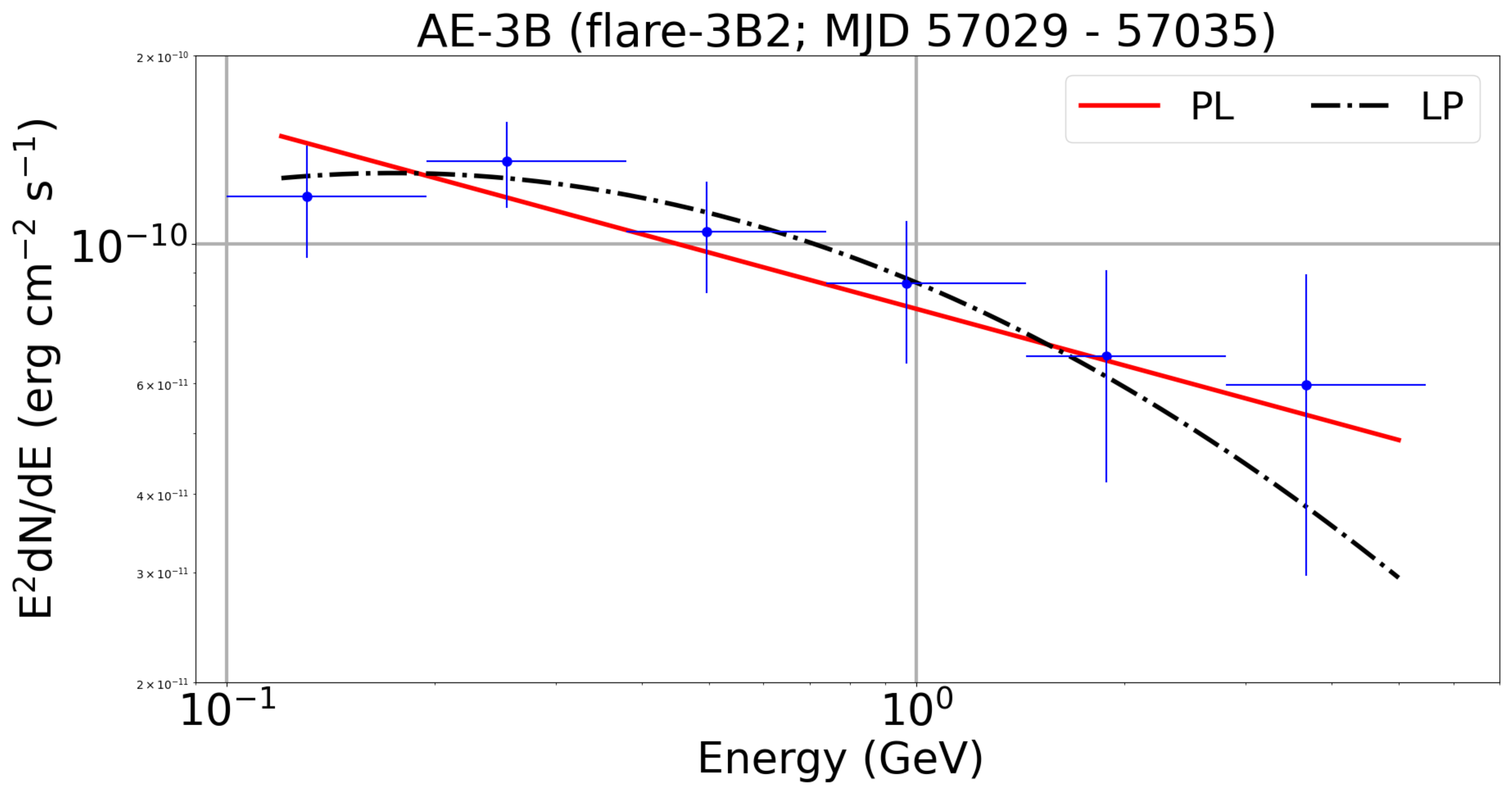}
\includegraphics[height=1.90in,width=2.6in]{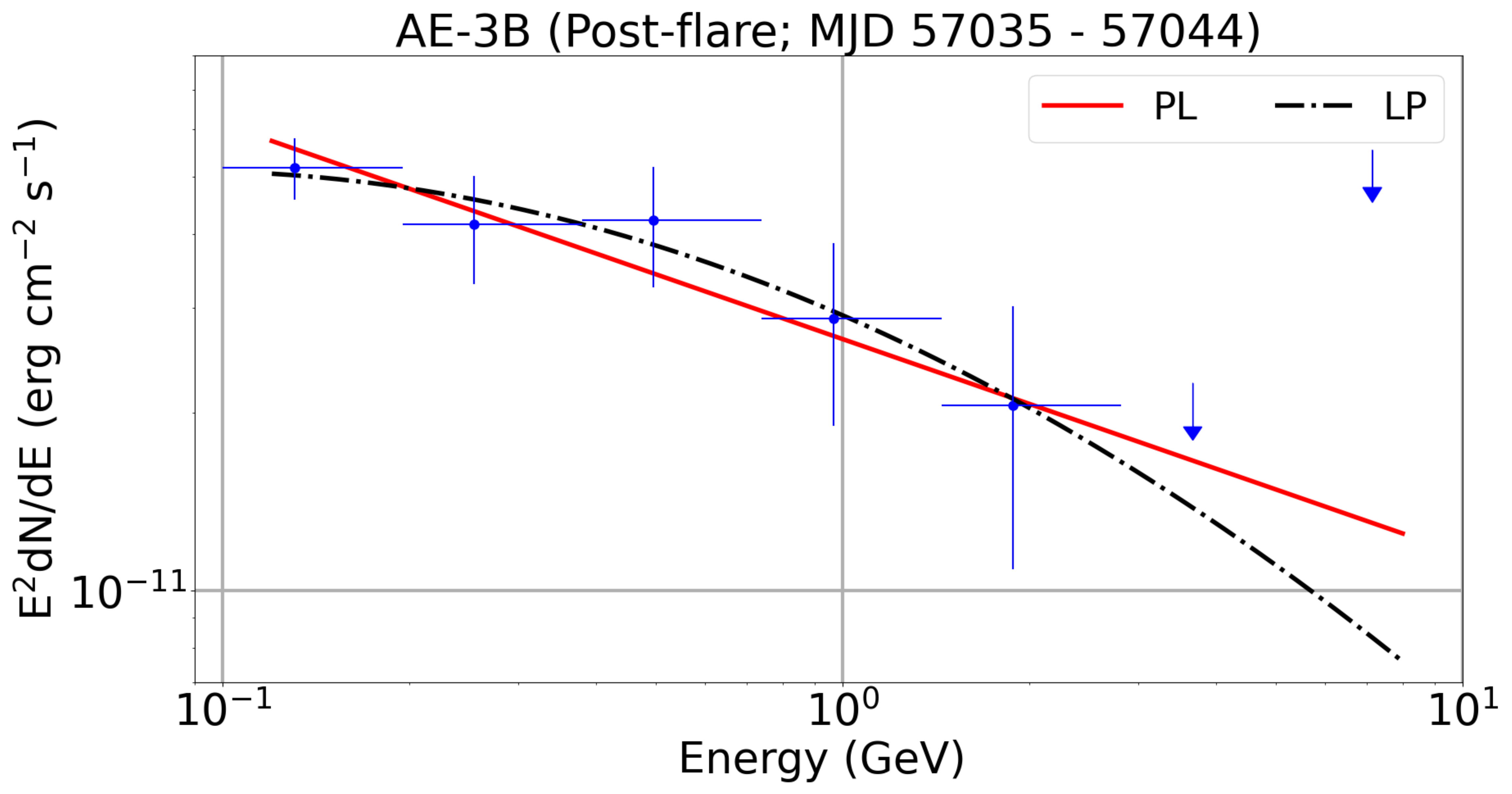}

\caption{Gamma-ray SEDs of different phases (Pre-flare, flare-B1, Q1, flare-3B2, flare, and Post-flare) of AE-3B. PL, LP describe the Powerlaw, Logparabola model which are shown by solid red and dash-dot black line, respectively.}
\label{fig:A16}

\end{figure*}

\begin{figure*}
\centering
\includegraphics[height=1.90in,width=2.6in]{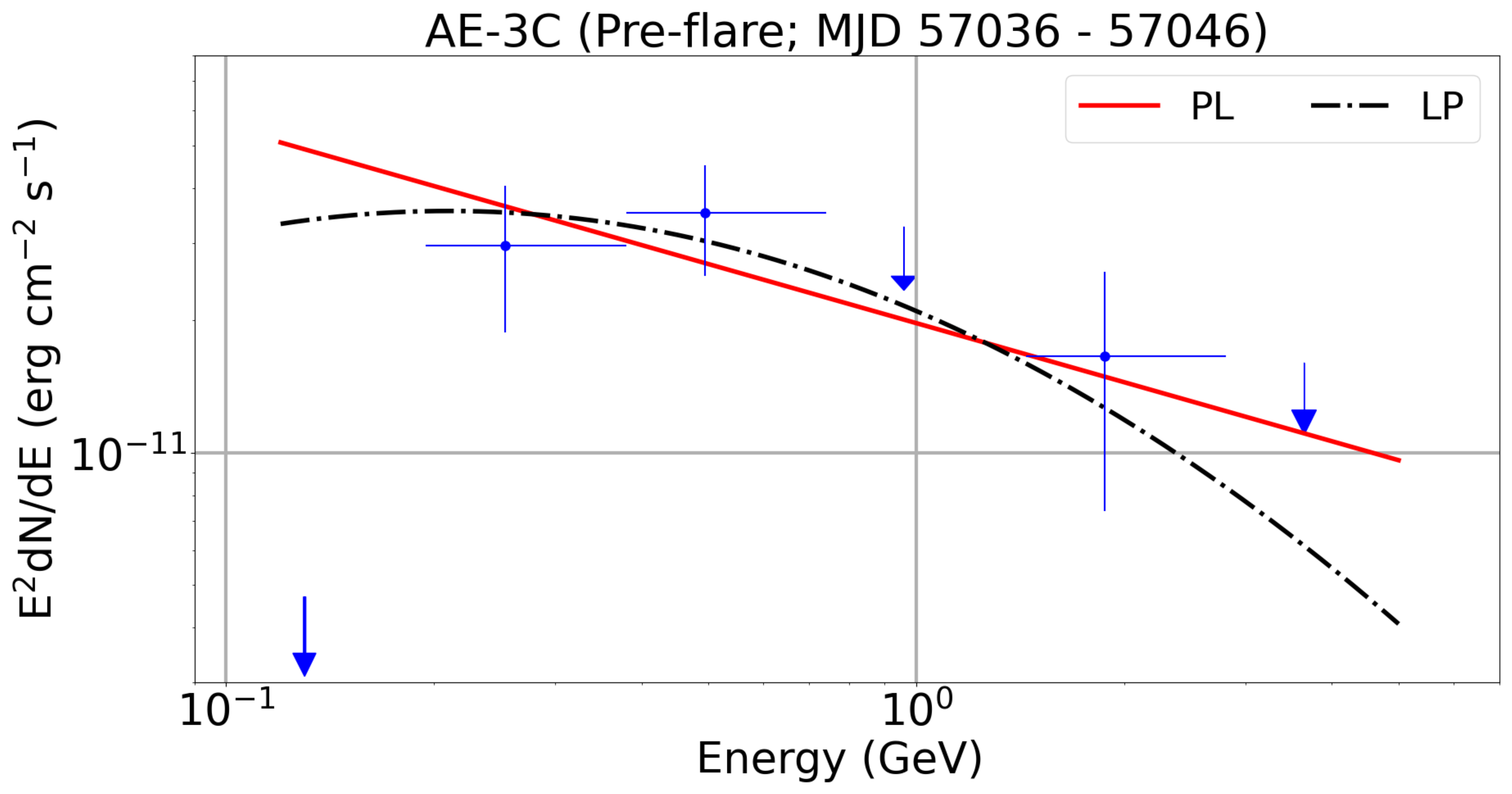}
\includegraphics[height=1.90in,width=2.6in]{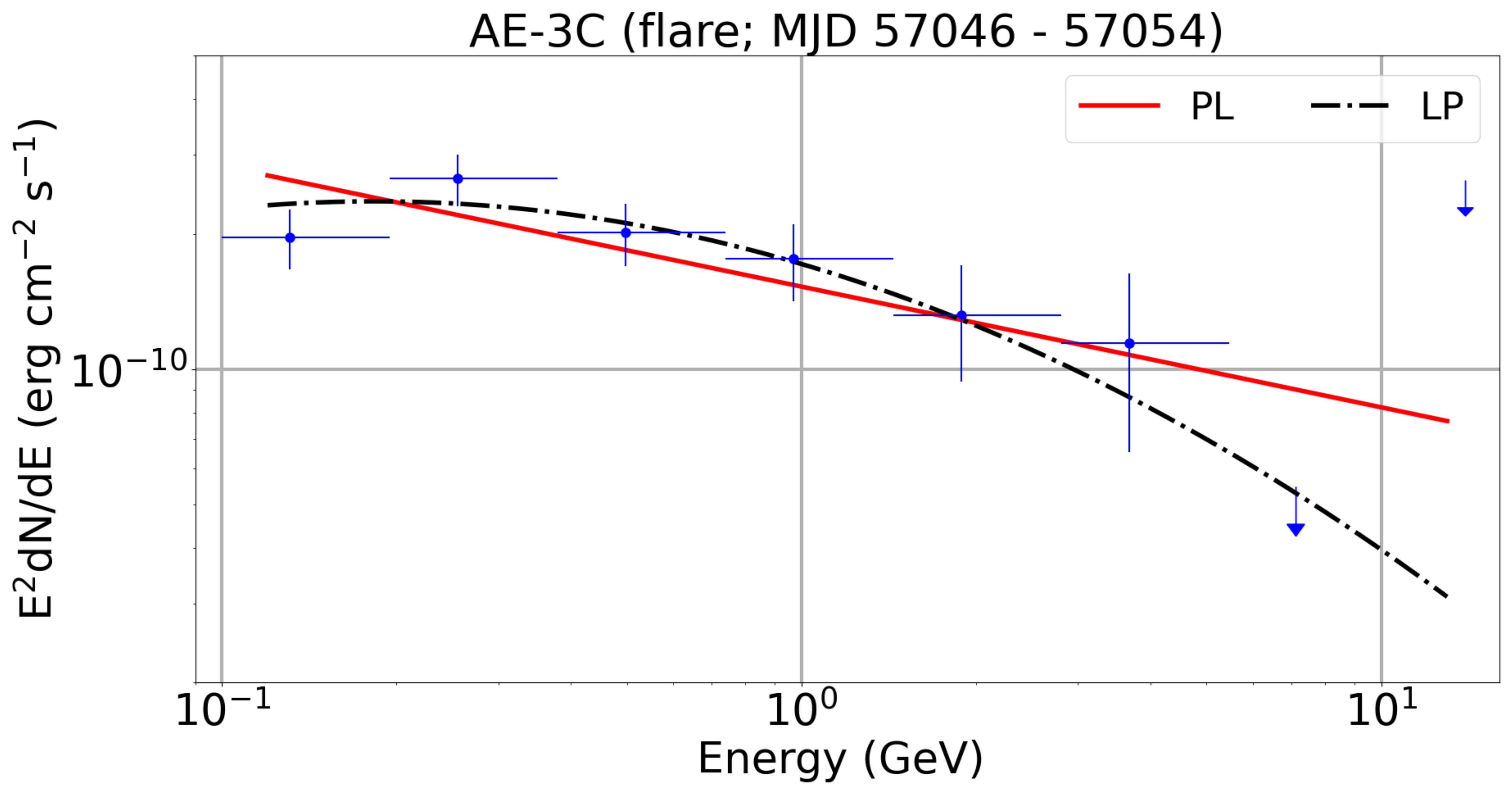}
\includegraphics[height=1.90in,width=2.6in]{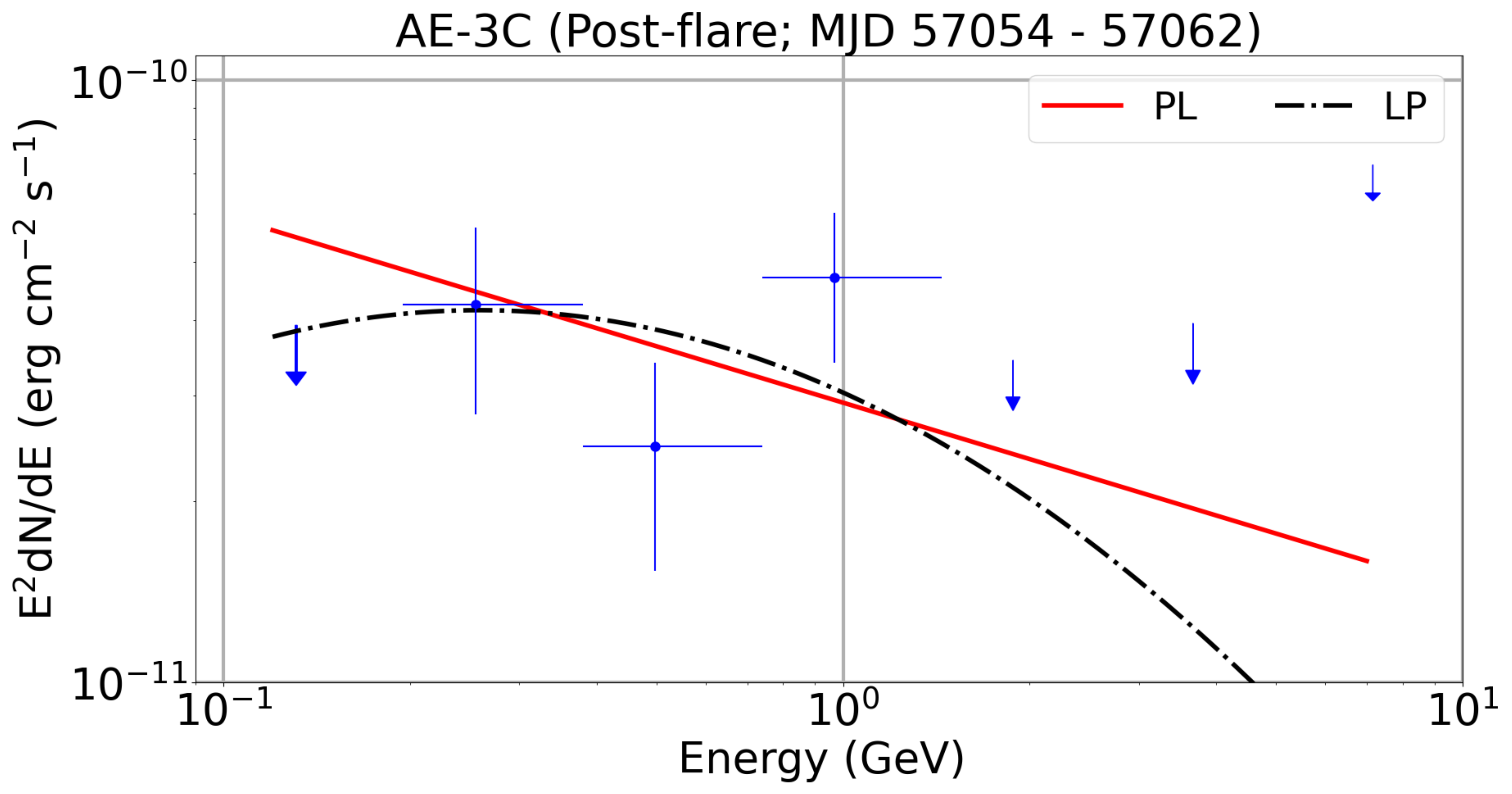}

\caption{Gamma-ray SEDs of different phases (Pre-flare, flare, and Post-flare) of AE-3C. PL, LP describe the Powerlaw, Logparabola model which are shown by solid red and dash-dot black line, respectively.}
\label{fig:A17}

\end{figure*}

\begin{figure*}
\centering
\includegraphics[height=1.90in,width=2.6in]{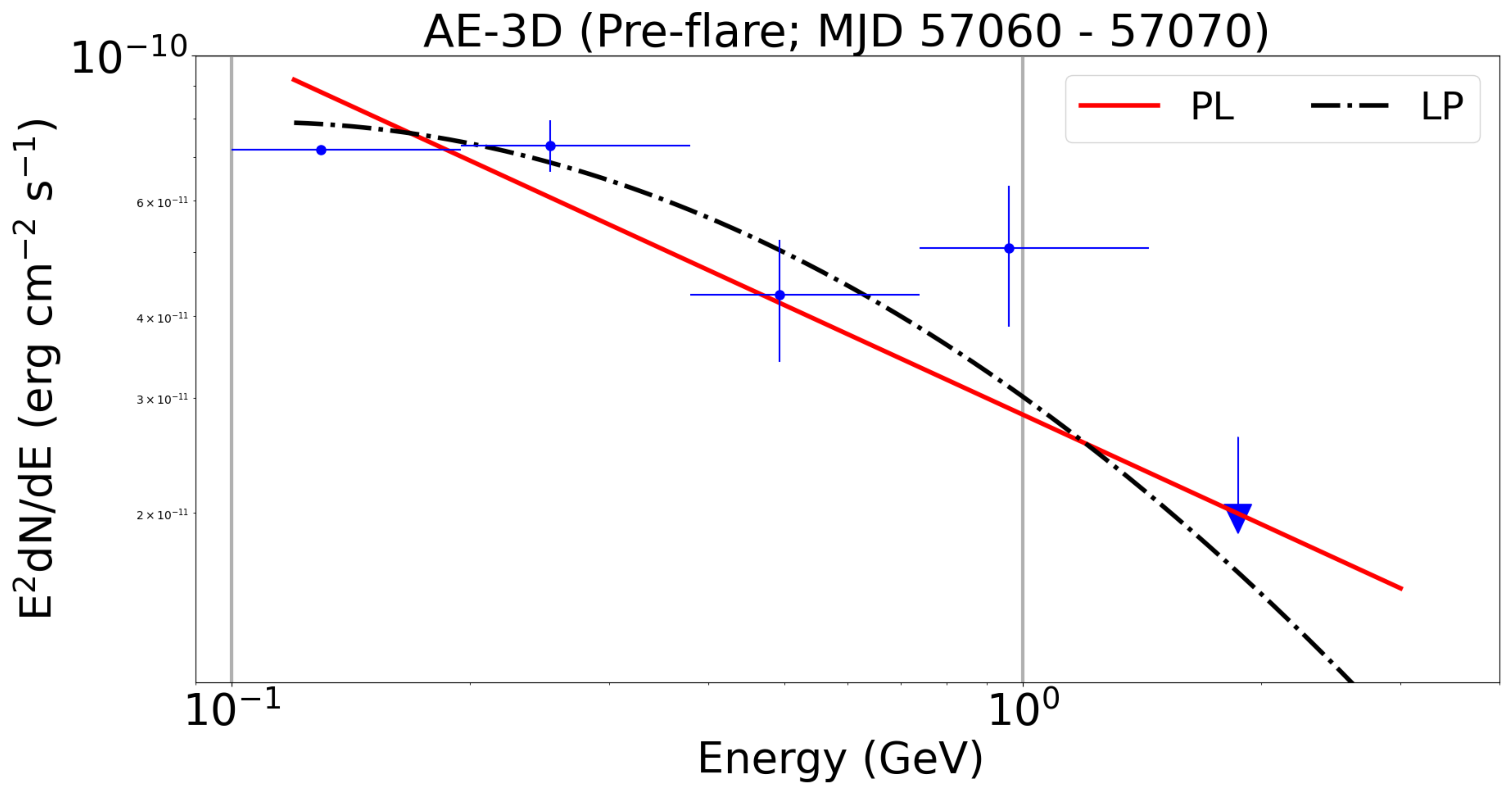}
\includegraphics[height=1.90in,width=2.6in]{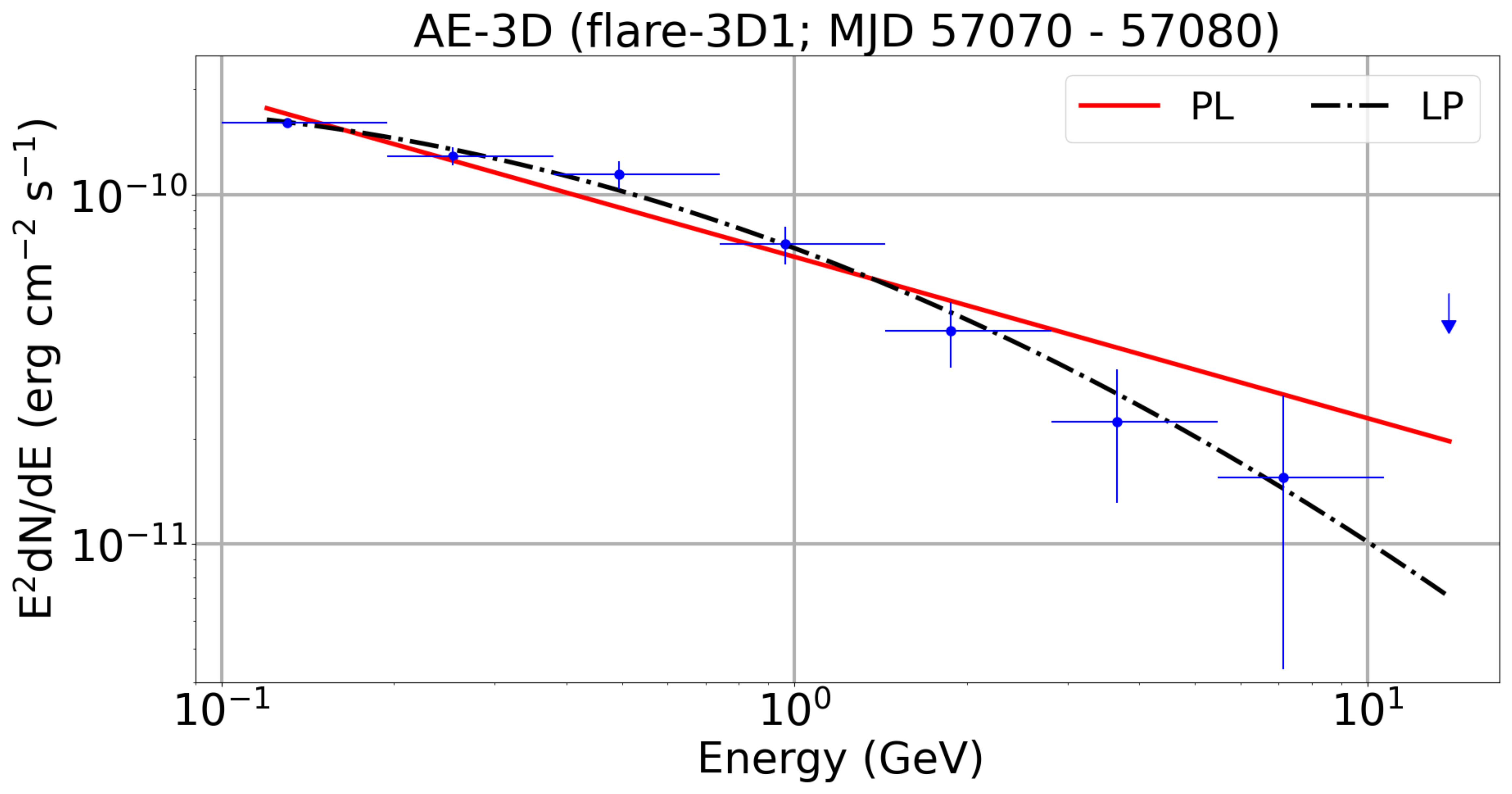}
\includegraphics[height=1.90in,width=2.6in]{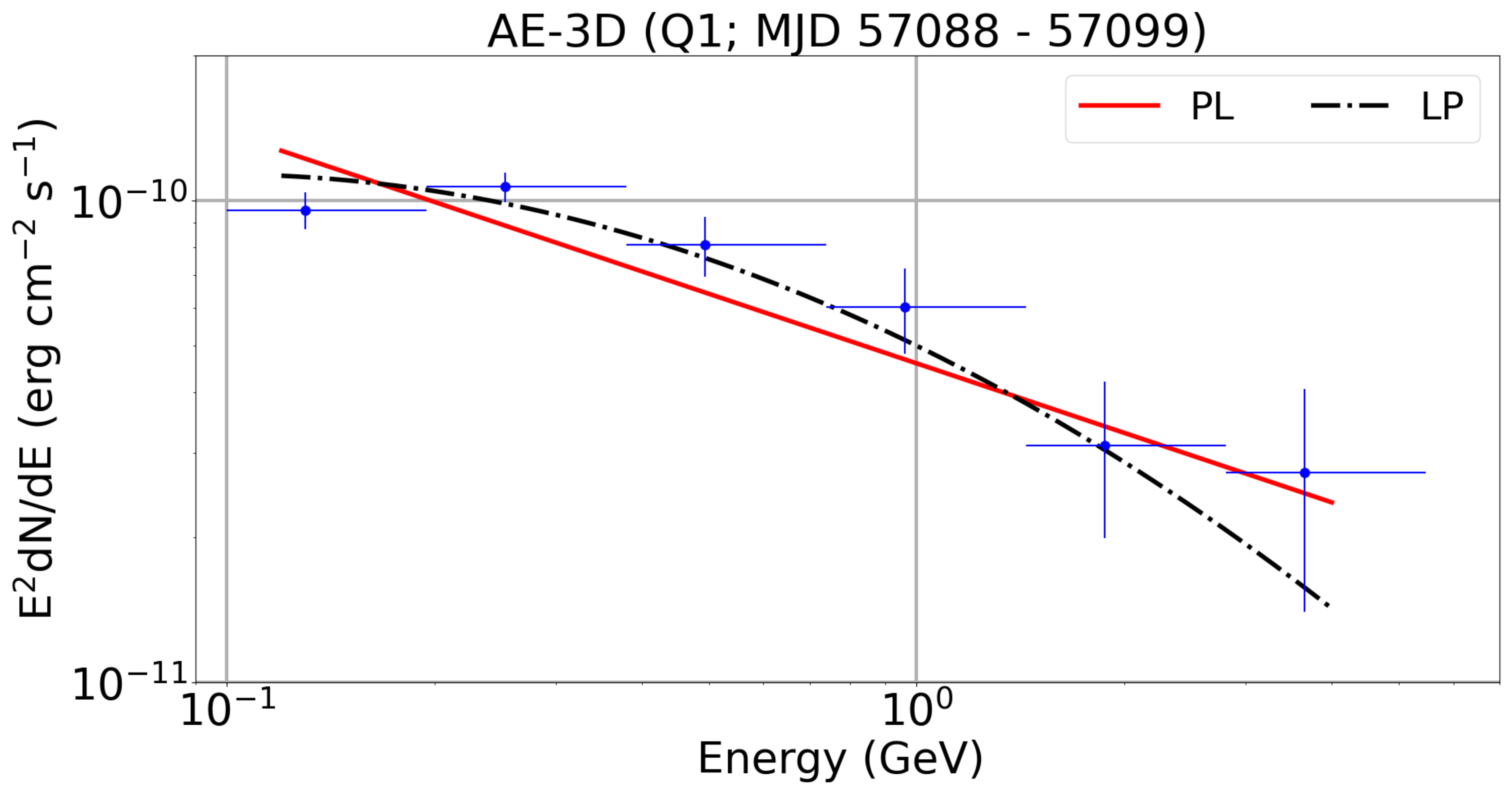}
\includegraphics[height=1.90in,width=2.6in]{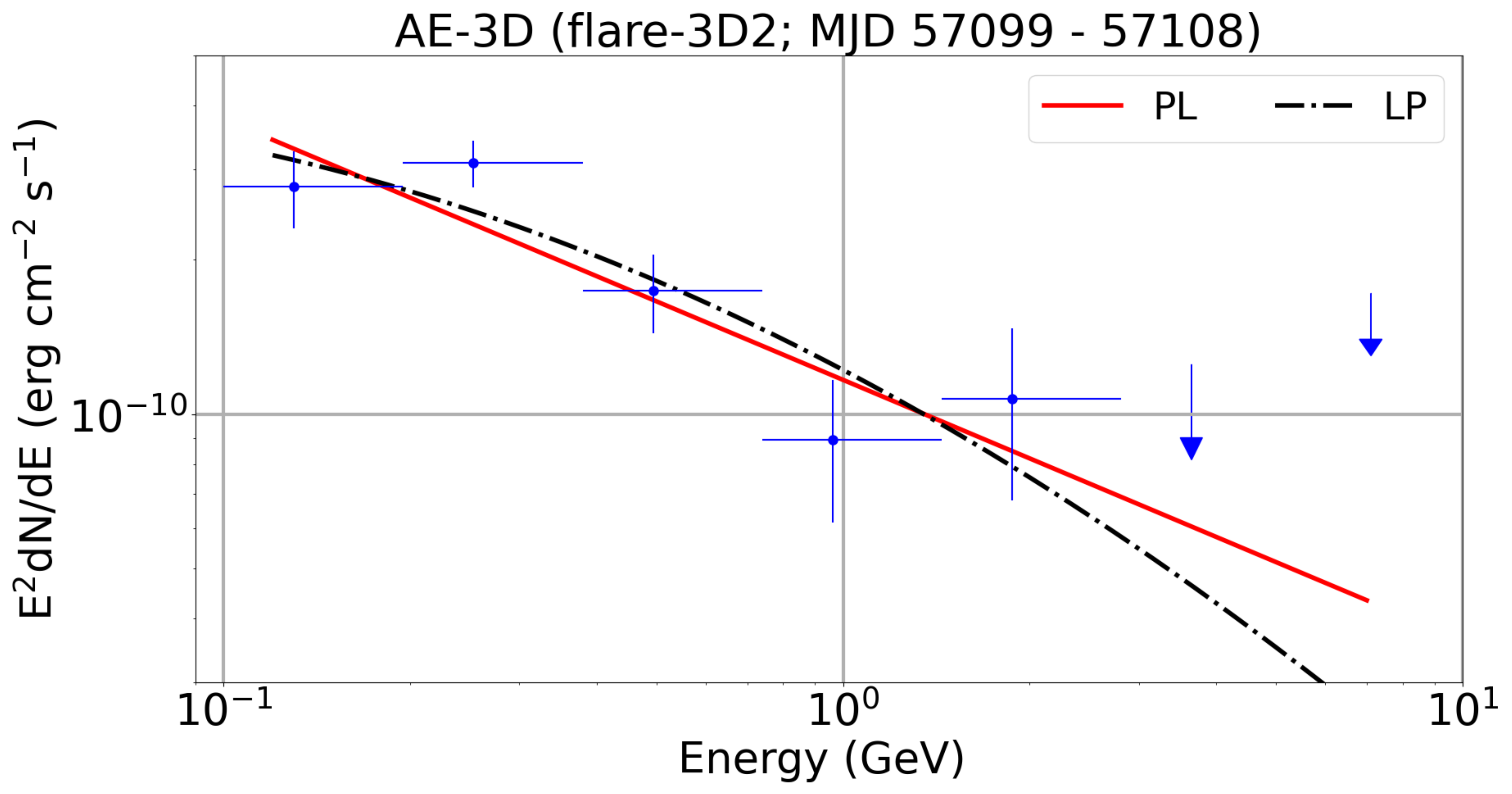}
\includegraphics[height=1.90in,width=2.6in]{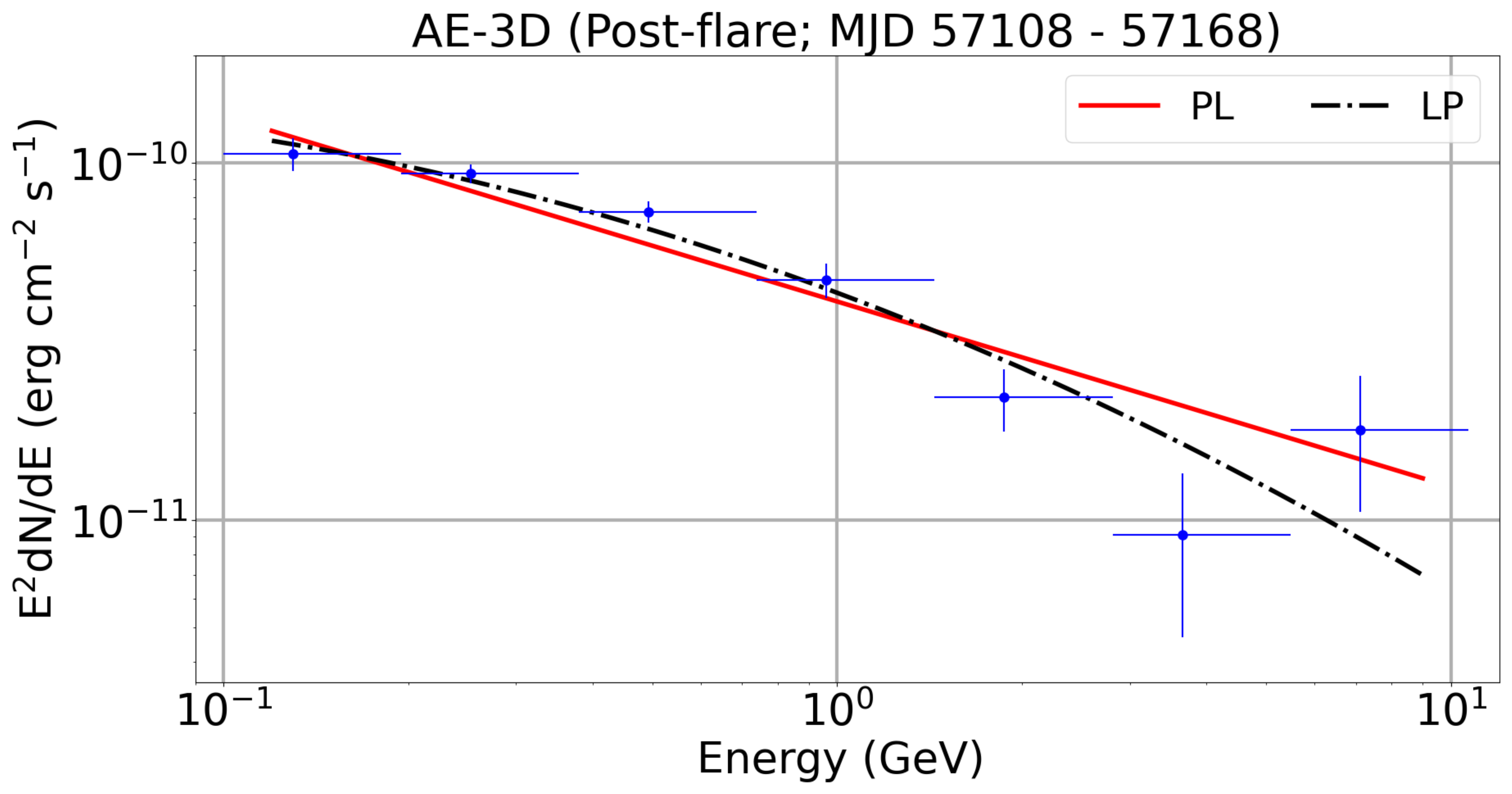}

\caption{Gamma-ray SEDs of different phases (Pre-flare, flare-3D1, Q1, flare-3D2, and Post-flare) of AE-3D. PL, LP describe the Powerlaw, Logparabola model which are shown by solid red and dash-dot black line, respectively.}
\label{fig:A18}
\end{figure*}

\newpage

\begin{table*}
\caption{Different Gamma-ray SED parameters of AE-2A phase (same as Table-\ref{tab:3}).} 
\label{tab:A1}
\centering
\begin{tabular}{ccccc rrrr}   
\hline\hline                        
 & & Powerlaw\\
\hline
Activity & $F_0$ & $\Gamma_{PL}$ & &  -log($\mathcal{L}$) \\ [1.5ex]
& [10$^{-6}$ ph cm$^{-2}$ s$^{-1}$] &  \\
\hline
Pre-flare &  0.22$\pm$0.02 & 2.50$\pm$0.10 &  - &  34253.27 & - \\
flare &  2.20$\pm$0.08 & 2.32$\pm$0.04 &  - &  23319.43 & - \\
Post-flare &  0.14$\pm$0.02 & 2.53$\pm$0.11 &  - &  45415.33 & - \\
\hline                          
 & & Logparabola\\ 
\hline
Activity & $F_0$ & $\alpha$ & $\beta$ & -log($\mathcal{L}$) & $\Delta$log($\mathcal{L}$) \\ [1.5ex]
& [10$^{-6}$ ph cm$^{-2}$ s$^{-1}$] &  \\
\hline
Pre-flare &  0.21$\pm$0.03 & 2.42$\pm$0.13 &  0.07$\pm$0.08 & 34252.88 & -0.39 \\
flare &  2.10$\pm$0.08 & 2.25$\pm$0.05 &  0.14$\pm$0.03 & 23307.70 & -11.73 \\
Post-flare &  0.12$\pm$0.02 & 2.23$\pm$0.18 &  0.30$\pm$0.14 & 45411.99 & -3.34 \\
\hline
\hline 
\end{tabular}
\end{table*}

\begin{table*}
\caption{Different Gamma-ray SED parameters of AE-2B phase (same as Table-\ref{tab:3}).} 
\label{tab:A2}
\centering
\begin{tabular}{ccccc rrrr}   
\hline\hline                        
 & & Powerlaw\\
\hline
Activity & $F_0$ & $\Gamma_{PL}$ & &  -log($\mathcal{L}$) \\ [1.5ex]
& [10$^{-6}$ ph cm$^{-2}$ s$^{-1}$] &  \\
\hline
Pre-flare &  0.24$\pm$0.02 & 2.53$\pm$0.09 &  - &  35302.70 & - \\
Flare-2B1 &  0.72$\pm$0.06 & 2.33$\pm$0.08 &  - &  9894.52 & - \\
Q1 &  0.23$\pm$0.05 & 2.28$\pm$0.15 &  - &  9985.95 & - \\
Flare-2B2 &  0.91$\pm$0.11 & 2.52$\pm$0.12 &  - &  5379.83 & - \\
Post-flare &  0.14$\pm$0.03 & 2.50$\pm$0.18 &  - &  4256.64 & - \\
\hline                          
 & & Logparabola\\ 
\hline
Activity & $F_0$ & $\alpha$ & $\beta$ & -log($\mathcal{L}$) & $\Delta$log($\mathcal{L}$) \\ [1.5ex]
& [10$^{-6}$ ph cm$^{-2}$ s$^{-1}$] &  \\
\hline
Pre-flare &  0.21$\pm$0.02 & 2.25$\pm$0.13 &  0.37$\pm$0.12 & 35295.69 & -7.01 \\
Flare-2B1 &  0.68$\pm$0.07 & 2.10$\pm$0.13 &  0.20$\pm$0.09 & 9891.07 & -3.45 \\
Q1 &  0.19$\pm$0.05 & 1.88$\pm$0.33 &  0.26$\pm$0.17 & 9984.32 & -1.63 \\
Flare-2B2 &  0.83$\pm$0.11 & 2.26$\pm$0.18 &  0.35$\pm$0.03 & 5376.51 & -3.32 \\
Post-flare &  0.14$\pm$0.02 & 2.45$\pm$0.22 &  0.04$\pm$0.12 & 4255.83 & -0.81 \\
\hline
\hline 
\end{tabular}
\end{table*}

\begin{table*}
\caption{Different Gamma-ray SED parameters of AE-3A phase (same as Table-\ref{tab:3}).} 
\label{tab:A3}
\centering
\begin{tabular}{ccccc rrrr}   
\hline\hline                        
& & Powerlaw\\
\hline
Activity & $F_0$ & $\Gamma_{PL}$ & &  -log($\mathcal{L}$) \\ [1.5ex]
& [10$^{-6}$ ph cm$^{-2}$ s$^{-1}$] &  \\
\hline
Pre-flare &  0.22$\pm$0.02 & 2.51$\pm$0.09 &  - &  41952.98 & - \\
flare &  0.71$\pm$0.04 & 2.34$\pm$0.05 &  - &  25506.21 & - \\
Post-flare &  0.36$\pm$0.03 & 2.51$\pm$0.07 &  - &  39585.82 & - \\
\hline                          
 & & Logparabola\\ 
\hline
Activity & $F_0$ & $\alpha$ & $\beta$ & -log($\mathcal{L}$) & $\Delta$log($\mathcal{L}$) \\ [1.5ex]
& [10$^{-6}$ ph cm$^{-2}$ s$^{-1}$] &  \\
\hline
Pre-flare &  0.21$\pm$0.02 & 2.38$\pm$0.13 &  0.12$\pm$0.08 & 41951.80 & -1.18 \\
flare &  0.67$\pm$0.04 & 2.15$\pm$0.08 &  0.17$\pm$0.05 & 25499.67 & -6.54 \\
Post-flare &  0.33$\pm$0.03 & 2.31$\pm$0.10 &  0.23$\pm$0.08 & 39580.66 & -5.16 \\
\hline
\hline 
\end{tabular}
\end{table*}

\begin{table*}
\caption{Different Gamma-ray SED parameters of AE-3B phase (same as Table-\ref{tab:3}).} 
\label{tab:A4}
\centering
\begin{tabular}{ccccc rrrr}   
\hline\hline                        
& & Powerlaw\\
\hline
Activity & $F_0$ & $\Gamma_{PL}$ & &  -log($\mathcal{L}$) \\ [1.5ex]
& [10$^{-6}$ ph cm$^{-2}$ s$^{-1}$] &  \\
\hline
Pre-flare &  0.42$\pm$0.03 & 2.48$\pm$0.09 &  - &  5828.98 & - \\
Flare-3B1 &  1.20$\pm$0.07 & 2.33$\pm$0.06 &  - &  13311.60 & - \\
Q1 &  0.50$\pm$0.05 & 2.55$\pm$0.10 &  - &  3600.68 & - \\
Flare-3B2 &  0.87$\pm$0.07 & 2.30$\pm$0.08 &  - &  8986.22 & - \\
Post-flare &  0.29$\pm$0.04 & 2.36$\pm$0.11 &  - &  15275.4 & - \\
\hline                          
& & Logparabola\\ 
\hline
Activity & $F_0$ & $\alpha$ & $\beta$ & -log($\mathcal{L}$) & $\Delta$log($\mathcal{L}$) \\ [1.5ex]
& [10$^{-6}$ ph cm$^{-2}$ s$^{-1}$] &  \\
\hline
Pre-flare &  0.46$\pm$0.04 & 2.33$\pm$0.10 &  0.13$\pm$0.07 & 5823.82 & -5.16 \\
Flare-3B1 &  1.10$\pm$0.07 & 2.39$\pm$0.07 &  0.17$\pm$0.06 & 13306.02 & -5.58 \\
Q1 &  0.50$\pm$0.04 & 2.44$\pm$0.10 &  0.16$\pm$0.10 & 3597.54 & -3.14 \\
Flare-3B2 &  0.83$\pm$0.07 & 2.14$\pm$0.12 &  0.13$\pm$0.07 & 8984.14 & -2.08 \\
Post-flare &  0.28$\pm$0.04 & 2.23$\pm$0.18 &  0.09$\pm$0.10 & 15274.89 & -0.51 \\
\hline
\hline 
\end{tabular}
\end{table*}

\begin{table*}
\caption{Different Gamma-ray SED parameters of AE-3C phase (same as Table-\ref{tab:3}).} 
\label{tab:A5}
\centering
\begin{tabular}{ccccc rrrr}   
\hline\hline                        
& & Powerlaw\\
\hline
Activity & $F_0$ & $\Gamma_{PL}$ & &  -log($\mathcal{L}$) \\ [1.5ex]
& [10$^{-6}$ ph cm$^{-2}$ s$^{-1}$] &  \\
\hline
Pre-flare &  0.25$\pm$0.04 & 2.45$\pm$0.14 &  - &  14534.75 & - \\
flare &  1.40$\pm$0.12 & 2.27$\pm$0.07 &  - &  5378.80 & - \\
Post-flare &  0.28$\pm$0.05 & 2.31$\pm$0.13 &  - &  11785.16 & - \\
\hline                          
 & & Logparabola\\ 
\hline
Activity & $F_0$ & $\alpha$ & $\beta$ & -log($\mathcal{L}$) & $\Delta$log($\mathcal{L}$) \\ [1.5ex]
& [10$^{-6}$ ph cm$^{-2}$ s$^{-1}$] &  \\
\hline
Pre-flare &  0.22$\pm$0.04 & 2.15$\pm$0.24 &  0.22$\pm$0.16 & 14534.76 & -0.01 \\
flare &  1.40$\pm$0.12 & 2.11$\pm$0.11 &  0.12$\pm$0.06 & 5376.60 & -2.20 \\
Post-flare &  0.25$\pm$0.05 & 2.05$\pm$0.24 &  0.17$\pm$0.13 & 11783.97 & -1.19 \\
\hline
\hline 
\end{tabular}
\end{table*}

\begin{table*}
\caption{Different Gamma-ray SED parameters of AE-3D phase (same as Table-\ref{tab:3}).} 
\label{tab:A6}
\centering
\begin{tabular}{ccccc rrrr}   
\hline\hline                        
 & & Powerlaw\\
\hline
Activity & $F_0$ & $\Gamma_{PL}$ & &  -log($\mathcal{L}$) \\ [1.5ex]
& [10$^{-6}$ ph cm$^{-2}$ s$^{-1}$] &  \\
\hline
Pre-flare &  0.41$\pm$0.04 & 2.56$\pm$0.10 &  - &  15826.67 & - \\
Flare-3D1 &  0.85$\pm$0.04 & 2.46$\pm$0.05 &  - &  33494.32 & - \\
Q1 &  0.63$\pm$0.05 & 2.48$\pm$0.07 &  - &  20612.02 & - \\
Flare-3D2 & 1.70$\pm$0.14 & 2.51$\pm$0.08 &  - &  7672.06 & - \\
Post-flare &  0.58$\pm$0.02 & 2.52$\pm$0.04 &  - &  94322.96 & - \\
\hline                          
 & & Logparabola\\ 
\hline
Activity & $F_0$ & $\alpha$ & $\beta$ & -log($\mathcal{L}$) & $\Delta$log($\mathcal{L}$) \\ [1.5ex]
& [10$^{-6}$ ph cm$^{-2}$ s$^{-1}$] &  \\
\hline
Pre-flare &  0.39$\pm$0.04 & 2.40$\pm$0.15 &  0.19$\pm$0.12 & 15824.85 & -1.82 \\
Flare-3D1 &  0.83$\pm$0.04 & 2.37$\pm$0.06 &  0.10$\pm$0.04 & 33490.66 & -3.66 \\
Q1 &  0.60$\pm$0.05 & 2.34$\pm$0.10 &  0.15$\pm$0.07 & 20609.42 & -2.60 \\
Flare-3D2 &  1.70$\pm$0.14 & 2.43$\pm$0.11 &  0.09$\pm$0.07 & 7671.28 & -0.78 \\
Post-flare &  0.57$\pm$0.02 & 2.44$\pm$0.05 &  0.08$\pm$0.03 & 94332.77 & 9.81 \\
\hline
\hline 
\end{tabular}
\end{table*}

\bsp	
\label{lastpage}
\end{document}